\documentclass[12pt]{JHEP3}

\usepackage{amsmath,amssymb,amsfonts,amsthm}
\usepackage{bm,bbm}
\usepackage{graphicx}
\usepackage{esint}
\usepackage{color}

\let\normalcolor\relax  % appears to be required when using color package!

\newcommand{\be}{\begin{equation}}
\newcommand{\ee}{\end{equation}}
\newcommand{\ba}{\begin{eqnarray}}
\newcommand{\ea}{\end{eqnarray}}
\def\bs{\begin{subequations}}
\def\es{\end{subequations}}
\def\a{\alpha}
\def\b{\beta}
\def\de{\delta}
\def\g{\gamma}
\def\la{\lambda}

\def\e{\epsilon}

\def\Om{\Omega}
\def\om{\omega}

\def\s{\sigma}
\def\vr{\varrho}
\def\vp{\varphi}

\def\cB{{\cal B}}

\def\cD{{\cal D}}
\def\cE{{\cal E}}
\def\cF{{\cal F}}

\def\cJ{{\cal J}}
\def\cK{{\cal K}}
\def\cL{{\cal L}}
\def\cM{{\cal M}}

\def\cO{{\cal O}}
\def\cP{{\cal P}}

\def\cS{{\cal S}}

\def\cV{{\cal V}}
\def\ds{d_{\rm S}}
\def\dh{d_{\rm H}}
\def\dw{d_{\rm W}}
\def\dl{d_{\rm L}}
\def\p{\partial}
\def\bp{\bar{\partial}}
\def\B{\Box}
\newcommand{\Eq}[1]{(\ref{#1})}
\def\com{\color{magenta}}
\def\cob{\color{blue}}
\newcommand{\oarX}[1]{\href{http://arxiv.org/abs/#1}{{\ttfamily\com #1}}}
\newcommand{\arX}[1]{\href{http://arxiv.org/abs/#1}{{\ttfamily\com arXiv:#1}}}
\newcommand{\doi}[2]{\href{http://dx.doi.org/#1}{\cob #2}}
\newcommand{\boxd}[1]{\boxed{\phantom{\Biggl(}#1\phantom{\Biggl)}}}
\def\lp{\ell_{\rm Pl}}
\def\rme{e}
\def\rmd{d}
\def\rmi{i}
\def\bd{\mathbbm{d}}
\def\x{q}
\def\bx{\bar{q}}

\def\y{{q'}}

\title{Geometry and field theory\\ in multi-fractional spacetime}

\author{Gianluca Calcagni\\
Max Planck Institute for Gravitational Physics (Albert Einstein Institute),\\
Am M\"uhlenberg 1, D-14476 Golm, Germany\\
E-mail: \email{calcagni@aei.mpg.de}}

\date{July 24, 2011}

\abstract{We construct a theory of fields living on continuous geometries with fractional Hausdorff and spectral dimensions, focussing on a flat background analogous to Minkowski spacetime. After reviewing the properties of fractional spaces with fixed dimension, presented in a companion paper, we generalize to a multi-fractional scenario inspired by multi-fractal geometry, where the dimension changes with the scale. This is related to the renormalization group properties of fractional field theories, illustrated by the example of a scalar field. Depending on the symmetries of the Lagrangian, one can define two models. In one of them, the effective dimension flows from 2 in the ultraviolet (UV) and geometry constrains the infrared limit to be four-dimensional. At the UV critical value, the model is rendered power-counting renormalizable. However, this is not the most fundamental regime. Compelling arguments of fractal geometry require an extension of the fractional action measure to complex order. In doing so, we obtain a hierarchy of scales characterizing different geometric regimes. At very small scales, discrete symmetries emerge and the notion of a continuous spacetime begins to blur, until one reaches a fundamental scale and an ultra-microscopic fractal structure. This fine hierarchy of geometries has implications for non-commutative theories and discrete quantum gravity. In the latter case, the present model can be viewed as a top-down realization of a quantum-discrete to classical-continuum transition.
}

\preprint{AEI-2011-037}
\preprint{\doi{10.1007/JHEP01(2012)065}{JHEP01(2012)065} \hspace{8.3cm} \arX{1107.5041}}

\begin{document}

%%%%%%%%%%%%%%%%%%%%%%%%%%%%%%%%%%%%%%%%%%%%%%%%%%%%%%%%%%%%%%%%%%%%%%%%%%%%%%%%%%%%%%%%%%%%%%%%%%%%%%%%%%%%%%%%%%%%%%%%%%%%%%%%%%%%%%%%%%%%%%%%%%%%%%%%%%%%%%%%%%%%%%%%%%%%%%%%%%%%%%%%%%%%%%%%%

\section{Introduction}

With the birth of quantum gravity models, it became clear that ordinary geometry is inadequate to describe the microscopic texture of spacetime. Using the concept of spectral dimension, borrowed from the spectral theory of fractal geometry, it was realized that many scenarios are characterized by a scale-dependent dimension. The spectral dimension $\ds$ is an indicator of the effective number of directions a pointwise probe feels when diffusing in a given ambient spacetime for a short amount of time. For ordinary manifolds, it corresponds to the integer topological dimension $D$, but for fractals this may not be the case. Several quantum gravity or quantum spacetime scenarios such as causal dynamical triangulations (CDT) \cite{AJL4,BeH}, asymptotically safe quantum Einstein gravity \cite{LaR5}, spin foams \cite{Mod08}--%,CaM,
\cite{MPM}, Ho\v{r}ava--Lifshitz gravity \cite{Hor3,SVW1}, $\kappa$-Minkowski non-commutative field theory \cite{Ben08}, and non-local super-renormalizable quantum gravity \cite{Mod11} are defined in ambient spacetimes with $D=4$ dimensions, but the spectral dimension at small scales differs from that value and $\ds<D$ (in many cases, $\ds\sim 2$) in the UV. The change of dimensionality with the scale is another typical property of fractals (more precisely, multi-fractals). Thus, it may be natural to regard all these models as different manifestations of the fact that the application of quantum mechanics to spacetime itself leads, in general, to a fractal geometry.

It is important to establish whether this is only an analogy or not, because the reduction of dimensionality is intimately related to the renormalization properties (meant as ultraviolet finiteness) of quantum gravity. In particular, there seems to be a conspiracy between UV finiteness and a spectral dimension $\ds\sim 2$ at small scales \cite{Car09}--%,Car10,
\cite{fra1}. However, quantum gravity research has not fully exploited the vast field of fractal geometry, and the spectral dimension alone is insufficient both to characterize a physical process as ``fractal'' and to control its geometric properties. First, other notions of dimension can and, actually, must be compared with $\ds$ in order to classify the geometry more precisely. Second, finding the spectral dimension at a given scale is a far cry from having dimensional flow under full control at all scales. Third, there exist other details in the geometric and topological structure of fractals which were never or seldom checked for in quantum gravity. In order to better understand the connection between quantum gravitational physics and fractal geometry, one can take two opposite perspectives. In one, a given model of quantum gravity can be chosen and its fractal properties checked. Unfortunately, due to technical difficulties, in the great majority of the cases the spectral dimension is the only computable fractal indicator. In the other perspective, one could start from fractal geometry itself and attempt to construct a theory of quantum gravity with the desired properties (dimensional flow from 2 to 4, UV finiteness, and so on).

The second approach was advocated and outlined in its main qualitative features in \cite{fra1}--%,fra2,
\cite{fra3}. To have a much closer contact with fractal geometry, a more rigorous programme has been initiated in \cite{frc1,fra4}. Exploiting the characterization of many fractals as systems governed by a fractional differential structure, we constructed a continuous space with Euclidean signature and whose Hausdorff and spectral dimensions are non-integer. This empty space, called fractional Euclidean space and denoted by $\cE_\a^D$, has a notion of distance, volume and dimension. In particular, it has topological dimension $D$ and Hausdorff and spectral dimensions $\dh=\ds=D\a$ (for non-anomalous diffusion), where $0<\a\leq 1$ is a fixed real parameter. Furthermore, it is endowed with symmetries, although not the usual rotation and translation group. In practice, on fractional Euclidean space one can ask the same questions and perform all the operations allowed in ordinary Euclidean space, but with a different calculus.

In this paper, we build upon the results of \cite{frc1} and carry out the agenda spelled out therein in greater detail. The idea is (i) to extend $\cE^D_\a$ to a space with Lorentzian signature, (ii) realize dimensional flow via tools of multi-fractal geometry, (iii) discuss an example of field theory and its renormalization properties, and (iv) generalize to models with even more realistic fractal properties. It will turn out that step (ii), which was not really made in \cite{fra1}--%,fra2,
\cite{fra3}, contains some pleasant news: The renormalization group (RG) flow and the multi-fractal construction are one and the same entity, but described with two different languages. This may be unsurprising, since the RG flow is based upon a scale hierarchy just like multi-fractals. What is perhaps surprising is the quantitative match: once symmetries are given, purely fractal geometric considerations lead to the same total action prescribed by an almost-traditional field theory analysis. On top of that, geometric requirements can fix the effective dimension of spacetime in the infrared (IR) to $\dh=\ds\sim 4$, provided the dimension in the UV is 2. Also, step (iv) will be crucial to probe scales even smaller than those at which the RG flow takes place, and will allow us to make contact with some features discovered numerically in causal dynamical triangulations. The overall physical picture has been shortly presented in \cite{fra4}. Gravity is not included, yet, but fractional Minkowski spacetime will be sufficient to illustrate the basic features of the proposal. We leave the fractional extension of general relativity to another publication. Below is the plan of the paper.
\begin{itemize}
\item {\it Section \ref{fms}.} Following the Euclidean construction of \cite{frc1}, we define a fractional ambient space with Lorentzian signature and describe its geometry (section \ref{preli}) and its Hausdorff and spectral dimensions (section \ref{hsw}). The choice of coordinate presentation and the role of the boundary, which were not discussed in \cite{frc1}, are here treated in detail. The fractional generalization of Lorentz transformations is presented in section \ref{lt}. 
\item {\it Section \ref{multis}.} Spacetimes with a multi-fractal structure are introduced. Exploiting the lore of multi-fractal geometry, the measure in the action is argued to be a linear superposition of contributions with fixed dimensionality (sections \ref{mul2} and \ref{mul3}). The dimension of spacetime changes with the scale, but not arbitrarily: the dimensions in the UV and in the IR are, in fact, intimately related (section \ref{dims}).
In particular, we shall analyze the role of two and four dimensions at, respectively, small and large scales. Observational constraints on the flow of the dimension near the infrared limit are discussed in section \ref{obse}.
\item {\it Section \ref{st}.} An example of classical fractional field theory is provided by a real scalar field $\phi$, whose ultraviolet finiteness can be easily probed via a standard power-counting argument (section \ref{pc}). Generic classical fractional actions are constructed in section \ref{afft}, where the equation of motion of $\phi$ is also derived. The choice of integer or fractional Lorentz symmetries in the Lagrangian density lead to two independent models, respectively, the integer symmetry scenario (section \ref{friss}) and the fractional symmetry scenario (section \ref{frass}). The Green function inverting the kinetic operator is calculated in section \ref{pro}. The power-counting renormalizability of the models is discussed at several points in the section, in particular in section \ref{sdd}.
\item {\it Section \ref{pola}.} To get in closer touch with fractal geometry, fractional theories are extended to the case where the measure is a linear superposition of fractional measures of complex order. Combining this superposition in a real quantity, one obtains a measure with logarithmic oscillations (section \ref{rcfo}). The average of the measure over a log-period corresponds to the real-order fractional measure. The oscillations are due to a discrete symmetry (sections \ref{lodsi} and \ref{logos}), and require a redefinition of the Hausdorff and spectral dimension in line with the definitions employed in fractal geometry (section \ref{osdi}).
\item {\it Section \ref{ra}.} After summarizing the physical picture in section \ref{hier}, we outline a research agenda focussed on the quantum theory, the inclusion of gravity and cosmological applications (section \ref{opis}). Section \ref{appli} is devoted to connections with doubly special relativity, non-commutative spacetimes and quantum gravity approaches. Multi-fractional field theories can be regarded either as stand-alone models of quantum gravity or as effective descriptions of other theories in certain regimes. In the bulk of the paper we assume the first attitude. Considering instead the second case, we advance possible applications of multi-fractional geometries to other, independent models of quantum gravity in section \ref{fuzz}.
\end{itemize}

%%%%%%%%%%%%%%%%%%%%%%%%%%%%%%%%%%%%%%%%%%%%%%%%%%%%%%%%%%%%%%%%%%%%%%%%%%%%%%%%

\subsection{Comparison with early proposals}\label{compa}

All the present material is novel with possibly two minor exceptions. The first is section \ref{fms}, which makes heavy use of the results of \cite{frc1}. These, however, are here immediately extended to an ambient spacetime with Lorentzian signature. The second exception is the very idea that spacetime be fractal. Before embarking ourselves in the construction of multi-fractional spaces with dimensional flow, it is useful to draw an exhaustive comparison between our proposal and other spacetime models in non-integer dimension which appeared in the early literature. This comparison, which was premature for the abstract Euclidean fractional space of fixed dimensionaly presented in \cite{frc1}, can give the reader a bird's eye view of the state of the art of fractal spacetimes and the status of our theory within.

Most of previous ``fractal'' field theoretical proposals assumed spacetime to have a non-integer but constant, non-dynamical dimensionality, thus fixing the attention to $4-\e$ dimensions with $0<\e\ll1$ (low-lacunarity regime). Here we mainly focus on the rather few papers attempting to realize the texture of spacetime via a mathematical fractal or fractal-inspired construction.
\begin{itemize}
\item Quantum field theories in non-integer dimension were considered in \cite{Wil73} via a simple analytically continued integration. They were regarded as abstract models created for the purpose of shedding some light into four-dimensional field theory, and the non-integer dimension formally appearing in manipulations was not associated with a fundamental modification of spacetime geometry. The hope was to unravel properties of the renormalization group flow which do not depend on dimensionality. At variance with these dimensionally-continued field theories, the philosophy of later publications \cite{Sti77}--%,Svo87,Ey89a,
\cite{Ey89b} was that spacetime has a non-integer but fixed, non-dynamical dimensionality. The attention was focussed on $\dh=4-\e$ for obvious empirical reasons. In other words, these models share many aspects with dimensional regularization but the parameter $\e$ is taken to be physical and non-vanishing, albeit small. (i) In \cite{Sti77} a mathematical justification to dimensional continuation was given, with applications to statistical mechanics and field theory. An axiomatic description of a metric space with non-integer dimension was proposed. Neither is this a vector space nor is it embedded in a vector space of integer dimension $D$, so it can hardly be compared with our framework. The lack of a manageable set of natural coordinates did not allow the author of \cite{Sti77} to explore the model much, and objects such as a ``Pythagorean'' coordinate distance, volumes, and the symmetry group of the space were not investigated. Also, this construction becomes more and more complicated with the increase of the topological dimension \cite{PSt}. Yet, it was possible to define an invertible Fourier transform (an open problem in fractal geometry, but solved in fractional spaces \cite{frc1}) and a natural Laplacian operator. (ii) In \cite{Svo87}, particle physics was defined directly on sets with general Borel measure of
 fixed dimensionality, i.e., a Lebesgue--Stieltjes measure with possibly fractal support. Renormalization properties in electrodynamics were considered for a general measure with low lacunarity, $\dh=4-\e$. Convergence of the Feynman diagrams is better than in four dimensions, as it was checked by looking at their superficial degree of divergence. This was presented as a ``new regularization method,'' meaning that physical applications should be sought out only at low lacunarity. (iii) In \cite{Ey89a,Ey89b}, scalar-field theory in Euclidean signature was constructed on low-lacunarity Sierpinski carpets with Hausdorff dimension $\dh=4-\e$. Because of the explicit fractal construction, the scaling property of the measure is discrete and the system displays a certain symmetry, called discrete scale invariance, which we shall discuss in section \ref{lodsi}. The propagator on a fractal lattice was computed using techniques which would have been later developed for the spectral theory on fractals \cite{Kig01}. (iv) The interesting UV properties of field theories in fractional spacetimes and the breaking of parity and time reversal therein, all topics we shall amply discuss, were also appreciated in \cite{Gol08}, where it was noticed that fractional and curvature effects are similar in the limit of almost-integer dimension. Even in this case, attention was limited to $\dh=4-\e$.\footnote{Breaking of time reversal and non-conservation of probability \cite{fra2} in fractional quantum mechanics were discussed also for fractional generalizations of the Schr\"odinger equation \cite{PSt,Nab04}--%,WX,DX,OMA,
\cite{Iom09}.}
\item In contrast with all these approaches, field theories on a genuinely and deeply ``anomalous'' spacetime ($\dh$ much different from 4) have received less attention, despite their promising applications in modern cosmology and quantum gravity. In \cite{FeH}, it was observed that the fine-scale structure of a quantum mechanical particle path is very irregular and described by a nowhere-differentiable curve.\footnote{The Hausdorff dimension of the path is 2 in a classical spacetime \cite{AbW} and smaller than 2 in quantum spacetimes with a minimal length \cite{NN}. In the latter case, $\dh$ can be even negative, corresponding to a Planckian regime where quantum fluctuations of the spacetime texture are large and the particle path is an empty set.} Ord drew inspiration from this fact to propose a model of quantum mechanics in fractal spacetime \cite{Ord83}. The idea that spacetime be a multi-dimensional fractal was also hinted at in \cite{NSc}. Physical descriptions and implications of fractal spacetimes have begun to be focussed on a geometrical perspective in \cite{fra1}--%,fra2,
\cite{fra3} within a Lebesgue--Stieltjes approach. Away from the low-lacunarity regime, which misses all the potentialities of fractal spacetimes, the Lebesgue--Stieltjes formalism \cite{fra1,fra3} can say little unless one specializes to specific measures. General absolutely continuous measures away from $\dh\sim 4$ were discussed in \cite{fra2}, but their special status as naive ``fractal'' measures did not allow us to make much progress with an adequate level of rigorousness.
\item In \cite{fra1,fra2}, fractional calculus served as a motivation for the introduction of general Lebesgue--Stieltjes models of spacetime, but it was not used in its full power. The framework of these papers shares only the main qualitative characteristic with the present one (namely, anomalous scaling of the measure and better renormalization properties of field theories) and it does not possess the richness of physical implications we can appreciate here. In particular, while in the models of \cite{fra1}--%,fra2,
\cite{fra3} it is not clear how to construct a rigorous definition of an invertible unitary transform between configuration and momentum space, this is can be done for fractional spaces \cite{frc3} thanks to the factorization of the coordinates in the measure.
\item Through the notion of distance and the calculations of ball volumes, we saw in \cite{frc1} and will see in section \ref{fms} that geometric coordinates $\x$ can be interpreted as a coordinate system intrinsic to the fractional space, and the mapping $x\to \x(x)$ relates the embedding (or extrinsic) viewpoint to a geometric (or intrinsic) viewpoint. These two equivalent pictures were conjectured, without giving details, at the end of \cite{Svo87}. In \cite{fra1,fra2}, a study of the deformed Poincar\'e algebra of a Lebesgue--Stieltjes model with arbitrary measure reached the same conclusion, naming the embedding and geometric pictures, respectively, conservative (as a system made of two dissipative parts) and dissipative (as a system dissipating energy-momentum in a bulk).
\item When this article was being finalized, the author became aware of an independent proposal for a fractal spacetime \cite{Not93}--%,Not97,
\cite{Not08}. While ordinary spacetime is described by smooth differentiable manifolds, abandoning the requirement of differentiability naturally leads to a fractal geometry characterized by non-absolute scales which can only be measured relatively to one another. Starting point, motivation, terminology and tools greatly differ from ours. Central is the so-called principle of scale relativity. Motivated by fractal geometry, coordinate frames are made explicitly dependent on the scale and, crucially, scales must transform according to certain very natural laws generalizing ordinary contractions/dilations. This principle can be applied to a number of systems, not only in physics, leading to a severe modification of our perception of Nature. In the realm of physics, no formal theory has been constructed upon scale relativity, but some of the consequences of this principle presents intriguing similarities with our fractional approach, including a breaking of parity symmetry and an almost obvious dimensional flow. A careful comparison between fractional spacetime formalism and the scale relativity proposal may be of mutual benefit. For instance, in the external scale picture of section \ref{mul2} we gave a few $\a(\ell)$ profiles as toy examples realizing a running from some critical value $\a_*$ to the integer charge $\a=1$. The external scale $\ell$ is the very same continuous scale labelling Nottale's fractal coordinates, and by the simple scaling arguments of scale relativity one can prescribe the scale dependence of the fractal dimension. In our notation, these fractal coordinates are indeed the geometric variables $q^\mu(\ell)=q^\mu[\a(\ell)]$ and eq.~(14) of \cite{Not08} reads
\be
\a(\ell)=1+\frac{\a_*-1}{1+(\ell/\ell_*)^{\a_*-1}}\,.
\ee
When $\ell\gg\ell_*$, $\a(\ell)\to 1$, while at small scales $\a(\ell)\to\a_*$. On one hand, implementation of scale relativity arguments could fix some loose points of fractional spacetime models and sharpen the overall physical interpretation. On the other hand, fractional models could provide the missing theoretical framework wherein to embed the scale relativity principle.\footnote{Such a theoretical framework was also proposed in \cite{BA07}, where non-differentiable manifolds were defined in a rather abstract fashion. Physical applications and consequences of these ``fractal manifolds,'' and the connection with standard fractal-geometry tools, are presently unclear to us.}
\item The dependence on the scale (or resolution) can be implemented at the level of fields rather than coordinates. Such is the philosophy of wavelet field theory \cite{Alt06,Alt07} which, not surprisingly, has better UV properties than ordinary field theory. We believe that also this approach converges to the same physics of fractional and fractal models.
\item Dimensional flow may be realized also in non-fractal scenarios. (i) Field theory actions with exotic measures were the subject of \cite{HW}. There, spacetime was described by a set of ``continuous'' coordinates and its dimension formally constrained by a variational principle. Despite the fractal-inspired motivation, there is no obvious point of contact with fractal geometric scenarios. The cosmological ``decrumpling'' model of \cite{MaN} also considered dimension to be a dynamical field. (ii) Another unrelated appearance of an exotic Hausdorff dimension in field theory is in four-dimensional gravity with a quantum conformal factor \cite{AMM}. As a trace anomaly effect, the conformally invariant IR fixed point of this model is associated with an anomalous dimension greater than the topological one. (iii) The universe described in \cite{ADFLS}--%,Anc2,
\cite{MSt} has a  crystal-like ``layered'' structure governed by a hierarchy of scales along the topological directions. Due to the fact that transitions from one dimension to another are rather sharp, away from the transition points these models strongly resemble dimensionally regularized spacetimes (see also \cite{Shi10}). As we have just seen, and contrary to what advertized in \cite{ADFLS,Anc2}, the paradigm that the effective dimensionality of spacetime depends on the probed scale is far from being new. At any rate, fractional spacetime theory is quite different from the crystal-world proposal except at the IR fixed point. In particular, in our framework there is no geometric reason why events in a $\dh=2$ regime should be planar (two-dimensional fractals may not be embedded in a plane: Brownian motion is an example), and gravitational waves should be produced even in $\dh=3$ dimensions, mainly for the reasons advanced in \cite{SVW0}.\footnote{To this list of non-fractal references, we should also add the effective field-theory fractional equations of motion of \cite{Zav00}--%,LiM,Lim05,
\cite{Her07}. These models seem to be purely mathematical.}
\item The measured value of the dimension of spacetime may be slightly smaller than four due to quantum fluctuations and to the intrinsic finite resolution of experiments. Then, the infinitesimal covering in the definition of Hausdorff dimension cannot be physically realized in the real world. This effect was studied in \cite{SvZ}, where no assumption was made on the true dimensionality of spacetime, and it is unrelated from field theories living in fractal geometry. Heuristic finite-resolution effects of quantum fluctuations have been also considered in \cite{Maz08}. The change of dimension in the multi-fractal flow considered here is far more dramatic than this type of corrections, which can be safely ignored.
\end{itemize}
To the best of our knowledge, the following features have been developed here for the first time: (i) the construction of a fractal-like structure in Lorentzian signature; (ii) an explicit realization of dimensional flow in explicitly multi-fractal spacetime structures; (iii) the construction of a system whose UV and IR dimensionality are deeply related to each other rather than being fixed phenomenologically;  (iv) a detailed clarification of what we mean by ``fractal spacetime'' and the fractal interpretation of fractional spacetime models; (v) a description of the symmetries underlying fractional spacetimes under the perspective of fractal geometry, and their consequent identification with the isometry group in the language of field theory; (vi) the construction and physical interpretation of fractional spacetimes with oscillatory measures and the associated emergence of discrete scales. Point (iii) was vaguely foreseen in \cite{fra2}, (i) and (v) were initiated in \cite{fra1,fra2}; the first part of (iv) was completed in \cite{frc1}. Point (vi) also opens up a novel connection with non-commutative spacetimes, as we shall see below.

Comparing with the proposal in \cite{fra1}--%,fra2,
\cite{fra3}, there are many differences and an appreciable amount of novelty. There, the simplified setting of an absolutely continuous measure captured the main qualitative features of the central idea, but its level of rigor was not satisfactory. On the other hand, here we use the formalism of fractional calculus, through which we control all the elements of the proposal more strictly and in greater detail. The results of \cite{fra1,fra2} partially lie on a Weyl-type integral, a special case of fractional integral. Whenever a point of contact is possible, we shall compare the present model with those results and clarify how issues of the latter are solved here. 

We stress that, while the great majority of the works cited above considered dimensional flow only as a general concept or paradigm, our aim here and in \cite{frc1} is to construct a hands-on theory of spacetime and fields where one can have as much control as possible. The concreteness of the model and its close contact with the lore of fractal geometry will allow us to go well beyond the isolated feature of anomalous scaling.

%%%%%%%%%%%%%%%%%%%%%%%%%%%%%%%%%%%%%%%%%%%%%%%%%%%%%%%%%%%%%%%%%%%%%%%%%%%%%%%%%%%%%%%%%%%%%%%%%%%%%%%%%%%%%%%%%%%%%%%%%%%%%%%%%%%%%%%%%%%%%%%%%%%%%%%%%%%%%%%%%%%%%%%%%%%%%%%%%%%%%%%%%%%%%%%%%%%%%%%%%%%%%%%%%%%%%%%%%%%%%%%%%%%%%%%%%%%%%%%%%%

\section{Fractional Minkowski spacetime}\label{fms}

The extension of fractal geometry to an ambient spacetime with Lorentzian signature is, to the best of our knowledge, a topic virtually untouched in the literature \cite{fra2}. Nevertheless, after having defined fractional geometry in Euclidean space \cite{frc1}, we can move to a Lorentzian spacetime straightforwardly. Curvature is not considered.

%%%%%%%%%%%%%%%%%%%%%%%%%%%%%%%%%%%%%%%%%%%%%%%%%%%%%%%%%%%%%%%%%%%%%%%%%%%%%%%%%%%%%%%%%%%%%%%%

\subsection{Definition}\label{preli}

We define \emph{fractional Minkowski spacetime} $\cM_\a^D$ of order $\a$ as a $D$-dimensional embedding Minkowski spacetime $M^D$ endowed with a set of rules ${\rm Calc}^\a=\{\p^\a, I^\a,\bd\}$ of integro-differential calculus (symbols denote derivative, integration and external differential), a measure $\vr_\a$ with a given support, a natural norm $\| \cdot \|$, and a Laplacian $\cK$:
\be
\boxd{\cM_\a^D = (M^D,\,{\rm Calc}^\a,\,\vr_\a,\,\| \cdot \|,\,\cK)\,.}
\ee

\subsubsection{Embedding and calculus}\label{emca}

The embedding $M^D$ is Minkowski spacetime in $D$ topological dimensions, with ``mostly plus'' signature $({-},{+},\cdots,{+})$. The embedding coordinates $x^\mu$ are labeled by Greek indices $\mu,\nu,\dots$ running from $0$ to $D-1$. The time direction will be sometimes denoted as $t=x^0$.

The embedding is, actually, only a choice of metric once a differential and metric structures are defined. The rules ${\rm Calc}^\a$ of differential geometry are given by fractional calculus \cite{Pod99,KST}. This is reviewed in \cite[section 2]{frc1}; here we recall the main definitions in one dimension. Given a real coordinate variable $x$ defined on an interval $[x_0,x_1]$, the left fractional integral of order $\a$ of a function $f(x)$ is
\be\label{I}
(I^\a f)(x) :=\frac{1}{\Gamma(\a)}\int_{x_0}^{x} \frac{\rmd x'}{(x-x')^{1-\a}} f(x')\,,
\ee
where
\be\label{0a1}
0<\a\leq 1\,.
\ee
Similarly, the right fractional integral of order $\a$ is
\be\label{baI}
(\bar{I}^\a f)(x) :=\frac{1}{\Gamma(\a)}\int_{x}^{x_1} \frac{\rmd x'}{(x'-x)^{1-\a}} f(x')\,,
\ee
where integration now is from $x$ to the end of the interval. Also, the left and right Caputo derivatives of order $\a$ are defined as
\ba
(\p^\a f)(x) &:=& (I^{1-\a}\p f)(x)\nonumber\\
&=&\frac{1}{\Gamma(1-\a)}\int_{x_0}^{x} \frac{\rmd x'}{(x-x')^{\a}} \p_{x'}f(x')\,,\label{pan}\\
(\bp^\a f)(x) &:=& (\bar{I}^{1-\a}\p f)(x)\nonumber\\
&=&-\frac{1}{\Gamma(1-\a)}\int_{x}^{x_1} \frac{\rmd x'}{(x'-x)^{\a}} \p_{x'}f(x')\,,\label{bpan}
\ea
where $\p$ is the ordinary first-order partial derivative. Sometimes we will indicate the integration terminals explicitly as subscripts, $\p^\a=\p^\a_{x_0,x}$ and $\bp^\a=\bp^\a_{x,x_1}$. Under the transformation
\ba\label{pari1}
x\to x_0+x_1-x\,,
\ea
the left operators are mapped into right operators:
\be\label{pari2}
(\bar{I}^\a f)(x)=(I^\a F)(x_0+x_1-x)\,,\qquad (\bp^\a f)(x)=(\p^\a F)(x_0+x_1-x)\,,
\ee
where $F(x):=f(x_0+x_1-x)$.

One can extend $\a$ to other ranges and define other differential operators (e.g., the Riemann--Liouville derivative). Different sets of fractional operators can correspond to inequivalent fractional spacetimes. In \cite{frc1} we have justified the use of the Caputo derivative via several arguments; the main one is that differential geometry and tensor calculus are considerably simplified. 
 
When $x_0=-\infty$, the left operators are called Liouville differintegrals, while for $x_1=+\infty$ they are called Weyl differintegrals; they will be all denoted by a subscript $\infty$ to the left of the symbol. When regarded as an approximation in the limit $t\gg t_0$, the Liouville operator is employed in mechanics to describe ``steady state'' systems, that is, systems which evolved well after the initial transient phase at $t_0$. To get the Weyl differintegral from the Liouville differintegral, it is sufficient to set $x_0=-x_1$ in \Eq{pari2} and then take the limit $x_0\to-\infty$:
\be\label{pari3}
({}_{\infty}\bar{I}^\a f)(x)=({}_{\infty}I^\a F)(-x)\,,\qquad ({}_{\infty}\bp^\a f)(x)=({}_{\infty}\p^\a F)(-x)\,,
\ee
where $F(x):=f(-x)$.

Some examples of fractional derivatives and integrals are the following. The Caputo, Liouville and Weyl derivatives of a constant are zero,
\be\label{pro1}
\p^\a 1=\bp^\a 1={}_{\infty}\p^\a 1={}_{\infty}\bp^\a 1 = 0\,,
\ee
while the left derivatives of a power law are
\ba
\p^\a (x-x_0)^\b &=& \frac{\Gamma(\b+1)}{\Gamma(\b-\a+1)}(x-x_0)^{\b-\a}\,,\qquad \b\neq 0\,,\label{speci}\\
{}_{\infty}\p^\a (x-x_*)^\b &=& (-1)^{-\a}\frac{\Gamma(\b+1)}{\Gamma(\b+1-\a)}\frac{\sin(\pi\b)}{\sin[\pi(\b-\a)]} (x-x_*)^{\b-\a}\,,\label{218}
\ea
for any $x_*$. The second expression is ill defined for $\b=\a$. Otherwise, it is real under certain conditions on the values of $\a$ and $\b$ and the sign of $x-x_*$; consistently, it vanished for $\b=0$. The $\a$-th order left integrals of power laws are given by the analytic continuation of the above formul\ae\ for $\a\to-\a$ for any $\b$:
\ba
I^\a (x-x_0)^\b &=&\frac{\Gamma(\b+1)}{\Gamma(\b+\a+1)}(x-x_0)^{\b+\a}\,,\label{speci2}\\
{}_{\infty}I^\a (x-x_*)^\b &=& \frac{(-1)^{-\a}\Gamma(-\a-\b)}{\Gamma(-\b)}(x-x_*)^{\b+\a}\,.\label{218I}
\ea
The eigenfunctions of the Liouville derivative operator are exponentials,
\be\label{espo}
{}_{\infty}\p^\a \rme^{\la x} = \la^\a\rme^{\la x}\,,
\ee
but those of Caputo derivatives with finite $x_0$ are Mittag-Leffler functions \cite{HMS}:
\be\label{mile}
\p^\a E_\a[\la(x-x_0)^\a] = \la E_\a[\la(x-x_0)^\a]\,.
\ee
Analogous formul\ae\ can be obtained for the right derivative by making use of eqs.~\Eq{pari2} and \Eq{pari3}. For example, the right version of eqs.~\Eq{speci}, \Eq{218} and \Eq{espo} are
\ba
\bp^\a (x_1-x)^\b &=&\frac{\Gamma(\b+1)}{\Gamma(\b-\a+1)}(x_1-x)^{\b-\a}\,,\qquad \b\neq 0\,,\label{specir}\\
{}_{\infty}\bp^\a (x-x_*)^\b &=&\frac{\Gamma(\b+1)}{\Gamma(\b+1-\a)}\frac{\sin(\pi\b)}{\sin[\pi(\b-\a)]} (x-x_*)^{\b-\a}\,,\label{218r}\\
{}_{\infty}\bp^\a \rme^{\la x} &=& (-\la)^\a\rme^{\la x}\,.\label{espo2}
\ea

\subsubsection{Measure}

The measure and the integration range of the action can be fixed once and for all by some simple arguments, which we develop in one dimension. 

The action will be defined via the right integral $\bar{I}^\a$. To extend the integration range as much as possible while keeping the fractional measure well defined, the extremum $x$ must remain finite while taking the limit $x_1\to+\infty$. The fractional space has then a boundary at some finite $x$. Without any loss of generality, we can set $x=0$ in the action integral. In fact, a translation $x\to y=x+x_*$ changes the coordinate \emph{presentation} of fractional operators, but it does not change the physics. For instance, the one-dimensional Weyl integral of a function becomes
\ba
{}_{\infty}\bar{I}^\a f &=& \frac{1}{\Gamma(\a)}\int_{x}^{+\infty}\frac{\rmd x'}{(x'-x)^{1-\a}}\,f(x')\nonumber\\
               &=& \frac{1}{\Gamma(\a)}\int_{y-x_*}^{+\infty}\frac{\rmd x'}{(x'+x_*-y)^{1-\a}}\,f(x')\nonumber\\
               &\ \stackrel{y'=x'+x_*-y}{=}\ & \frac{1}{\Gamma(\a)}\int_{0}^{+\infty}\frac{\rmd y'}{{y'}^{1-\a}}\,f(y'+y-x_*)\,.\nonumber
\ea
In the new reference frame, the coordinate dependence of $f$ and of the boundary have changed, but this only modifies the way the fractional one-dimensional space is embedded in $\mathbb{R}$. As an embedding coordinate change, this operation is always possible and one can further set $y=x_*$. For the Liouville integral the measure range is in the negative semi-axis, but this would be yet another change of presentation; so we can just pick the Weyl integral. The combination ${}_{\infty}I^\a+{}_{\infty}\bar I^\a$ would hide the presence of the singularity at $x=0$ by a fictitious integration on the whole line. 

In $D$ topological dimensions, each embedding direction $x^\mu$ is associated with a ``fractional charge'' $\a_{\mu}$ (the subscript $\mu$ is not a vector index). The simplest case is of an isotropic fractional spacetime, where all $\a_{\mu}$ are equal \cite{fra1,fra2,frc1}. In particular, the time direction $t=x^0$ is on an equal footing with spatial coordinates, $\a_0=\a$. In another configuration worth mentioning for its applications in non-commutative spacetimes (\cite{fra4,ACOS} and section \ref{ncft}), time is an integer direction, $\a_0=1$, in which case the spatial part of the measure carries the whole effect of dimensional flow. Keeping $\a_0$ general, the action in fractional Minkowski space is
\be\label{act0}
\boxd{S_\a = \int_0^{+\infty} \rmd^D\x\,\cL=\int_0^{+\infty}\rmd\vr_{\a}(x)\, \cL=\int_0^{+\infty}\rmd^Dx\, v_{\a}(x)\, \cL\,,}
\ee
where $\cL$ is a Lagrangian density, the integral is
\ba
\int_0^{+\infty} \rmd^D\x &=& \prod_{\mu=0}^{D-1}\int_0^{+\infty} \rmd\x^\mu\\
                          &=& \frac{1}{\Gamma(\a_0)}\int_0^{+\infty} \frac{\rmd t}{t^{1-\a_0}}\prod_{\mu=1}^{D-1}\frac{1}{\Gamma(\a)}\int_0^{+\infty} \frac{\rmd x}{x^{1-\a}}\,,
\ea
and the measure along each direction is
\be\label{finfra}
\boxd{\x^\mu:=\vr_\a(x^\mu):=\frac{(x^\mu)^\a}{\Gamma(\a+1)}\,, \qquad x^\mu\geq 0\,.}
\ee
To keep notation light, we wrote $\a$ instead of $\a_\mu$ in eqs.~\Eq{act0} and \Eq{finfra}, with the understanding that the time coordinate may scale differently.
%Generally, the subscript $\a$ in \Eq{act0} labels a set of $D$ fractional charges $\{\a_{\mu}\}$.

While the measure of Minkowski spacetime is the ordinary Lebesgue measure $\vr(x)=\otimes_\mu x^\mu$, $\cM_\a^D$ is equipped with the Lebesgue--Stieltjes measure
\be
\vr_\a(x):=\vr_{\a_0}(t)\bigotimes_{\mu =1}^{D-1} \vr_\a(x^\mu)\,,
\ee
whose scaling property is
\be\label{scavr2}
\vr_\a(\la x) = \la^{\a_0+(D-1)\a}\vr(x)\,,\qquad \la>0\,.
\ee

In eq.~\Eq{finfra} we have written the measure in the $\mu$ direction as a coordinate $\x^\mu$, dubbed geometric or fractional in \cite{frc1}. Depending on the differential calculus associated with the action integral, $\{\x^\mu\}$ can be regarded as the natural coordinate system spanning fractional spacetime \cite[section 3.3]{frc1}. For instance, the fractional fundamental theorem of calculus states that the Caputo derivative is the left inverse of the fractional integral defined on the same sector (left or right) and the same interval \cite[section 2.3.3]{frc1}. If we based the calculus on $\cM_\a^D$ only on one sector, the right integral would be associated with the Weyl derivative
\bs\label{dabo}\be\label{da1}
\boxd{\bp^\a_\mu:={}_\infty\bp^\a_{x^\mu}=\frac{\bp^\a}{\bp^\a\x^\mu}:=-\frac{1}{\Gamma(1-\a)}\int_{x^\mu}^{+\infty}\frac{\rmd {x'}^\mu}{({x'}^\mu-x^\mu)^\a}\p_{\mu}\qquad \textrm{(no sum over $\mu$)}\,.}
\ee
Because of eq.~\Eq{218r}, eq.~\Eq{finfra} does not define a geometric coordinate with respect to this derivative, meaning that $\bp^\a_\mu \x^\nu\neq \de_\mu^\nu$ (actually, it is ill defined). However, $\x^\mu$ is the geometric coordinate associated with the left derivative $\p^\a$ with terminal $x_0=0$,
\be\label{da2}
\boxd{\p^\a_\mu:= \p^\a_{0,x^\mu}=\frac{\p^\a}{\p^\a\x^\mu}:=\frac{1}{\Gamma(1-\a)}\int_0^{x^\mu}\frac{\rmd {x'}^\mu}{(x^\mu-{x'}^\mu)^\a}\p_{\mu}\qquad \textrm{(no sum over $\mu$)}\,.}
\ee\es
By virtue of eqs.~\Eq{pro1} and \Eq{speci}, 
\be
\p^\a_\mu\x^\nu=\delta_\mu^\nu\,.
\ee
Since both derivatives \Eq{da1} and \Eq{da2} will appear in the same theory (one in the action, the other in the equations of motion \cite[section 2.3.5]{frc1}), we are at liberty of choosing either in the action. In order to have a well-defined geometric coordinate system, the natural fractional differential is constructed via the left derivative (\cite[sections 2.4, 3.3]{frc1} and references therein),
\be\label{diffmu}
\boxd{\bd: = \bd\x^\mu\, \p^\a_\mu\,,\qquad \bd\x^\mu=(\rmd x^\mu)^\a\,,\qquad [\bd]=0\,,}
\ee
with $\x^\mu$ and $\p^\a_\mu$ given by eqs.~\Eq{finfra} and \Eq{da2}. The symbol $\p^\a_\mu$ may be actually regarded both as the partial fractional derivative with respect to $x^\mu$ and as the one with respect to $\x^\mu$, $\p^\a_\mu = \bd/\bd\x^\mu$. Therefore, we can define a geometric integral \cite{frc1,Tar12}
\be\label{geomi2}
\fint_0^\x := \frac{1}{\Gamma(\a)}\int_{0}^{x} \left(\frac{\rmd x'}{x'}\right)^{1-\a}\,,
\ee
so that the action \Eq{act0} can be fully recast in geometric coordinate formalism:
\be\label{act1}
S_\a = \fint_0^{+\infty}\bd^{D}\x\,\cL\,.
\ee

\subsubsection{Boundary}\label{boun}

Fractional spacetime $\cM_\a^D$ corresponds to the first orthant $\x^\mu>0$. Due to the presence of the boundary, translation and rotation invariance are globally broken, since these transformations would change the aspect of $\cM_\a^D$ ``looking from far away.'' However, as embedding transformations they do not affect the local symmetries of $\cM_\a^D$, which can still be investigated.

This gives a sharp physical interpretation for the model defined by eqs.~\Eq{finfra}, \Eq{dabo} and \Eq{act1}. Despite being an integro-differential operator, the fractional derivative is an intrinsically local operator in the sense of fractional differential geometry \cite{BeA3,BeA4} (see also \cite[section 3.1]{frc1}). If $\cM_\a^D$ possesses local symmetries and boundary effects are negligible well ``inside'' fractional spacetime, then the structures of $\fint$, $\bp^\a$ and $\p^\a$ are mutually compatible. To show this, consider the one-dimensional interval $[x_0,x_1]$ and the behaviour of the differintegrals $\p^\a_{x_0,x}$ and $\bp^\a_{x,x_1}$, $\a\neq 0$, far from the terminals (see \cite{Pod99}, sections 2.7.5 and 2.7.6, for left operators). The terminals are at $x=x_0$ and $x=x_1$. For any finite $x_0$, the asymptotic behaviour of $\p^\a_{x_0,x}$ away from $x_0$ can be obtained by taking either $x\gg 1$ at fixed $x_0$ or $-x_0\gg 1$ at fixed $x$:
\be\label{lim0}
\p^\a_{x_0,x}\ \stackrel{|x|\gg |x_0|}{\sim}\ \p^\a_{0,x}\,,\qquad {\rm or}\qquad \p^\a_{x_0,x}\ \stackrel{|x_0|\gg |x|}{\sim}\ \p^\a_{x-x_0,x}\,.
\ee
Actually, the second asymptotic limit is tantamount to sending $x_0\to-\infty$ with $x$ finite, and corresponds to the Liouville derivative. In our case, the lower terminal (or boundary) is finite and equal to $x=x_0=0$, and the first limit is the natural one for the fractional derivative inside the action. On the other hand, away from a finite $x_1$ the right differintegrals behave as
\be\label{lim1}
\bp^\a_{x,x_1}\ \stackrel{|x|\gg |x_1|}{\sim}\ \bp^\a_{x,0}\,,\qquad {\rm or}\qquad \bp^\a_{x,x_1}\ \stackrel{|x_1|\gg |x|}{\sim}\ \p^\a_{x,x+x_1}\,.
\ee
For us, the upper terminal is $x=x_1=+\infty$, and the second expression coincides with the Weyl differintegral in this limit. Thus, \emph{the theory defined by eqs.~\Eq{finfra}, \Eq{dabo} and \Eq{act1} can be physically interpreted as a local model of fractional spacetime where boundary effects are negligible}.

What happens near the boundary? Consider now a $D=1$ model where $x_1$ is finite:
\be\nonumber
S_\a' = \fint_{0}^{\x(x_1)}\bd\x(x)\,\cL\,.
\ee
Let $f(x)$ be an analytic function on $[x_0,x_1]$ with integral singularities at $x=x_0$ and $x=x_1$, i.e., such that it can be written equivalently as $f(x)=(x-x_i)^{\b_i} f_i(x)$, with $\b_i>-1$, $f_i(x_i)\neq 0$, and $i=0,1$. Expanding $f(x)$ in Taylor series around either terminal $x=x_i$,
\be\nonumber
f(x)=\sum_{n=0}^{+\infty} \frac{f^{(n)}_i(x_i)}{n!} |x-x_i|^{\b_i+n}\,,
\ee
and retaining the leading term,
\ba
(\p^\a f)(x) &\approx& \frac{\Gamma(\b_0+1)f_0(x_0)}{\Gamma(\b_0-\a+1)}(x-x_0)^{\b_0-\a}\,,\nonumber\\
(\bp^\a f)(x)&\approx& \frac{\Gamma(\b_1+1)f_1(x_1)}{\Gamma(\b_1-\a+1)}(x_1-x)^{\b_1-\a}\,,\nonumber
\ea
we obtain
\bs\be
\lim_{x\to x_0} (\p^\a f)(x)=\left\{\begin{matrix} 0                    &\quad {\rm if}~ \a<\b_0\\
                                                   \Gamma(\a+1)f_0(x_0) &\quad {\rm if}~ \a=\b_0\\
                                                   \infty               &\quad {\rm if}~ \a>\b_0\end{matrix}\right.\,,
\ee
and
\be
\lim_{x\to x_1} (\bp^\a f)(x)=\left\{\begin{matrix} 0                    &\quad {\rm if}~ \a<\b_1\\
                                                    \Gamma(\a+1)f_1(x_1) &\quad {\rm if}~ \a=\b_1\\
                                                    \infty               &\quad {\rm if}~ \a>\b_1\end{matrix}\right.\,.
\ee\es
Evaluation of differintegrals near terminal points lead to a collapse of their definition (even when the result is finite) and differential calculus thereon becomes inadequate. This conclusion is pleasantly expected in the light of the physical picture presented at the end of the paper. Roughly speaking, when all points of $\cM_\a^D$ are ``too close to the boundary'' the space and its boundary become one and the same. Such qualitative description reminds the topology of totally disconnected and post-critically finite fractals \cite{KiL}, and in such regime fractional calculus, as a continuum approximation of these fractals, breaks down. Moreover, the use of Liouville/Weyl operators does not allow one to consider transient regimes because the limits $|x_{0,1}|\to \infty$ correspond to looking far away from the terminal conditions, eqs.~\Eq{lim0} and \Eq{lim1}. To capture these regimes, one should take finite upper and lower boundaries. Later we will see that one can (and, actually, should) perform another extension of the model, in order to get phenomena which are transient in a different sense: promoting real fractional operators to complex ones. 

\subsubsection{Metric and distance}\label{dis}

Coordinate transformations between the Cartesian systems $\{x^I\}$ and a generic curvilinear system $\{y^\mu\}$ are governed by the fractional generalization of vielbeins:
\be\label{xye}
e_\mu^I:=\p_\mu^\a \x^I(y)\,,
\ee
which are $D\times D$ orthonormal matrices, $e_I^\mu e_\mu^J = \delta_I^J$. From these, one can define the \emph{fractional metric} \cite{CSN1}
\be
g_{\mu\nu}:= \eta_{IJ} e_\mu^I e_\nu^J \,,
\ee
where $\eta_{IJ}={\rm diag}(-1,1,\cdots,1)$ is the Minkowski metric. The fractional line element is then
\be\label{linel}
\rmd s^\a := \left[g_{\mu\nu} (\rmd x^\mu)^\a\otimes (\rmd x^\nu)^\a\right]^{\frac12}\,,
\ee
or, in geometric notation,
\be\label{gele}
\bd {\rm s}^2 = g_{\mu\nu} \bd\x^\mu\otimes \bd\x^\nu\,.
\ee
For $\cM^D_\a$, the metric is just the Minkowski metric, $g_{\mu\nu}=\eta_{\mu\nu}$. The spatial distance between two points in fractional spacetime is then the $2\a$-norm
\be\label{dista}
\Delta_\a(x,y) := \left\{[\Delta(x^\mu,y^\mu)]^\a [\Delta(x_\mu,y_\mu)]^\a\right\}^{\frac{1}{2\a}} :=\left(\sum_{\mu=1}^{D-1}|x^\mu-y^\mu|^{2\a}\right)^{\frac{1}{2\a}}\,,
\ee
where $\Delta(x^\mu,y^\mu)=|x^\mu-y^\mu|$. This is a norm only if $\a\geq 1/2$, i.e., when the triangle inequality holds. Therefore, we can further restrict $\a$ to lie in the range
\be\label{newra}
\boxd{\tfrac12\leq\a\leq1\,.}
\ee 
As already emphasized in \cite{frc1}, one should not confuse eq.~\Eq{dista} with the choice of a $p$-norm (all topologically equivalent) in a given space: as $\a$ changes, so does the geometry of space.

The last ingredient completing the definition of fractional spacetime is the Laplacian $\cK$, entering the kinetic term of a scalar field theory living in $\cM_\a^D$. Before discussing this operator, we turn to two fundamental geometric properties of fractional spacetime: its fractal dimensions and the symmetry group.

%%%%%%%%%%%%%%%%%%%%%%%%%%%%%%%%%%%%%%%%%%%%%%%%%%%%%%%%%%%%%%%%%%%%%%%%%%%%%%%%%%%%%%%%%%%%%%%%

\subsection{Hausdorff, spectral and walk dimensions}\label{hsw}

In fractal geometry, there exist many definitions of ``dimension'' \cite{Fal03} (reviewed in \cite{frc1}). For ordinary spacetimes, there is no benefit in making distinctions among these definitions because they all agree in recognizing the topological dimension $D$ as the number of degrees of freedom experienced by an observer in the measurement of distances, in diffusion processes, and so on. However, the topological dimension is a bad indicator of the geometry of fractional spaces, and it is necessary to resort to the fractal machinery. This enters the picture in a simplified way, since the background is continuous. Full details are given in \cite{frc1}.

The first useful indicator is the Hausdorff dimension $\dh$. Its operational definition on a smooth set is the scaling law for the volume $\cV^{(D)}$ of a $D$-ball $\cB_D$ of radius $\de$, which is
\be
\vr[\cB_D(\de)]=\cV^{(D)}(\de) \propto \de^{\dh}\,.
\ee
Then, in the limit of infinitely small radius,
\be\label{opdeH}
\dh := \lim_{\de\to 0}\frac{\ln \cV^{(D)}(\de)}{\ln \de}\,.
\ee
$\dh$ tells ``how many directions'' the observer feels in configuration space by making static measurements. From the scaling property \Eq{scavr2}, one can already infer that $\dh=\a_0+(D-1)\a\leq D$. Fractional models are characterized by a Hausdorff dimension smaller than or equal to the topological dimension $D$ of embedding spacetime.

A dynamical probe of dimensionality, on the other hand, consists in letting a test particle diffuse starting at point $x$ and ending at point $x'$. In flat fractional spacetime, this random walk is governed by a diffusion equation for the heat kernel $P(x,x',\s)$,
\be\label{dife}
(\cD_\s^\b-\cK_x^{\rm E})P(x,x',\s)=0\,,\qquad P(x,x',0)=\de_\a(x,x')\,,
\ee
where $\s$ is diffusion time (a parameter not to be confused with physical or coordinate time), $\cD_\s^\b$ is a diffusion differential operator of order $\b$, $\cK_x^{\rm E}$ is the Laplacian (acting on the $x$ dependence) defined on the Euclideanized ambient space, and $\de_\a(x,x')=v_\a^{-1}(x)\de(x-x')$ is the Dirac distribution in fractional space. The operators in \Eq{dife} can be chosen as
\ba
\cD_\s^\b &=&\p_\s,\,\p^\b_\s,\,{}_\infty\bp^\b_\s\,,\\
\cK^{\rm E}&=& \de^{\mu\nu}\left(\p_\mu\p_\nu+\frac{\p_\mu v_\a}{v_\a}\p_\nu\right)\,,
\ea
with $\a_0$ possibly different from the other charges. The kinetic operator is such that its eigenfunctions (Bessel functions of the first kind times a power) are the expansion basis of the fractional gemeralization of the ``Fourier'' transform \cite{frc3}. The heat kernel $P(x,x,\s)$ at coincident points $x=x'$ and averaged in the space volume is called return probability. In fractional momentum space (see \cite{frc1} and section \ref{pro}),
\be\label{kkkin}
\cP(\s) \propto \s^{-\frac{D\a\b}{2}}\,.
\ee
The spectral dimension of fractional spacetime is then
\ba
\ds &:=& -2\frac{\rmd\ln \cP(\s)}{\rmd\ln\s}\,,\label{spedi}\\
    &=& \b[\a_0+(D-1)\a]=\b\dh\,.\label{newspe}
\ea
Finally, the walk dimension is the ratio $\dw:= 2\dh/\ds$. If $\dh\neq\ds$, the diffusion law is said to be anomalous. For fractals, $\ds\leq\dh$ and $\dw\geq 2$, while diffusion with $\dw<2$ is typically associated with jump processes.

These matters were discussed in \cite{frc1}, where the Hausdorff and spectral dimensions have been calculated for fractional isotropic Euclidean space. By definition, both the Hausdorff and spectral dimensions of a Lorentzian spacetime are calculated in Euclidean signature, so the results of \cite{frc1} do not need any modification, except the separation of the fractional charge in time direction from the others. This poses, however, a caveat. Normal (or Gaussian) diffusion takes place if the derivative order of $\cK^{\rm E}$ is twice the order of $\cD_\s^\b$, i.e., when $\b=1$. So, Gaussian processes are produced by $\cD_\s^\b=\p_\s$ and $\cK^{\rm E}$ given above. Qualitative arguments suggest that Gaussian diffusion would be also achieved with $\cD_\s^\b=\p^\a_\s$ and a kinetic term of the form such as $\cK^{\rm E}_\a=\sum_\mu{}_\infty\bp_\mu^{2\a}$. In this case, however, for $\a_0\neq \a$ diffusion would be anomalous, since anisotropy in the fractional charge induces non-trivial couplings in the Laplacian $\cK^{\rm E}_\a$ in order to match the scaling dimension of derivatives of different order.

If $\cM_\a^D$ is non-anomalous and $\b=1$, then the Hausdorff and spectral dimensions coincide, eq.~\Eq{newspe}.
In particular, $\ds\leq\dh$ if $\b\leq 1$, while if $\b>1$ fractional Minkowski spacetime cannot be considered a fractal \cite{frc1}. As a general result, on anisotropic models diffusion processes are anomalous. Note that the measure in momentum space is the same distribution $\vr_\a$ as in configuration space, so the Hausdorff dimension of fractional momentum space is $\dh$ for any $\b$.

%The allowed forms of the kinetic operator $\cK$ are dictated by symmetry, which is now time to discuss.

%%%%%%%%%%%%%%%%%%%%%%%%%%%%%%%%%%%%%%%%%%%%%%%%%%%%%%%%%%%%%%%%%%%%%%%%%%%%%%%%%%%%%%%%%%%%%%%%

\subsection{Fractional Poincar\'e transformations}\label{lt}

The scaling property of the Lebesgue--Stieltjes measure of fractional calculus, eq.\ \Eq{scavr2}, is associated both with a non-integer dimension and with a self-similar non-trivial structure. Self-similarity implies the same structure at all scales. So, fractional models should be compared with self-similar fractals, while models with dimensional flow are multi-fractal structures, which have different self-similar properties in ranges centered at different scales of magnification. However, in fractional spacetime the role of isometries is far more significant than that of similarities, as discussed in section 4.2 of \cite{frc1}. Clearly, a particular presentation of fractional calculus breaks all Poincar\'e symmetries, via the definition of measures which are neither translation nor rotation invariant. This may cause to believe that fractional systems, and fractal systems in general, are of little or no physical significance as classical and quantum field theories, where Lorentz invariance is an essential ingredient. For a generic model with absolutely continuous Lebesgue--Stieltjes measure $\rmd\vr(x)=v(x)\,\rmd^Dx$, it was shown that Poincar\'e algebra was deformed even if the action itself is Poincar\'e invariant \cite{fra2}. The key ingredient to obtain this result was to start with a Poincar\'e-invariant Lebesgue--Stieltjes measure, i.e., regard $v(x)$ as a Lorentz scalar. However, confusion arises when one chooses a particular profile for the measure weight $v(x)$: any such profile, which is needed as a concrete realization of anomalous scaling, explicitly breaks Poincar\'e symmetries. For instance, a profile such as $v(x)=|{\bf x}|^{-(D-1)\a}$ is rotation invariant but neither translation nor boost invariant. Fractional models display explicit profiles $v(x)$ by definition, and the attitude of \cite{fra1}--%,fra2,
\cite{fra3} is no longer tenable. Therefore, it is necessary to reexamine the issue of symmetries.

In ordinary relativistic field theories, one observes that the line element (cross-product symbol $\otimes$ omitted)
\be\label{gele0}
\rmd s^2=\eta_{\mu\nu}\rmd x^\mu \rmd x^\nu
\ee
is preserved by the isometry group defined by the Poincar\'e transformations
\be\label{potra}
{x'}^\mu = \Lambda_\nu^\mu x^\nu + a^\mu\,,
\ee
where $\Lambda_\nu^\mu$ are $D\times D$ constant matrices such that
\be\nonumber
\Lambda_\nu^\mu \Lambda^\nu_\rho = \Lambda_\nu^\mu (\Lambda_\nu^\rho)^{-1} = \delta^\mu_\rho\,,
\ee
and $a^\mu$ is a constant vector. Then,
\be\nonumber
\rmd {s'}^2 = \eta_{\mu\nu}\rmd {x'}^\mu \rmd {x'}^\nu = \eta_{\mu\nu}\Lambda_\rho^\mu \Lambda^\nu_\s \rmd x^\rho \rmd x^\s=\eta_{\rho\s}\rmd x^\rho \rmd x^\s\,.
\ee
The Lorentz transformations $\Lambda_\nu^\mu=\p {x'}^\mu(x)/\p x^\nu$ include spatial rotations and spacetime boosts ($\det\Lambda=+1$), as well as improper discrete transformations such as time reversal and parity ($\det\Lambda=-1$). The $\Lambda$ are actually frame transformations, since they act in internal space and vielbeins transform as
${e'}_\mu^I=\Lambda_J^{I}e_{\mu}^J$, but in homogeneous spacetimes one can choose one and the same frame for every point (the so-called Fermi frame), and tangent and Minkowski spaces are identified.

One can construct actions which are Poincar\'e invariant under proper transformations. Lagrangians $\cL[\p_x,\vp^i(x)]$ of some fields $\vp^i$ are defined to be proper scalars. On the other hand, the Lebesgue measure $\sqrt{-g}\,\rmd^D x$ is invariant, too, because $\det\Lambda =1$. In fact, in first-order formalism the volume element can be written as
\be\label{vofo}
|\det e| \rmd^D x=\frac{1}{D!} e^{I_0}\wedge \dots e^{I_{D-1}} \e_{I_0\cdots I_{D-1}}\,,
\ee
where $e^I:=e^I_\mu\rmd x^\mu$ and $\e^{I_0\cdots I_{D-1}}$ is the Levi-Civita symbol, which is an internal pseudo-tensor: $\epsilon'_{I_0\cdots I_{D-1}}=\det\Lambda\Lambda_{I_0}^{J_0}\dots \Lambda_{I_{D-1}}^{J_{D-1}}\epsilon_{J_0\cdots J_{D-1}}$. Applying a Lorentz transformation, one finds that $|\det e'|\rmd^D x'=\det\Lambda |\det e|\rmd^D x$. Thus, an action and equations of motions defined in a given coordinate frame $\{x\}$ will be functionally the same in another frame $\{x'\}$ related to the other by a proper Poincar\'e transformation \Eq{potra}. Observers defined in local inertial frames experience natural phenomena governed by the same set of equations: physical laws depend neither on the position nor on the orientation of the observer's frame.

In the case of fractional Minkowski space, the line element is given by eq.~\Eq{linel}
or, in geometric notation,
\be\label{gele2}
\bd {\rm s}^2 = \eta_{\mu\nu}\, \bd\x^\mu\, \bd\x^\nu\,.
\ee
Consistently with eq.~\Eq{xye}, this suggests to define \emph{fractional Poincar\'e transformations} which are linear in geometric coordinates:
\be\label{fpotra}
\boxd{{\x'}^\mu = \tilde\Lambda_\nu^\mu \x^\nu +\tilde a^\mu\,,\qquad \tilde\Lambda_\nu^\mu\tilde\Lambda_\mu^\rho=\delta_\nu^\rho\,,}
\ee
where $\tilde a^\mu$ is a constant vector and
\be\label{magic}
\tilde\Lambda_\nu^\mu=\frac{\p^\a {\x'}^\mu}{\p^\a \x^\nu}=\frac{\p {\x'}^\mu}{\p \x^\nu}%=\frac{\p({x'}^\mu-{x_0'}^\mu)^\a}{\p(x^\nu-x_0^\nu)^\a}
\ee
are $D\times D$ constant matrices, associated with the ordinary $SO(D-1,1)$ group. Equation \Eq{fpotra} is in accordance with \cite[section 4.2]{frc1}, where it was argued that fractional Euclidean space $\cE_\a^D$ is characterized by the group of affine transformations
\be\nonumber
{\x'}^\mu = {\rm A}^\mu_\nu \x^\nu+\tilde a^\mu\,.
\ee

The line element \Eq{gele2} is preserved under the transformations \Eq{fpotra}. Crucial to this result is the fact that the fractional differential, made of Caputo derivatives, is zero on a constant. If we had used Riemann--Liouville calculus, we would not have been able to write \Eq{fpotra} in such a simple form.  

We now discuss whether eq.~\Eq{fpotra} is the only transformation preserving the fractional line element for general $\x$. Let us first recall a textbook exercise for the integer case \cite{Wei72}. Under a general non-singular coordinate transformation $x^\mu\to {x'}^\mu$, the line element \Eq{gele0} changes as
\be\nonumber
\rmd {s'}^2 = \eta_{\mu\nu} \rmd{x'}^\mu \rmd{x'}^\nu=\eta_{\mu\nu}\p_\rho {x'}^\mu \p_\s{x'}^\nu\rmd{x}^\rho \rmd{x}^\s\,,
\ee
but imposing $\rmd{s'}^2=\rmd{s}^2$, one gets $\eta_{\rho\s}=\eta_{\mu\nu}\p_\rho{x'}^\mu\p_\s{x'}^\nu$. Differentiating with respect to $x^\tau$, one has
\be\nonumber
0=\eta_{\mu\nu}\left(\frac{\p^2 {x'}^\mu}{\p x^\tau\p x^\rho}\frac{\p {x'}^\nu}{\p x^\s}+\frac{\p {x'}^\mu}{\p x^\rho}\frac{\p^2 {x'}^\nu}{\p x^\tau\p x^\s}\right)\,;
\ee
adding and subtracting the same equation with, respectively, $\tau\leftrightarrow \rho$ and $\tau\leftrightarrow \s$, one obtains an equation with six terms. They cancel one another except two identical, giving twice the first term. Since $\eta_{\mu\nu}$ and $\p {x'}^\nu/\p x^\s$ are non-singular, one ends with the condition
\be\nonumber
\frac{\p^2 {x'}^\mu}{\p x^\tau\p x^\rho}=0\,,
\ee
whose solution is eq.~\Eq{potra}. In the fractional case, the same argument does not go through as smoothly. The failure of the Leibniz rule 
\be\label{leib}
\p_x[f(g)]=\frac{\p f}{\p g}\, \p_x g
\ee
and its replacement with
\ba
(\p^\a f)(x)%=\sum_{j=1}^{+\infty}\binom{\a}{j} (\p^j f)(x) I_x^{j-\a}\{1\}=\frac{\a}{\Gamma(2-\a)}(x-x_0)^{1-\a}(\p f)(x)+
&=& \frac{1}{\Gamma(1-\a)}\frac{f(x)-f(x_0)}{(x-x_0)^{\a}}\nonumber\\
&&+\sum_{j=1}^{+\infty} \frac{\sin[\pi(j-\a)]}{\pi(j-\a)}\frac{\Gamma(1+\a)}{\Gamma(1+j)}(x-x_0)^{j-\a} (\p^j f)(x)\label{leru2}
\ea
forbid mixed derivatives to combine in a simple way. At most, one can recognize eq.~\Eq{fpotra} as a sufficient but not necessary condition for the line element to be invariant. The fact that we have infinite terms all of different order, however, makes it likely that eq.~\Eq{fpotra} is also necessary, unless miraculous cancellations take place. %To see this, it is very instructive to apply eq.~\Eq{leru2} in the case where $f(x)=q'(x)$ is a fractional Poincar\'e transformation. Thanks to eq.~\Eq{magic}, the $j$-th derivative of ${\x'}^\mu$ is proportional to $\p_{\x^\nu} {\x'}^\mu$:
%\be\nonumber
%\p^j_\nu {\x'}^\mu= \frac{x^{\a-j}}{\Gamma(\a+1-j)} \frac{\p^\a {\x'}^\mu}{\p^\a \x^\nu}= \frac{x^{\a-j}}{\Gamma(\a+1-j)}\tilde\Lambda_\nu^\mu\,.
%\ee
%Then,
%\ba
%\p^\a_{x^\nu} {q'}^\mu &=& \left[\sum_{j=1}^{+\infty}\binom{\a}{j} \frac{1}{\Gamma(\a+1-j)^2}\right]\tilde\Lambda_\nu^\mu\nonumber\\
%&=& \left(1-\frac{\sin\pi\a}{\pi\a}\right)\frac{\p {\x'}^\mu}{\p \x^\nu}\,,\label{leru4}
%\ea
%implying
%\be
%\boxd{\bd{\x'}^\mu =\left(1-\frac{\sin\pi\a}{\pi\a}\right) \frac{\p {\x'}^\mu}{\p \x^\nu} \bd{\x}^\nu\,.}
%\ee
%This expression states the equivalence, up to a factor (non-singular for $\a\neq0$), of the fractional derivative $\p^\a$ with the ordinary derivative $\p_\x$ when acting upon a coordinate trasformation of the form \Eq{fpotra}. A non-constant transformation would have not led to this simple result.% Anyway, for $x\lesssim 1$ the dominant term in eq.~\Eq{leru2} is, formally, $j=1$, and $\p^\a\propto\p_\x$; at this point the argument above can be repeated with $\rmd x^\mu$ replaced by $\bd\x^\mu$

Just like for integer transformations, the orthogonality relation in \Eq{fpotra} implies that
\be
(\tilde\Lambda_0^0)^2=1+\sum_{i=1}^{D-1} (\tilde\Lambda^0_i)^2\geq 1\,,\qquad (\det \tilde\Lambda)^2=1\,,
\ee
so one can distinguish between proper and improper transformations. We dub the semi-direct product of translations and fractional Lorentz transformations on the embedding coordinates $x^\mu$ the \emph{fractional Poincar\'e group} $\Pi_\a$.

Clearly, fractional transformations $\tilde\Lambda_\nu^\mu$ neither act linearly on embedding coordinates $x^\mu$ nor are simply given by the elements of $\Lambda_\nu^\mu$ to the power of $\a$: $(\rmd {x'}^\mu)^\a=(\Lambda_\nu^\mu\rmd x^\nu)^\a$ under an integer transformation, while $\bd{\x'}^\mu=\tilde\Lambda_\nu^\mu\bd\x^\nu =\tilde\Lambda_\nu^\mu(\rmd x^\nu)^\a$. We can find an approximate relation between integer and fractional Lorentz transformations when $\a=1-\e$, $\e\ll 1$. Noting that
\be
\x = x+x(1-\gamma-\ln x)\e+O(\e^2)\,,%\x = (x-x_0)+(x-x_0)[1-\gamma-\ln(x-x_0)]\e+O(\e^2)\,,
\ee
where $\gamma$ is Euler's constant, fractional Poincar\'e transformations reduce to integer ones up to correction terms:
\be
\tilde\Lambda_\nu^\mu=[1+(\gamma-1)\e]\Lambda_\nu^\mu+O(\e^2,x\ln x)\,.
\ee
The fact that fractional symmetries are intrinsically non-linear resembles the situation in non-commutative $\kappa$-Minkowski spacetime, which is invariant under deformed, non-linear Poincar\'e symmetries \cite{LRNT}--%,LNR,LuR,LRZ,MaR,AAmD,
\cite{AAAMT}. The relation between fractional and non-commutative theories is discussed elsewhere \cite{ACOS}.

Because of the boundary at $x^\mu=0$, the fractional Poincar\'e transformations \Eq{fpotra} are not global symmetries of fractional Minkowski spacetime. Yet, they are the guiding principle to write an invariant action up to boundary terms. We show first that the fractional measure, derivative and differential are invariant under the fractional isometry group $\Pi_\a$, up to boundary terms. The task is easy in geometric coordinates. The proof for the integration in eq.~\Eq{act0} is formally the same as in ordinary calculus, with $e^I$, $\Lambda_{I}^{J}$ and $x$ in eq.~\Eq{vofo} and below replaced by their fractional counterparts. The lower terminal in \Eq{act0} is translated by $\tilde a^\mu$ but this does not affect the physics well inside the spacetime bulk, according to section \ref{preli}.

For consistency, writing the measure as in eq.~\Eq{act1} should yield the same result. To check this, it is sufficient to verify that the fractional Caputo derivative transforms as a covariant vector. This is true, thanks to eq.~\Eq{magic}:
\be
\frac{\p^\a}{\p^\a {\x'}^\mu}=\frac{\p \x^\nu}{\p {\x'}^\mu}\frac{\p^\a}{\p^\a {\x}^\nu}=\tilde\Lambda^\nu_\mu\frac{\p^\a}{\p^\a {\x}^\nu}\,.
\ee
In particular, operators of the form $\eta^{\mu\nu}\phi\p^\a_\mu\p^\a_\nu\phi$ are Lorentz covariant in a fractional sense, if $\phi$ is a scalar.

Looking at the definition of $\p_\mu^\a$, this result would not have been obvious. For $0<\a<1$, the left derivative can be written in terms of the geometric coordinate $\x$ (index $\mu$ omitted):
\ba
(\p^\a f)(x) &=& \frac{1}{\Gamma(1-\a)}\int_0^x \frac{\rmd x'}{(x-x')^\a} \p_{x'}f(x')\nonumber\\
						 &=& -\frac{1}{\Gamma(1-\a)}\int_0^x \frac{\rmd y}{y^\a} \p_{y}f(x-y)\nonumber\\
						 &=& -\frac{1}{\Gamma(1-\a)}\int_0^\x \frac{\rmd \y}{\y} \p_{\y}\tilde f(\x,\y)\,,
\ea
where $\tilde f(\x,\y)=f\{[\Gamma(1+\a)\x]^{1/\a}-[\Gamma(1+\a)\y]^{1/\a}\}$. For the Weyl derivative, one obtains a similar expression with the upper terminal $x$ replaced by $+\infty$ and $\tilde f(\x,\y)=f\{[\Gamma(1+\a)\x]^{1/\a}+[\Gamma(1+\a)\y]^{1/\a}\}$. Performing a fractional Poincar\'e transformation and changing integration variable leads to a seemingly non-vectorial expression, just like acting naively on $\p/\p x^\mu$ with eq.~\Eq{potra}.
%\ba
%({}_{\infty}\bp^\a f)(x) &=& -\frac{1}{\Gamma(1-\a)}\int_{x}^{+\infty} \frac{\rmd x'}{(x'-x)^\a} \p_{x'}f(x')\nonumber\\%
%						 &=& -\frac{1}{\Gamma(1-\a)}\int_0^{+\infty} \frac{\rmd y}{y^\a} \p_{y}f(x+y)\nonumber\\
%						 &=& -\frac{1}{\Gamma(1-\a)}\int_0^{+\infty} \frac{\rmd \y}{\y} \p_{\y}\bar f(\x,\y)\,,%\nonumber\\
%\ea
%where $\bar f(\x,\y)=f\{[\Gamma(1+\a)\x]^{1/\a}+[\Gamma(1+\a)\y]^{1/\a}\}$.

To summarize, the explicit coordinate dependence $x^\mu$ of fractional operators is just a presentation in the embedding and fractional covariance must be defined in the space of geometric coordinates $\x^\mu$.

%%%%%%%%%%%%%%%%%%%%%%%%%%%%%%%%%%%%%%%%%%%%%%%%%%%%%%%%%%%%%%%%%%%%%%%%%%%%%%%%%%%%%%%%%%%%%%%%%%%%%%%%%%%%%%%%%%%%%%%%%%%%%%%%%%%%%%%%%%%%%%%%%%%%%%%%%%%%%%%%%%%%%%%%%%%%%%%%%%%%%%%%%%%%%%%%

\section{Dimensional flow}\label{multis}

So far, we have discussed various ingredients for the construction of a spacetime characterized by a measure of fixed fractional order. This was done for the purpose of simplifying the description of an unfamiliar type of geometry to the bone. However, a major goal of this proposal is to introduce the often-advertized idea that geometry changes with the scale or, in more colorful words, to give spacetime a multi-fractal structure. In \cite{fra1}--%,fra2,
\cite{fra3}, it was assumed that the parameter $\a$ would somehow flow from some finite $\a=\a_*<1$ to $\a=1$, without however giving any detail about how this flow takes place.

One could continue to keep $\a$ fixed and develop the concrete example of a scalar theory, were it not for a simple but maybe surprising fact. Namely, \emph{a multi-fractional classical structure is intimately related with the quantum structure of this class of field theories}, and it determines its renormalization group properties. The reason is that the presence of all possible fractional operators in the classical action determines a hierarchy of scales. This hierarchy can be interpreted both as the very definition of multi-fractality in the fractional context and as a self-consistency requirement for renormalization. Not only these interpretations are not mutually exclusive, but they practically amount to one and the same. Hence the logical necessity to first present the definition of multi-fractionality, and then construct consistent classical actions. Here we begin to carry out this programme.

One possibility would be to promote $\a$ to a Lorentz scalar, a function of spacetime coordinates:
\be\label{ax}
\a\to \a(x^\mu)\,.
\ee
The dimension of spacetime would become dynamical, a possibility already considered in the past \cite{HW}. One should add a kinetic term and a potential for this new field into the total action, which would be augmented by an extra contribution $S[\a(x)]$. However, the system would quickly become intractable due to the complicated integration measure, not to mention the transcendental dependence on $\a$ in $S[\a(x)]$. We also ask ourselves whether we wish a model to become fractional at small spatial scales or at early times, or both: the first case is a universe becoming fractal below a certain critical scale at any given time, while the second corresponds, roughly speaking, to a spacetime becoming fractal near the big bang. Here we face physically inequivalent scenarios, depending on how we define the action. For example, a realization of a spacetime becoming fractional at early times would be to allow the fractional order of time direction to be $\a_0=1$, while that of spatial directions to be only time dependent, $\a_{i}\to \a(t)$. Then, one would have no problems in integrating, if not for the fact that time and fractional spatial derivatives would no longer commute. If only ordinary derivatives were present in the classical action, the only minor complication would be to integrate the measure by parts when deriving the equations of motion. 

%%%%%%%%%%%%%%%%%%%%%%%%%%%%%%%%%%%%%%%%%%%%%%%%%%%%%%%%%%%%%%%%%%%%%%%%%%%%%%%%%%%%%%%%%%%%%%%%

\subsection{External scale or renormalization group picture}\label{mul2}

The main problem arising with eq.~\Eq{ax} is that geometric coordinates would be defined through transcendental expressions and, in general, the fate of fractional Poincar\'e transformations is not clear. In a Lorentz-covariant framework, the presence of a characteristic scale can spoil the symmetries of the system, unless one introduces the latter with care. This suggests that treating $\a$ as a field may be unsatisfactory. In alternative, one can adopt the perspective of critical systems and regard $\a$ as an order parameter. More precisely, one can parametrize $\a$ not with spacetime coordinates, but with an external scale parameter governing dimensional flow. Physically, one can introduce a critical length/time scale $\ell_*$ below which the system flows to a deep fractional regime, and above which it occupies the whole embedding space. Assuming that $\a$ acquires a finite critical value $\a_*$ at the bottom of this regime, we have
\be\label{asias2}
\a(\ell)\sim\left\{ \begin{matrix} \a_*&\quad {\rm as}~\ell \lesssim \ell_*\\
                                 1&\quad {\rm as}~\ell \gg \ell_*\end{matrix}\right. \,.
\ee
Notice that a similar behaviour would appear in a renormalization group picture, where the parameter $\a$ runs with the energy scale. If interpreted as a fundamental scale (which is true in fractional theories of real order), $\ell_*$ may be associated with the Planck scale. In this context, as in modern approaches such as asymptotic safety gravity, it does make sense to consider scales below $\ell_*$, because the latter is not a cut-off of the theory. Anyway, we refrain
from the identification $\ell_*=\lp$ because in the complex-order theory the scale hierarchy will go through a little revolution, and the Planck scale will be pushed further deep in the UV spacetime structure.

Defining the adimensional multi-fractional external time $\s:=\ell/\ell_*$, we can replace the limits in eq.~\Eq{asias2} with
\be\label{asias}
\a(\s)\sim\left\{ \begin{matrix} \a_*&\quad {\rm as}~\s \to 0\\
                                 1&\qquad {\rm as}~\s \to +\infty\end{matrix}\right. \,.
\ee
For instance, profiles realizing \Eq{asias} are
\be\label{varie}
\a(\s)=\frac{\a_*+\s}{1+\s}\,,\qquad \a(\s)=\a_*+(1-\a_*)\tanh\s\,,\qquad \a(\s)=\a_*+(1-\a_*){\rm erf}\s\,,
\ee
where erf is the error function. The latter is solution of the diffusion equation in Euclidean space with the Heaviside distribution as initial condition.

Let eq.~\Eq{act0} be an action characterized by a measure $\vr_\a(x)$ of dimension $\dh(\a)$. In particular, $\dh(\a)$ is given by eq.~\Eq{newspe}. In the external time picture, the fundamental action is decorated with an extra integration over external time,
\be\label{ext}
\boxd{S=\int_0^{+\infty}\rmd\s g(\s) S_{\a(\s)}\,,}
\ee
where $g(\s)$ is a one-parameter measure. Since $\s$ is an external time parameter, spacetime covariance is respected if $S_\a$ is covariant. The spacetime whose measure is a superposition of fractional Minkowski measures will be called \emph{multi-fractional Minkowski spacetime}, denoted by $\cM^D_*$.

%%%%%%%%%%%%%%%%%%%%%%%%%%%%%%%%%%%%%%%%%%%%%%%%%%%%%%%%%%%%%%%%%%%%%%%%%%%%%%%%%%%%%%%%%%%%%%%%

\subsection{Multi-fractional spacetime}\label{mul3}

The external-scale \emph{Ansatz} \Eq{ext} admits a neat interpretation in terms of multi-fractal geometry. Just like fractals, multi-fractals do not have a precise definition but, intuitively, they are sets with scale-dependent fractal properties, on which mass distributions (i.e., measures) do not obey a simple power law (see, e.g., \cite[chapter 17]{Fal03} and \cite{Har01}). We have seen that the Hausdorff dimension of a smooth set is conveniently determined by the scaling law of the measure of balls of infinitely small radius, eq.~\Eq{opdeH}.\footnote{For a non-smooth set embedded in $\mathbb{R}^D$, eq.~\Eq{opdeH} still defines a ``fractal dimension'' but, in general, it does not coincide with $\dh$.} One can make this definition local and dependent on the center $x$ of the ball, provided the following limit exists:
\be\label{opdel}
\dl := \lim_{\de\to 0}\frac{\ln \vr[\cB_D(x,\de)]}{\ln \de}\,,
\ee
a quantity called \emph{local dimension} or \emph{H\"older exponent} of $\vr$ at $x$. Consider now the set $\cF_d$ of the points where the local dimension exists and equals $d$:
\be
\cF_d =\left\{x\in\mathbb{R}^D~:~\dl=d\right\}\,.
\ee
In other words, ball measures centered at every $x$ in $\cF_d$ scale as $\cV^{(D)}(\de)=\vr[\cB_D(x,\de)]\sim \de^d$ as $\de\to 0$. As $d$ varies, so does the set $\cF_d$ and its Hausdorff dimension. The problem of calculating this dimension, called \emph{fine multi-fractal} (or \emph{singularity}) \emph{spectrum}, is the subject of fine multi-fractal analysis:
\be\label{spect}
f_{\rm H}(d):= \dh(\cF_d)\,.
\ee
The fine spectrum $f_{\rm H}$ gives information about the limiting behaviour of $\vr$ at any point.

Self-similar sets well illustrate the methods of multi-fractal spectral analysis. We recall some basic definitions from \cite{Fal03}, discussed in \cite{frc1} in the context of fractional spaces. Consider a set of $N$ maps $\cS_i\,:\,\mathbb{R}^D\to\mathbb{R}^D$, $i=1,\dots,N\geq 2$, such that
\be\label{simi}
\Delta[\cS_i(x),\cS_i(y)]\leq \la_i\Delta(x,y)\,,\qquad x,y\in\mathbb{R}^D\,,\qquad 0<\la_i<1\,,
\ee
where the distance $\Delta$ between two points is $\Delta(x,y)=|x-y|$ in ordinary integer geometry. Any such map is called \emph{contraction} and the number $\la_i$ is its \emph{ratio}. If equality holds, $\cS_i$ is a \emph{contracting similarity} or simply a similarity. By definition, self-similar deterministic fractals are invariant under contraction maps and can be expressed as the union of the images of $\cS_i$ \cite{Hut81},
\be\label{sss}
\cF= \bigcup_{i=1}^N \cS_i(\cF)\,.
\ee
Suppose the strong separation condition holds, i.e., there exists a closed set $U$ such that $\cS_i(U)\subset U$ for all $i=1,\dots,N$ and $\cS_i(U)\cup \cS_{j\neq i}(U)=\emptyset$. $\cF\subset U$ is constructed taking sequences of similarities and the intersection of sets $U_k=\cS_{i_1}\circ\cdots\circ \cS_{i_k}(U)$. If $|U|=1$, then the diameter of the $k$-th iteration set is the product of similarity ratios, $|U_k|=\la_{i_1}\dots \la_{i_k}$. Let $0<g_i<1$ be $N$ probabilities (or mass ratios, or weights) such that $\sum_i g_i=1$. One can imagine to distribute a mass on sets $U_k$ by dividing it repeatedly in $N$ subsets of $U_k$, in the ratios $g_1\,:\cdots:\,g_N$. This defines a \emph{self-similar measure} $\vr$ with support $\cF$, such that $\vr(U_k)=g_{i_1}\dots g_{i_k}$ and, for all sets $A\subseteq\cF$ \cite{Hut81},
\be\label{ssim}
\vr(A)=\sum_{i=1}^N g_i\,\vr[\cS_i^{-1}(A)]\,.
\ee
The case $N=+\infty$ defines so-called infinite self-similar measures, describing fractals with an infinite number of similarities \cite{RiM,YRL}. Given a real parameter $u$, we define the \emph{singularity} (or \emph{correlation}) \emph{exponent} $\theta(u)$ as the real number such that \cite{Gra83}--%,HeP,HJKPS,
\cite{Rie95}
\be\label{mfssm}
\sum_{i=1}^N g_i^u \la_i^{\theta(u)}=1\,.
\ee
The correlation exponent exists and is unique, since $0<\la_i,g_i<1$. As a function of $u$, $\theta$ is decreasing and $\lim_{u\to \pm\infty}\theta(u)=\mp\infty$. Finally, the \emph{generalized dimensions} are
\be
d(u):= \frac{\theta(u)}{u-1}\,,\qquad u\neq 1\,,
\ee
and a non-singular definition, which we do not report here, is employed for $u=1$. Fractals characterized by just one dimension at all scales are special cases of multi-fractals. For deterministic fractals, the probabilities are all equal to $g_i=1/N$. In all fractals with equal contracting ratios $\la_i=\la$, the generalized dimensions all coincide with the capacity $d_{\rm C}$ of the set (e.g., \cite{frc1}), which is also the Hausdorff dimension. From eq.~\Eq{mfssm},
\be\label{mfssm2}
N \frac{\la^{\theta(u)}}{N^u} =1\qquad \Rightarrow\qquad d(u)=-\frac{\ln N}{\ln\la}=:d_{\rm C}=\dh\,.
\ee

For self-similar measures \Eq{ssim}, and under some weak assumptions, the spectrum \Eq{spect} enjoys a number of properties \cite{Fal03}. First, the support of the fine spectrum is in a certain finite interval $[d_{\rm min},d_{\rm max}]$, on which $f_{\rm H}$ is given by the Legendre transform of $\theta$:
\be
f_{\rm H}(d)= \inf_{u\in\mathbb{R}} [\theta(u)+d u]\,.
\ee
For $d\notin [d_{\rm min},d_{\rm max}]$, $\cF_d$ is the empty set. Second, $f_{\rm H}$ is a concave function of $d$. The maximum of the spectrum is at $d=d(0)$ and equals the dimension of the support of the measure, ${\max}_df_{\rm H}(d)=f_{\rm H}[d(0)]=\theta(0)=\dh({\rm supp}\vr)$. In this case, eq.~\Eq{mfssm} becomes the well-known algebraic condition for the Hausdorff dimension of a self-similar set:
\be\label{ssm}
\sum_{i=1}^N \la_i^{\dh({\rm supp}\vr)}=1\,.
\ee
The support of the measure can be a fractal; for instance, the multi-fractal Cantor set is a mass distribution over the Cantor set, which is its support. Third, at $u=1$, for eq.~\Eq{mfssm} $\theta(1)=0$ and the spectrum equals the Hausdorff dimension of the measure: $f_{\rm H}[d(1)]=d(1)=\dh(\vr)$.\footnote{The Hausdorff dimension of a measure is
\be\nonumber
\dh(\vr) := {\rm inf}\left\{\dh E~:~E \textrm{ is a Borel set with } \vr(E)>0\right\}\,.
\ee
Intuitively, it is the dimension of a set on which a significant part of the mass distribution $\vr$ is concentrated \cite{Fal03}.}

There are a plethora of physical phenomena described by multi-fractal geometry. Some multi-fractal systems can be approximated by fractional dynamics. For example, in the case of the fractional Fokker--Planck--Kolmogorov equation, one simply replaces fractional derivatives with sums of derivatives of any allowed fractional order \cite{Zas1}. Multi-fractional Brownian motion is another instance \cite{Fal03,Har01,PLV}--%,ALV1,ALV2,
\cite{Mis08}.

Let us go back to fractional spacetime $\cM^D_\a$ of fixed order $\a$, and consider for simplicity the isotropic case $\a_\mu=\a$. In \cite{frc1}, we have argued that its Euclidean analogue $\cE_\a^D$ can be characterized by two similarity maps acting on geometric coordinates. For any direction $\mu$,
\be\label{Rdsimx}
{\rm S}_1(\x^\mu) := \la\x^\mu\,,\qquad {\rm S}_2(\x^\mu) := (1-\la)\x^\mu+\la\,,
\ee
where $\la$ is arbitrary and chosen to be the same along all directions. Since $\la$ is arbitrary, this set is trivially self-similar; yet, indeed it is self-similar. A mass would be equally distributed on, say, $N=2^D$ subsets, with probabilities $g_i=1/N$. In geometric coordinates, the scaling is $\la=1/2=N^{-1/D}$. Applying eq.~\Eq{mfssm2}, $\dh=-\ln N/\ln\la=D$, and $\dh=D\a$ for embedding coordinates. From the point of view of the fractional embedding spanned by the $x$, the scaling ratio is
\be
\la=\left(\frac12\right)^{\frac{1}{\a}}=\left(\frac1N\right)^{\frac{1}{D\a}}=g_i^{\frac{1}{D\a}}\,,
\ee
and the probabilities can be regarded as $\a$-dependent for a given $\la$:
\be\label{galp}
g_i=g_\a:=\la^{D\a}\,.
\ee
The same conclusion holds for fractional Minkowski spacetime. 

Collecting these results, we determine the extension of the measure $\vr_\a$ to a multi-fractional measure with the self-similar structure \Eq{ssim}. A non-degenerate set of probabilities $\{g_i\}$ must be introduced to weight the sub-copies of $\cM_\a^D$ differently, with the index $i$ running on a given set (with more than two elements, possibly). Each copy is labelled by $\a$, which plays the role of generalized dimension at a given scale. Thus, and by eq.~\Eq{galp}, the labelling of the probabilities is assumed by $\a$. For discrete $\a$, one should exchange the $\a$-order integration in the fractional model with a sum of integrals over all possible ranges in $\a$:
\be\label{discr}
\boxd{S=\sum_\a g_\a\, S_\a\,,\qquad \tfrac12\leq\a\leq 1\,.}
\ee
The external time/RG picture \Eq{ext} is nothing but a continuum version of \Eq{discr}. 

The coefficients $g_\a$ are probabilities from the point of view of fractal geometry, and coupling constants from the perspective of field theory. Assuming the Lagrangian to be the same for all contributions, the $g_\a$ are dimensionful in order for $S$ to be dimensionless. Then, they determine the scale at which geometry changes. Consider a simplified isotropic model with one such (length) scale $\ell_*$, where $D$-dimensional integrations (whatever the choice of fractional calculus) are given by
\be
I_D=I_D^{\a_1}+\ell_*^{D(\a_1-\a_2)}I_D^{\a_2}\,,\qquad [I_D]=-D\a_1\,,\qquad \tfrac12\leq\a_1<\a_2\leq 1\,.
\ee
The measure is binomial and volumes are made of two pieces. The volume of a $D$-ball of radius $R$ in a space endowed with this structure is
\be
\cV^{(D)}(R)= \ell_*^{D\a_1}\left[\Om_{D,\a_1} \left(\frac{R}{\ell_*}\right)^{D\a_1}+\Om_{D,\a_2}  \left(\frac{R}{\ell_*}\right)^{D\a_2}\right]\,,
\ee
where $\Om_{D,\a}$ is the volume of a unit ball. Depending on the size of the ball (i.e., on the scale one is probing), either term will dominate over the other, thus giving two types of scaling laws. For a small ball ($R\ll\ell_*$), $\cV^{(D)}\sim R^{D\a_1}$, while $\cV^{(D)}\sim \tilde R^{D\a_2}$ for a large ball ($R\gg\ell_*$), where $\tilde R=R \ell_*^{-1+\a_1/\a_2}$ is the radius of the ball measured in macroscopic length units (units effectively change with the scale in a multi-fractional setting).

%%%%%%%%%%%%%%%%%%%%%%%%%%%%%%%%%%%%%%%%%%%%%%%%%%%%%%%%%%%%%%%%%%%%%%%%%%%%%%%%%%%%%%%%%%%%%%%%

\subsection{Dimensionality and the role of $\dh=2$ and $D=4$}\label{dims}

An important consequence of the geometric picture outlined in section \ref{dis} is that, for a given topological dimension $D\geq 1$, not all fractional measures are possible. In the (at least) spatially isotropic case, the Hausdorff dimension $\dh=\a_0+(D-1)\a$ is associated with $2\a$-norms only if $\a,\a_0\geq 1/2$. If $\a_0\neq 1$, this implies
\be
D\leq 2 \dh\,,
\ee
while for $\a_0=1$ one has
\be\label{dstar}
D\leq 2 \dh-1\,.
\ee
For instance, dimensions $\dh(\a)<2$ are not associated with $2\a$-norms, unless $D\leq 3$. The interesting dimension $\dh(\a)=2$ is achieved at the critical value
\be\label{cripo}
\a_*=\left\{ \begin{matrix} \frac{2}{D}&\quad {\rm if}~\a_0=\a\\
                             \frac{2-\a_0}{D-1}&\quad {\rm if}~\a_0\neq\a\end{matrix}\right. \,.
\ee
Imposing $\a_*\geq 1/2$, the critical value $\a_*$ exists for $D\leq 4$ in the fully isotropic case or in the spatially isotropic case with $\a_0\neq 1$. This excludes an integer time direction if a two-dimensional regime is required. See table \ref{tab1}.

Furthermore, if one also assumes that $\a,\a_0\leq 1$, there is a lower bound for $D$, 
\be\label{dgd}
D\geq \dh\,,
\ee
stating that fractional spacetime is embedded in the abstract space $M^D$. Thus, the dimensionality of the critical point with the requirement of the existence of a natural norm provides a guiding principle in determining the maximal dimension in the infrared. A less clear indication of this feature was noticed in \cite{fra2} for the structure of a scalar propagator. The reader may wonder whether one can also produce geometries where $D<\dh$, as it happens in certain quantum gravity models. In section \ref{fuzz} we will comment on this.

\TABLE{\hspace{3cm}%\begin{table}%[ht]
%\begin{ruledtabular}
\begin{tabular}{|c|c|c|c|}\hline
  		    & $\dh<2$ & $\dh=2$ & $\dh>2$  \\\hline\hline
$\a_0<1$  & $D\leq3$             & $D\leq 4$             & $D\leq 2\dh$          \\
$\a_0=1$  & $D\leq2$             & $D\leq 3$             & $D\leq 2\dh-1$        \\\hline      			 
\end{tabular}%\end{ruledtabular}
\caption{\label{tab1} Allowed topological dimension of the embedding for a fractional spacetime with Hausdorff dimension $\dh(\a_0,\a)$ and natural $2\a$-norm.}
\hspace{3cm}}%\end{table}

In the embedding picture considered in this paper, one can now reinterpret the parameter $\a$ also as a measure of maximal dimensional dispersion. In fact, the difference between the maximum and minimum topological dimension for a given $\dh$ is $\Delta D_{\rm max}=2\dh-\dh=\dh=D\a$ for the fully isotropic model, so that
\be\label{exc1}
\frac{\Delta D_{\rm max}}{D}=\a\,.
\ee
On the other hand, the inverse of $\a$ is the dispersion between the maximum and minimum allowed Hausdorff dimension, $\Delta \dh=D-0=D=\dh/\a$:
\be\label{exc2}
\frac{\Delta \dh}{\dh}=\frac1\a\,.
\ee
Also, the difference between the maximum allowed topological dimension and the actual $D$ is $\Delta D=2\dh-D=D(2\a-1)$, hence
\be\label{exc3}
\frac{\Delta D}{D}=2\a-1\,.
\ee
The special role of $D=4$ consists in allowing, in the same dimensional flow, a phenomenologically viable macroscopic scenario and an ultraviolet configuration with natural $2\a$-norm and Hausdorff dimension equal to 2. The requirement of having a geometric norm throughout the dimensional evolution is crucial. One should be careful in drawing the conclusion that $D=4$ is favoured, since $\dh=2$ was only suggested by quantum gravity arguments (and by a preliminary analysis of the UV finiteness of the theory in $D=4$, section \ref{sdd}). However, there exists a mysterious relation between the fundamental constants of Nature which further supports the very special role of $\dh=2$ as the characteristic dimension of a quantum theory of gravity. Including Planck's constant $\hbar$, the electron charge $e$, Newton's constant $G$, and the speed of light $c$, one can construct a dimensionless constant in $D$ dimensions as $C=\hbar^{3-D} e^{D-2} G^{2-\frac{D}{2}} c^{D-5}$ \cite{Bar83}. The dimensional arguments leading to this combination are unchanged if one replaces the topological dimension with the Hausdorff dimension. Replacing also $\hbar$ with the Planck length
\be\label{plen}
\lp:=\sqrt{\frac{\hbar G}{c^3}} \approx 1.6163 \times 10^{-35}\,\mbox{m}\,,
\ee
the same constant can be recast as
\be\label{cint}
C=\lp^{2(3-\dh)} e^{\dh-2} G^{\frac{\dh}{2}-1} c^{2(2-\dh)}\,.
\ee
In $\dh>3$, the Planck length appears in negative powers, a hint that perturbative quantum gravity is non-renormalizable in these dimensions. At $\dh=3$ the Planck length disappears (and, in fact, perturbative gravity is renormalizable in three dimensions). Remarkably, in $\dh=2$ the fundamental constant coincides with (the square of) the Planck length, $C=\lp^2$, while all the other couplings disappear.

This argument highlights the peculiar status of $\dh=2$ in quantum theories of matter and gravity.\footnote{Noting, as it is usually done, that $[G]=2-D$ vanishes in two dimensions and then deducing that RG properties of gravity are special in $D=2$ is not quite the same. Beside gravity, eq.~\Eq{cint} involves also other forces of Nature.} In the present context of dimensional flow, we can take it as a hint of the special role of $D=4$. Other observations select $D=4$ topological dimensions as special \cite{Bar83}. Just to mention a few examples (which all assume one time direction), stable planetary orbits appear only in $D=4$; stable Bohr atoms, in $D\leq 4$; in $D=4$ the number of generators of spatial translations and rotations is the same, with consequences for electromagnetism; wave signals propagate free of reverberation and distortion only in $D=4$; spacetimes with $D\leq 3$ do not contain gravitational waves; chaos may appear in dynamical systems only in $D\geq 4$; and others. It would be interesting to embed these arguments in a multi-fractional model and apply them to the Hausdorff dimension.

%%%%%%%%%%%%%%%%%%%%%%%%%%%%%%%%%%%%%%%%%%%%%%%%%%%%%%%%%%%%%%%%%%%%%%%%%%%%%%%%%%%%%%%%%%%%%%%%

\subsection{Experimental bounds on dimensional flow}\label{obse}

The Hausdorff dimension is a physical observable. The volume law of, say, a mass distribution concentrated in a spherical region of size $\sim R$ is represented in a log-log plot,
\be
\ln \cV(R) = \ln\Om_{D,\a}+D\a \ln R\,,
\ee
where $\Om_{D,\a}$ is the volume of a $D$-ball with unit radius. One can, in principle, obtain independent measurements of volumes and radii and determine both the dimensionality of the mass distribution (tilt of the line) and the solid-angle factor (offset). With adequate technology, one can translate this type of experiments in the realm of spacetime geometry. Experimental constraints on fractional models can be obtained from observations where gravitational effects are almost or completely negligible. These can be, mainly, of three types: particle physics tests, equivalence principle/Lorentz invariance tests, and post-Newtonian tests. We saw in \cite[section 3.5]{frc1} that different presentations of fractional geometry yield different output values for, e.g., volume measurements, but they enjoy the same scaling properties for a given $\a$. Thus, if we assume that at scales about and above those probed by accelerators we are already out of the multi-fractional regime, we can consider a $\dh=D-\e$ expansion for a fixed-$\a$ configuration, and focus our attention to the parameter $\e$. This falls short of constraining the characteristic scale $\ell_*$ at which the UV critical point is approached, since these experiments correspond to scales $\ell\gg\ell_*$. Anyway, later we will also give a bound on $\ell_*$ from particle physics.

In table \ref{tab2} we report the expressions of the unit volume $\Om_{D,\a}$ at $D=2,3,4$, where
\be
\a=1-\frac{\e}{D}\,.
\ee
Calculations where $\dh=D-\e$ correspond, in fractal geometry, to regimes with low lacunarity, i.e., where fractal space is almost translation invariant \cite{GMBA}.

\TABLE{\hspace{2.5cm}%\begin{table}%[ht]
%\begin{ruledtabular}
\begin{tabular}{|c||c||c|}\hline
$D$ & $\Om_{D,1-\e/D}$            & $\Om_{D-\e,1}$ \\\hline\hline
2   & $\pi(1-0.42\e)$             & $\pi(1-0.36\e)$ \\
3   & $\frac{4\pi}{3}(1-0.54\e)$  & $\frac{4\pi}{3}(1-0.22\e)$\\
4   & $\frac{\pi^2}{2}(1-0.63\e)$ & $\frac{\pi^2}{2}(1-0.11\e)$ \\\hline      			 
\end{tabular}%\end{ruledtabular} %$\frac{4\pi}{3}-2.26\e$  $\frac{4\pi}{3}+1.77\e$
\caption{\label{tab2} Volume $\Om_{D,\a}$ of unit $D$-balls in various dimensions, for $\a\sim 1$. The corrections in traditional dimensional regularization are shown in the last column \cite{frc1}.}
\hspace{2cm}}%\end{table}\hspace{2cm}}%\end{table}

The coefficients in front of $\e$ are all of the same order of magnitude in fractional models and in spacetimes with dimension modified according to the dimensional-regularization scheme \cite{BG72}--%,tV,
\cite{Lei75}, so we can accept bounds on the latter and apply them to the former. For instance, measurements of the anomalous magnetic moment $g-2$ of the muon can compare the theoretical prediction in four dimensions with that in spacetimes with dimension modified according to the dimensional-regularization scheme. The order of magnitude of the upper bound on $\e$ was estimated as $|\e|<10^{-5}$ in \cite{ScM}, while from the anomalous magnetic moment of the electron, $\e\sim 10^3 |g_{\rm theor}-g_{\rm exp}|$ \cite{Svo87,SvZ} and one finds $|\e|<10^{-7}$ \cite{ZS}. Taking the latest experimental determination of the muon $g-2$ \cite{JN}, where $g_{\rm exp}-g_{\rm theor}\sim 10^{-11}$, we have
\be\label{boue0}
|\e|<10^{-8}\,,\qquad \ell\sim 10^{-15}\,{\rm m}\,.
\ee
Going to atomic scales, experimental measurements of the Lamb shift in hydrogen yield \cite{ScM,MuS}
\be\label{boue1}
|\e|<10^{-11}\,,\qquad \ell\sim 10^{-11}\,{\rm m}\,,
\ee
tighter than the previous bounds.

As far as intermediate-scale experiments are concerned ($\ell < 10^{3}~{\rm m}$), one could reinterpret tests of the equivalence principle \cite{HLa}. Of course, at mesoscopic scales above particle physics and below planetary we know that the equivalence principle holds with great accuracy, and local physics is described by Minkowski spacetime (plus eventual corrections of general relativity, measurable already at the size of the LHC accelerator). However, it is a legitimate question to ask what the bounds would be on a $4-\e$ geometry. So, one can use fractional equations correcting Euclidean geometry and apply them to the data of these experiments. The equivalence principle is closely related to Lorentz invariance, and tests of the latter would provide parallel constraints on fractional effects \cite{Mat05,JLM}. Upper bounds on $\e$, and hence on the dimensional flow at such mesoscopic scales, should be quite stringent.

These bounds should be compared with others done at larger scales. Anomalous correlation functions result in deviations from Newton's law and a precession of planetary orbits. Taking into account the theoretical prediction of general relativity for the precession of Mercury, any dimensional effect should be smaller than the experimental error, thus yielding \cite{ScM,MuS,JY}
\be\label{boue2}
|\e|<10^{-9}\,,\qquad \ell\sim 10^{11}\,{\rm m}\,,
\ee
less tight than eq.~\Eq{boue1}. A similar bound is obtained from pulsar measurements at a distance $\delta\sim10^4\,{\rm ly}$ from us \cite{JY}, so that one can regard \Eq{boue1} as valid not only here and today, but also in a spacetime sphere of size $\delta$.

Finally, a fit of the black-body spectrum of the cosmic microwave background gives an upper bound on $\e$ at spacetime scales comprised between the decoupling era and today \cite{CO}:
\be\label{boue3}
|\e|<10^{-5}\,,\qquad \ell\sim 14.4\,\mbox{Gpc}\,.
\ee
The best-fit value of $\e$ is strictly positive, in agreement with the direction of dimensional flow in fractional theories.

All the above estimates rely on a number of assumptions which require a careful scrutiny in the present framework of fractional spacetime. Among these assumptions, we mention naive implementations of dimension effects (via dimensional regularization, which is of a non-dynamical nature), the use of unmodified Einstein or Schr\"odinger equations, and integer time direction (in our language, fractional charge $\a_0=1$). If the parameter $\e$ is rendered dynamical, as it would naturally be in multi-fractional dimensional flow, then the above particle-physics bounds are no longer reliable and, in fact, a determination of spacetime dimension becomes much more difficult. Examples are measurements of oscillations of neutral $B$ mesons and of the muon $g-2$ \cite{She09}: At mass scales $M> 300\div 400~ {\rm GeV}$, any $2<\dh<5$ is compatible with experiments. This translates into a rough upper bound for $\ell_*$:
\be
\ell_*<10^{-18}\,\mbox{m}\,.
\ee

To carry out a complete revision of the results presented in this section, one will have first to extend the model to curved spacetimes with gravity. This is part of the theoretical programme here proposed for future studies; in parallel, the improvement of data could allow an update on some of the above results. Lamb shift measurements seem to be the most promising for the tightest absolute constraint. However, one should not underestimate the bound \Eq{boue3}, which, if confirmed, would constrain the end of the multi-fractional era of the Universe at times prior to the formation of the cosmic microwave background.

%%%%%%%%%%%%%%%%%%%%%%%%%%%%%%%%%%%%%%%%%%%%%%%%%%%%%%%%%%%%%%%%%%%%%%%%%%%%%%%%%%%%%%%%%%%%%%%%%%%%%%%%%%%%%%%%%%%%%%%%%%%%%%%%%%%%%%%%%%%%%%%%%%%%%%%%%%%%%%%%%%%%%%%%%%%%%%%%%%%%%%%%%%%%%%%%%%%%%%%%%%%%%%%%%%%%%%%%%%%%%%%%%%%%%%%%%%%%%%%%%%

\section{Scalar field theory}\label{st}

A Lorentz scalar is a favourite guinea pig with which to check the properties of a field theory in a given geometry setting. We studied a real scalar field in \cite{fra2} for a generic absolutely continuous Lebesgue--Stieltjes measure. Here we do just that but focussing on fractional measures and power-counting renormalization properties. The latter are important when fractional theories are regarded as fundamental, in which case one should check that they are ultraviolet finite.

%%%%%%%%%%%%%%%%%%%%%%%%%%%%%%%%%%%%%%%%%%%%%%%%%%%%%%%%%%%%%%%%%%%%%%%%%%%%%%%%%%%%%%%%%%%%%%%%

\subsection{Power-counting renormalizability}\label{pc}

A standard power-counting argument is sufficient to understand qualitatively the relation between coupling dimensionality and renormalization properties of a field theory \cite{Pol92}--%,Ram97,
\cite{Wei95}. The reader already acquainted with it can skip this section. When constructing perturbation theory, one must take into account all possible gauge-inequivalent interactions order by order in the effective low-energy action. Some of the couplings diverge when the regulator in the regularization scheme is removed. However, if the operators $\cO\sim\int\rmd\vr\,\cO_d$ associated with these couplings $g_{\rm ren}$ are already present at the tree level, with bare couplings $g$ of dimension $\dh-d$ (where $d\geq 0$ is the scaling dimension of $\cO_d$), one can absorb the divergence into an effective coupling which is defined to be finite when the regulator is removed. Contrary to the bare couplings $g$, the effective couplings $g_{\rm eff}=g+g_{\rm ren}$ are what one physically measures. If this procedure works order by order, the theory is said to be perturbatively renormalizable, and hence physically predictive. This means that the number of physical couplings we measure at any perturbative order is finite. 

In the renormalization group picture, the physical action stems from the bare action by integrating out momentum modes greater than a certain energy cut-off scale $E$, and then removing the cut-off. An operator is said to be relevant if its associated coupling $g$ has positive scaling dimension. On the other hand, operators with dimensionless coupling are called marginal, while operators with $[g]<0$ are irrelevant. In terms of dimensionless constants
\be
\tilde g= g E^{d-\dh}\,,\qquad [\tilde g]=0\,,
\ee
the operator $\cO$ associated with $g$ scales as
\be
\cO=  g \int \rmd\vr\, \cO_d \sim \tilde g \left(\frac{k}{E}\right)^{d-\dh}\,,
\ee
where $k$ is the momentum. Therefore, relevant operators are important at low energies ($k/E\ll 1$). Marginal operators are equally important at all scales. Their detailed behaviour is not obvious and these can be, case by case, either marginally relevant or irrelevant. Irrelevant operators become important in the ultraviolet ($k/E\gg 1$) but, contrary to what the name suggests, some of them can also alter macroscopic physics. Since, typically, there is a finite number of relevant operators and also of marginal operators, macroscopic physics is described only by few observables. In fractional theories, this would not be the case if we included the infinite class of operators with fractional derivatives. Imposition of symmetries will drastically reduce this infinite multiplicity.
 
If divergences are present, they correspond to local operators of dimensionality increasing with the order of the perturbation expansion. Suppose the bare action $S$ contains only relevant operators; then, only a finite number of relevant operators (those which did not appear in $S$) will enter the effective action, and any divergence will be absorbed in the finite number of couplings $\{ g\}$. For instance, in electromagnetism the electron mass and charge have non-negative dimension in natural units, and one can formally absorb the divergences just in these two coupling constants, which are then determined by experiments. Conversely, if even one irrelevant operator appears in $S$, one can construct new irrelevant operators at each order. Explicit calculations can determine whether their couplings are finite or not. If they diverge, the perturbative approach looses predictivity because we can absorb all the divergences only by adding an infinite number of operators to the action.

A theory is said to be \emph{power-counting renormalizable} if 
\be\label{powcr}
[g]\geq 0
\ee
for all bare couplings $ g$. This condition is not sufficient to guarantee that the theory be renormalizable in the sense of the full renormalization group flow, but it provides a good guiding principle in many situations. If a model is not power-counting renormalizable, then it will likely be non-renormalizable unless remarkable divergence cancellations happen. An example is ordinary and supersymmetric gravity, where these cancellations do happen \cite{BCFIJ} and explicit calculations are necessary to settle the issue.

The relation between the good UV behaviour of a theory and the absence of irrelevant operators can be understood by looking at the \emph{superficial degree of divergence} of a Feynman diagram. Consider a one-particle-irreducible Feynman sub-graph with $L$ loops, $I$ internal propagators and $V$ vertices. The superficial degree of divergence $\de$ is the canonical dimension of all these contributions: given a UV energy cut-off $E$, the divergent part of the diagram scales as $E^\de$. If $\de=0$, one has at most superficial logarithmic divergences and the theory is power-counting renormalizable. When $\de<0$ for every sub-diagram in a Feynman graph, the graph is convergent; if only a finite number of Feynman diagrams diverge superficially, the theory is power-counting super-renormalizable.

We can count divergences in the case of an ordinary scalar field theory in $D$ dimensions:
\be
S=\int\rmd^D x\,\left(\frac12\phi\p_\mu\p^\mu\phi-\frac12m^2\phi^2-\frac{g_p}{p}\phi^p\right)\,.
\ee
Each loop integral over momenta gives $[\rmd^D k]=D$, while the propagator $\tilde G(k^2)=-1/(k^2+m^2)$ has $[\tilde G]=-2$. Interaction vertices do not carry dimensionality and, overall, $\delta = DL-2I$. Since $I\geq L$, the maximum superficial degree of divergence can be $L(D-2)$. $L$ is the number of independent momenta, given by $I$ minus the number of relations they satisfy among themselves: these are $V-1$ (one for each vertex, up to the total momentum conservation), so that 
\be
L=I-V+1\,.
\ee
This result is often called Euler's theorem for graphs. With only mass and a $\phi^p$ interaction, for each vertex there are $p$ lines, so that $pV=N+2I$, where $N$ is the number of external legs in the diagram. Replacing $L$ and $I$ with these expressions, one obtains
\be\label{sudedi}
\de = DL-2I = D-[g_p] V-\left(\frac{D}{2}-1\right)N\,,
\ee
where we used $[g_p]=D-p(D-2)/2$. This formula can be also derived by dimensional arguments. A diagram with $N$ external lines can be generated by a $g_N\phi^N$ term, so that its scaling dimension is $[g_N]$. On the other hand, with only the $\phi^p$ interaction term available, the divergent part of the diagram scales as $g_p^V E^\de$. Therefore, we have $[g_N]=[g_p] V+\de$, eq.~\Eq{sudedi}. 

If $N$ is the maximum power in the potential, the superficial degree of divergence is $\de =[g_N](1-V)$. For the theory to be power-counting renormalizable, it must be $[g_N]\geq 0$, implying 
\be\label{scap0}
N\leq p_D:=\frac{2D}{D-2}\,.
\ee
In two dimensions, $\de$ does not depend on the number of external legs ($N$ is unconstrained) and the greater the number of vertices the more convergent is the diagram. In four dimensions, the $\phi^4$ theory is renormalizable while higher powers of $\phi$ are responsible for an infinite number of divergent diagrams. In general, $\delta$ is bounded by the dimension of operators which already appear in the bare action.

%%%%%%%%%%%%%%%%%%%%%%%%%%%%%%%%%%%%%%%%%%%%%%%%%%%%%%%%%%%%%%%%%%%%%%%%%%%%%%%%%%%%%%%%%%%%%%%%%%%

\subsection{Actions of fractional field theories}\label{afft}

\subsubsection{Role of symmetries}
 
We are mainly interested in the interplay between the renormalization group flow and the symmetries of the model. In ordinary field theories, it is natural to impose the symmetries of the measure also to the Lagrangian density $\cL$. These symmetries are preserved along the renormalization group flow, and they protect the theory from an infinite multiplicity of non-covariant relevant operators. In fractional field theories, on the other hand, we are in an unusual situation. The measure does change along the RG flow, and fractional Lorentz invariance is not constant with the scale: both the fractional charge $\a$ and the form of eq.~\Eq{fpotra} flow as well. We do need symmetries to protect the model from uncontrollable divergences, but the remarkable role of geometry in fractional scenarios strongly constrains the action even in the absence of RG-constant symmetries. Calling $\Pi_\a$ and $\Pi_\cL$ the symmetry group of, respectively, the measure and the Lagrangian density, we can construct two inequivalent classes of fractional field theories:
\begin{itemize}
\item \emph{Integer-symmetry scenario.} While the symmetry of the measure guarantees protection against arbitrary measure distributions, one can prescribe a constant symmetry for $\cL$ in the RG sense. Since in the infrared the Lagrangian should be Lorentz invariant, we assume $\Pi_\cL=\Pi_1$, the integer Poincar\'e group in $D$ dimensions.
\item \emph{Fractional-symmetry scenario.} In ordinary models, the symmetry of the measure and of the Lagrangian density are the same. If we impose $\Pi_\a=\Pi_\cL$, we obtain an action invariant under fractional Lorentz transformations. This symmetry varies along the RG flow but it forbids non-scalar operators at any given $\a$.
\end{itemize}
We consider these cases separately but we anticipate that physical arguments in favour of the integer-symmetry scenario will be advanced in sections \ref{friss} and \ref{pro}. 

As in section \ref{mul2}, take an action $S_{\a(\s)}$ with fixed fractional order at every given $\s$. The total Lagrangian can be split into a kinetic and potential part, $\cL=\cL_\cK-\cL_V$. By definition, the kinetic term is characterized by a dimensionless coupling (in the ultraviolet), a quadratic dependence on the field $\phi$, and a differential operator $\cK$:
\be\label{lk1}
\cL_\cK=\tfrac12\phi \cK \phi\,.
\ee
Symmetrized kinetic terms of the form ${\rm D}\phi\,{\rm D}\phi$ can be recast in the form \Eq{lk1} after integration by part. Later we shall select the important case 
\be\label{lk2}
\cL_\cK=-\tfrac12\p_\mu\phi\, \p^\mu \phi\,,
\ee
corresponding to
\be\label{ck2}
\boxd{\cK:=\eta^{\mu\nu}\left(\p_\mu\p_\nu+\frac{\p_\mu v_\a}{v_\a}\p_\nu\right) = \eta^{\mu\nu}\left(\p_\mu\p_\nu-\frac{1-\a}{x^\mu}\p_\nu\right)\,,\qquad [\cK]=2\,.}
\ee
In the last step, the choice $a_0\neq\a$ is implicitly allowed.

To get a dimensionless operator
\be
\cO_{\cK}=\int\rmd\vr_\a\, \cL_\cK\,,\qquad [\cO_{\cK}]=0\,,
\ee
the scaling dimension of the field should be 
\be\label{kei}
[\phi]=\frac{\dh-[\cK]}{2}\,.
\ee
As far as the potential term is concerned, we take power-law operators
\be\label{potpo}
\cO_p = \frac{g_p}{p} \int\rmd\vr_\a\,\phi^p\,,
\ee
where
\be\label{sned0}
[g_p]=\dh-\frac{p}{2}(\dh-[\cK])\,.
\ee
We will be interested in the left and right fractional Beltrami--Laplace operators
\be\label{kkg}
\cK_\g:=\eta^{\mu\nu}\p^\g_\mu\p^\g_\nu\,,\qquad \bar\cK_\g:=\eta^{\mu\nu}{}_\infty\bp^\g_\mu{}_\infty\bp^\g_\nu\,,
\ee
which define an infinite multiplicity of derivative operators:
\be\label{Oab}
\cO_{\a,\g,n}=g_{\a,\g,n}\int\rmd\vr_\a\, \phi(\cK_\g)^n\phi\,,\qquad \bar\cO_{\a,\g,n}=g_{\a,\g,n}\int\rmd\vr_\a\, \phi(\bar\cK_\g)^n\phi\,,
\ee
where $\a,\,\g>0$, $\g$ will be chosen later, $n\geq 1$ is integer, and $[g_{\a,\g,n}]=2(1-n)\g$. When $n=1$, $\cO_{\a,\g,1}$ and $\bar\cO_{\a,\g,1}$ will be denoted as $\cO_{\a,\g}$ and $\bar\cO_{\a,\g}$, respectively.

\subsubsection{Fractional Klein--Gordon equation}

The dynamics of fractional systems have been studied both in Hamiltonian and Lagrangian formalism. Fractional phase space and Hamilton equations are known for classical-mechanics systems with integer measure and fractional derivatives \cite{Rie1}--%,Rie2,DrY,Tar1,Tar2,Tar4,Tar5,Cre07,Tar10,
\cite{RAMB}, with integer derivatives and fractional measure \cite{El05a}, or with measure and derivatives both fractional \cite{MBR}--%,ElT1,
\cite{AbG}. Still in fractional mechanics, generalizations of the variational principle and Lagrangian equations of motion have been explored for one-dimensional actions with integer measure and fractional derivatives \cite{Rie1,Rie2,Cre07,Agr02,MuB}, fractional measure and integer derivatives \cite{El05a,El05b}--%,FT2,
\cite{FT3}, and measure and derivatives both fractional in one \cite{MBR,ElT1} and many dimensions \cite{ElT2}. Variational principle and equations of motion for Lebesgue--Stieltjes actions with absolutely continuous measure were considered in \cite{fra2,PSt,UO}, in connection with fractional dynamics and, in particular, field theory.
The same methods allow one to construct Noether currents and (non-)conservation laws \cite{fra2,PSt,Cre07,FT2,FT3} for these dissipative systems \cite{Rie1,Rie2,El05a,El05b,Pel91}.

Variation of fractional actions follow the rules of fractional integration by parts (e.g., \cite[section 2.3.5]{frc1}). Given two suitable functions $f_1$ and $f_2$, from \cite[eq.~(2.1.50)]{KST} 
\be\label{ibp}
\int_{x_0}^{x_1}\rmd x\, f_1\,\p^\a f_2 = \int_{x_0}^{x_1}\rmd x\, f_2\,\bp^\a f_1\,,
\ee
one obtains the integration by parts of fractional integrals of fractional integrands. Recalling that left and right derivatives have, respectively, lower terminal $x_0=0$ and upper terminal $x_1=+\infty$,
\ba
\bar I^\a_{0,\infty}\left\{f_2 \p^\g f_1\right\} &=& \frac{1}{\Gamma(\a)}\int_{0}^{+\infty} \rmd x\,x^{\a-1}\, f_2 \p^\g f_1\nonumber\\
                      &=& \frac{1}{\Gamma(\a)}\int_{0}^{+\infty} \rmd x\, f_1\,{}_{\infty}\bp^\g [x^{\a-1}f_2]\,.
\ea
Using the Leibniz formula, the last derivative term can be expanded as an infinite series if $\g\neq 1$. In many dimensions and for the scalar theory with potential $V(\phi)$, the action integral
\be\label{sag}
S_{\a,\g}=\int\rmd^Dx\,v_\a\,\cL(\phi,\cK_\g\phi,\bar\cK_\g\phi)
\ee
will depend on the kinetic operators $\cK_\g$ and $\bar\cK_\g$. Consider an infinitesimal variation $\de\phi$ which vanishes at the boundary $x=0,\infty$. Then, the equation of motion at given $\a$
\be
0= \frac{\de S_{\a,\g}}{\de\phi}
\ee
is
\be
0 = \frac{\p\cL}{\p\phi}+\frac{1}{v_\a}\left\{\bar\cK_\g\left[v_\a\frac{\p\cL}{\p(\cK_\g\phi)}\right]+\cK_\g\left[v_\a\frac{\p\cL}{\p(\bar\cK_\g\phi)}\right]\right\}\label{eomab}\,,
\ee
where, for convenience, we expressed the functional variations in terms of differential operators of order $2\g$. Varying with respect to operators of order $\g$ would not change the final result but would make its derivation more complicated.

If $S_{\a,\g}=\bar\cO_{\a,\g}-\cO_V$, eq.~\Eq{eomab} yields
\be\label{eomb}
0=\frac12\left[\bar\cK_\g\phi+\frac{1}{v_\a}\cK_\g(v_\a\phi)\right] -V'(\phi)\,,
\ee
where $V'=\p V/\p\phi$. In the particular case $\g=1$, the last two terms of eq.~\Eq{eomab} collapse into one another and the Euler--Lagrange equation reads
\be\label{eleb}
0 %&=&\frac{\p\cL}{\p\phi}-\frac{1}{v_\a}\p_\mu\left[v_\a\frac{\p\cL}{\p(\p_\mu\phi)}\right]\\
  =\frac{\p\cL}{\p\phi}+\frac{1}{v_\a}\B\left[v_\a\frac{\p\cL}{\p(\B\phi)}\right]\,,
\ee
where $\B=\p_\mu\p^\mu$, so that
\bs\label{kgeq1}\ba
0 &=&\B\phi+\frac{\p_\mu v_\a}{v_\a}\p^\mu\phi+\frac{\B v_\a}{2v_\a}\phi -V'(\phi)\\
  &=& \B\phi-\frac{1-\a}{x^\mu}\p_\mu\phi+(2-\a)(1-\a)\phi\left(\frac{1}{x^\mu}\eta_{\mu\nu}\frac{1}{x^\nu}\right) -V'(\phi)\,.\label{eom2}%\eta_{\mu\nu}\frac{1}{x^\mu x^\nu}\,.
\ea\es
Here contracted indices are summed over as usual; in particular, the next-to-last contribution is $\eta_{\mu\nu}(x^\mu x^\nu)^{-1}=-t^{-2}+x_1^{-2}+\dots+x_{D-1}^{-2}$. These equations can be extended to multi-fractional spacetimes simply by summing over $\a$ with weights $g_\a$:
\be
\boxd{0 =\sum_\a g_\a \left\{v_\a\frac{\p\cL}{\p\phi}+\bar\cK_\g\left[v_\a\frac{\p\cL}{\p(\cK_\g\phi)}\right]+\cK_\g\left[v_\a\frac{\p\cL}{\p(\bar\cK_\g\phi)}\right]\right\}\label{eomm}\,.}
\ee
For $\g=1$, eq.~\Eq{kgeq1} has a term proportional to $\B v_\a$, absent in \cite{fra1,fra2}. In the case of \Eq{lk2}, in fact, one has
\be\label{kgeq2}
0 =\cK\phi -V'(\phi)=\B\phi+\frac{\p_\mu v_\a}{v_\a}\p^\mu\phi -V'(\phi) = \B\phi-\frac{1-\a}{x^\mu}\p_\mu\phi-V'(\phi)\,.
\ee
and its multi-fractional generalization
\be\label{kgeq2mf}
\boxd{0 =\sum_\a g_\a \left\{\B\phi+\frac{\p_\mu v_\a}{v_\a}\p^\mu\phi -V'(\phi)\right\}\,.}
\ee

With respect to \cite{fra1,fra2}, there is an important difference: the weight $v_\a=\prod_\mu \p_\mu \x(x^\mu)$ is here regarded as a fixed coordinate profile, not a Lorentz scalar. A consequence of this fact is that the equations of motion are not Lorentz invariant (in a fractional sense, and for any $\g$) as in the Lebesgue--Stieltjes interpretation of \cite{fra1,fra2}, where no particular profile is chosen for $v_\a(x)$. Thus, we find ourselves in a situation where the action is invariant under certain symmetries while the equations of motion are not. The root of the problem is, of course, the residual $v_\a$ dependence after integrating by parts. Then, classical physics is not invariant under any of the symmetries enjoyed by the action, except in the IR limit where Lorentz invariance is restored up to $O(1-\a)$ terms.\footnote{One might try to reconcile this situation with symmetry requirements in field theory by regarding the Lebesgue--Stieltjes formalism of \cite{fra1,fra2} as a general framework with absolutely continuous measures, and fractional models with integer-order kinetic operators as explicit realizations (via a particular coordinate presentation of the differential structure) breaking the formal symmetry of the general formalism. This would be somewhat analogous to a choice of background metric in general relativity: while the mother theory is diffeomorphism invariant, explicit solutions break all or most of the symmetries. Similarly, fractional models may be interpreted as explicit realizations of a non-standard differential structure. In that case, however, the fractional Lorentz symmetry of the fractional measure would be regarded as accidental. Also, fractional models with fractional-order kinetic operators, and any of the geometric properties typical of fractional models considered here and in \cite{frc1}, would hardly stem from a presentation-independent framework. For these reasons, our attitude is not to regard fractional theories as a subclass of more general formulations.}

%%%%%%%%%%%%%%%%%%%%%%%%%%%%%%%%%%%%%%%%%%%%%%%%%%%%%%%%%%%%%%%%%%%%%%%%%%%%%%%%%%%%%%%%%%%%%%%%

\subsection{Integer-symmetry scenario}\label{friss}

The right derivative is not a derivative operator for the left fractional coordinates $\x^\mu$. Even if one is entitled to define the theory with one type of derivative, the other type will always pop in via eq.~\Eq{ibp}; this is at variance with the ordinary calculus of variations, where the same operator appears both in the action and in the equations of motion. 

%To overcome this obstacle, one %could either modify the theoretical structure or change attitude. In the former case, we could construct a vector space with twice the number of dimensions, $X=(\x,\bx)^T$ ($T$ indicates transpose), but with independent sectors, together with matrix differential and derivative operators acting on vector functions:
%\be\nonumber
%\mathcal{D}^\a:=\left(\begin{matrix} \p^\a & 0 \\ 0 & \bp^\a\\ \end{matrix}\right)\,,\qquad F(X):=\left(\begin{matrix} f(\x) \\ \bar{f}(\bx)\\ \end{matrix}\right)\,,
%\ee
%and so on. Then, eq.~\Eq{ibp} becomes a sort of scalar product,
%\be\nonumber
%\int_{x_0}^{x_1}\rmd x\, F^T \mathcal{D}^\a G = \int_{x_0}^{x_1}\rmd x\, G^T \mathcal{D}^{\a *} F\,,
%\ee
%where we defined a conjugation operation as
%\be\nonumber
%\mathcal{D}^{\a *} := U^\dagger \mathcal{D}^{\a} U\,,\qquad  U :=\left(\begin{matrix} 0 & 1 \\ 1 & 0\\ \end{matrix}\right)\,.
%\ee
%The left and right sectors talk to each other only in the equations of motion. This is a formal writing and, however useful as a compact notation, it adds nothing to the physical interpretation of the theory. Perhaps it will be interesting to further develop this model, which we shall drop in this paper. 
%A more economic way to go on is to change attitude. The first option is to 
An alternative is to specialize to the case $\g=1$ in eq.~\Eq{sag} and prescribe that the Lagrangian contains only ordinary derivatives, eqs.~\Eq{eleb}--\Eq{eom2}. This scenario has the double advantage of simplicity (removing any reference to independent sectors) and of constituting the correct \emph{Ansatz} for a field theory which is power-counting renormalizable at the two-dimensional critical point $\a_*$. It is indeed one choice of operator $\p^\g$ against infinitely many others parametrized by $\g$, but it is also the only one with both these characteristics.

In the spirit of the integer-symmetry class of scenarios, this is also the natural choice. Contrary to integer-order actions, however, $\phi\p^\mu\p_\mu \phi\neq -\p_\mu\phi\p^\mu\phi$ under integration by parts, because the non-trivial measure weight is responsible for extra friction terms. Therefore, the ordering of derivatives in the kinetic term is important to determine the dynamical properties of the model (also at the quantum level, through the propagator, as we shall see later). Here we take eq.~\Eq{ck2}, as in \cite{fra1,fra2} but contrary to \cite{fra3} (where the $\g=1$ special case of eq.~\Eq{kkg}, $\cK_1=\B$, was assumed).

The field and coupling \Eq{sned0} have dimension
\be\label{coco2}
[\phi]=\frac{\dh}{2}-1\,,\qquad [g_p]=\dh-p\frac{\dh-2}{2}\,.
\ee
The kinetic operator $\cO_\cK$ is marginal. The $\cO_p$ are marginal or relevant if, and only if,
\be\label{scap}
p \leq p_\a:=\frac{2\dh}{\dh-2}\,.
\ee
The total action at a given $\s$ is
\be\label{sas2}
S_{\a(\s)}=-\int\rmd\vr_{\a(\s)}\left[\frac12\eta^{\mu\nu}\p_\mu\phi\p_\nu\phi+\frac{g_p}{p}\sum_{p\leq p_{\a(\s)}}\phi^p\right]\,.
\ee
The fact that the dimension of the scalar field vanishes at $\dh=2$, eq.\ \Eq{coco2}, signals the presence of a critical point, as it happens, for instance, in Ho\v{r}ava--Lifshitz models. For general $D$, there exists such a point at $\a=\a_*=2/D$, where $\dh=2$ and all the $\cO_p$ are relevant \cite{fra1,fra2}. A critical point, eventually identified with a UV fixed point, is important for the power-counting renormalizability of the theory, proven in section \ref{sdd} (see also \cite{fra2}). This property supersedes any heuristic argument such as eq.\ \Eq{cint} in selecting $\dh=2$ as special.

In section \ref{dims}, we have seen that the existence of a norm throughout the dimensional flow constrains the geometry to configurations with $\a\geq 1/2$. We can make a conjecture and identify the critical point $\a_*$ with the lowest allowed $\a$; this is most natural in a multi-fractional scenario, where one should get continuous access to all admissible points in the flow. Then, if the lower limit $\a=1/2$ coincides with the critical value $\a_*=2/D$, one gets $D=4$. Therefore, the dimension of the embedding can be fixed to $D=4$ not because of phenomenological reasons, but by virtue of a combination of reasonable geometric requirements.
%Apart from the operators considered here, there are infinitely many more combinations of derivatives and powers of the scalar field. %Since the integer action is not invariant under one given symmetry group, the equations of motion will not enjoy any symmetry. This is apparent from eq.~\Eq{eom2}.

%%%%%%%%%%%%%%%%%%%%%%%%%%%%%%%%%%%%%%%%%%%%%%%%%%%%%%%%%%%%%%%%%%%%%%%%%%%%%%%%%%%%%%%%%%%%%%%%

\subsection{Fractional-symmetry scenario}\label{frass}

Instead of fixing the kinetic term to the integer Beltrami--Laplace operator $\cK_1$ or to \Eq{lk2}, one could take fractional differential operators. In doing so, we would accept the fact that fractional theories will generically feature different operators at the action and the dynamical level, but at the same time we would like to understand why. The answer is given by eqs.~\Eq{pari1}--\Eq{pari3}: \emph{The operators $\p^\a$ and $\bp^\a$ are related to each other by a reflection centered at $x_0+x_1$.} Debating whether to use $\p^\a$ or $\bp^\a$ in the action is tantamount to discriminating, in the integer case, between the operators $\rmd/\rmd x$ and $-\rmd/\rmd x$, and the statement that $\bp^\a\x=f(\x)\neq 1$ for some non-trivial $f(\x)$ is the analogue of $\rmd x/\rmd(-x)=-1\neq 1$. In the fractional case, powers of $-1$ make the issue and the formul\ae\ visually complicated, but the essence is the same. In this respect, the choice between a left or a right theory is merely political. If the integration domain is symmetric, $x_0=-x_1$, then also the reflection is symmetric and we usually describe it as a parity or a time reversal transformation. Also, the Weyl derivative ${}_\infty\bp^\a$ and the Caputo operator $\p^\a$ with lower terminal $x_0=0$ are formally conjugate to each other under a reflection at infinity.

Operators \Eq{Oab} are not invariant under fractional Poincar\'e transformations with charge $\a$ unless $\g=\a$. We keep only $\cO_{\a,\a,n}$, which are based on the kinetic operator
\be\label{Ka}
\bar\cK_\a=\eta^{\mu\nu}{}_\infty\bp^\a_\mu {}_\infty\bp^\a_\nu\,,\qquad [\bar\cK_\a]=2\a\,.
\ee
Then, for $\dh=D\a$,
\be\label{coco1}
\phi=\left(\frac{D}{2}-1\right)\a\,,\qquad [g_{\a,\a,n}]=-2\a(n-1)\,,\qquad [g_p]=\left(D-p\frac{D-2}{2}\right)\a\,.
\ee
The anisotropic case $\a_0\neq\a_{i}\neq\a_{j}$ is not possible for a scalar field, unless one introduces dimensionful couplings into the definition of the fractional d'Alembertian. These couplings should then appear also in the matrices $\tilde\Lambda_\nu^\mu$, acting on the geometric coordinates $\x^\mu:=(x^{\mu})^{\a_{\mu}}/\Gamma(\a_{\mu}+1)$. We do not consider this multi-scale Lorentz symmetry here.

Equation \Eq{Ka} or $\cK_\a$ are not the only second-order operators invariant under fractional Lorentz transformations. In fact, one can also take
\be\label{Ka2}
\tilde \cK_\a=\eta^{\mu\nu}\frac{\p}{\p \x^\mu} \frac{\p}{\p \x^\nu}\,,
\ee
which has the same scaling properties and yields a very similar model. However, in this case one could formulate the theory directly in $\x$ coordinates with no need of fractional calculus, and the connection with fractal geometry would be somehow lost. In the following we do not consider eq.~\Eq{Ka2}.

In the external time picture, the engineering dimension of the scalar field changes with $\s$, thus giving a sort of effective continuous renormalization group flow. In general, one should add all possible relevant operators, which would emerge anyway at the quantum level. %A difference between doing so in the flowing $\a$ picture and in the constant $\a$ picture is about units. In the first case, the conformal dimension of the scalar field (eq.~\Eq{kin}) varies with the scale, so that at each scale one makes measurements in natural units. In the second case, the dimension of the field is determined in the ultraviolet, where it is zero, and remains so at all scales; in the infrared, dimensionful coupling constants are reabsorbed into the definition of $\phi$ to obtain the usual scaling in $D$ dimensions, $[\phi]=D/2-1$. From the point of view of the low-energy effective theory, dimensional analysis is more transparent in the units adopted in this section.

From eq.~\Eq{coco1}, the only marginal operators are, when $\a\neq 0$, $\bar\cO_{\a,\a,1}=\bar\cO_{\a,\a}$ and $\cO_{p_D}$, with $p_D$ given by eq.~\Eq{scap0}, while for $\a=0$ all the operators are marginal. Relevant operators exist only if $\a\neq 0$ and are the $\cO_p$ with $p<p_D$. This is the same condition as in ordinary $D$-dimensional field theories: in four dimensions, $V\sim\phi^2$ and $V\sim\phi^3$ are relevant and $V\sim\phi^4$ is marginal. The total action with fractional symmetry at a given $\s$ is
\be\label{sas}
S_{\a(\s)}=\bar\cO_{\a(\s),\a(\s)}-\sum_{p\leq p_D}\cO_p=\int\rmd\vr_{\a(\s)}\left[\frac12\phi\,\bar\cK_{\a(\s)}\phi-\frac{g_p}{p}\sum_{p\leq p_D}\phi^p\right]\,.
\ee

The relevant operators are responsible from making the system flow from the UV fixed point. A sharp change in the two-point correlation function of $\phi$ would happen when $[\phi]=0$; this would typically signal a phase transition across a critical point. In turn, a critical point can be naturally identified with the UV fixed point, as in Ho\v{r}ava--Lifshitz models. This identification does not guarantee the existence of a perturbative UV fixed point (which can be inferred only by explicit loop calculations), but it is a positive hint in that direction.

Apparently, the fractional-symmetry scenario does not have a two-dimensional critical point in $D>2$. If $\dh=D\a$ and $D>2$, the phase transition would happen only when $\a=0$, corresponding to Pointland (zero-dimensional manifold, no spacetime, no dynamics). This does not imply that there is no non-trivial UV fixed point, but it makes its existence less clear. In section \ref{pola} we shall make a crucial extension of the theory such that the limit $\a\to 0$ will correspond to a spacetime with some residual geometric structure. However, for $\a<1/2$ there is no natural norm and the geometric construction of real-valued $\a$ models breaks up progressively towards the UV. To summarize, it may be possible to extend the flow down to the critical point at $\a=0$, but one would have to give up to geometric structure anyway. This extension may not be even sufficient: Regardless the range of $\a$, the argument of power-counting renormalizability fails in the present case, since the maximum allowed $p=p_D$ is finite. Therefore, we have fewer indications about the perturbative renormalization properties of the theory. This does not jeopardize the supposed good UV behaviour of fractal field models, as we shall see in section \ref{sdd}.

%%%%%%%%%%%%%%%%%%%%%%%%%%%%%%%%%%%%%%%%%%%%%%%%%%%%%%%%%%%%%%%%%%%%%%%%%%%%%%%%%%%%%%%%%%%%%%%%

\subsection{Green function}\label{pro}

The propagator of a real scalar field in Lebesgue--Stieltjes theories with absolutely continuous measure was computed in \cite{fra1,fra2} for the kinetic operator \Eq{lk2} but, because fractional measures are not Lorentz invariant, not many of the details of that calculation fit into the present framework. A cleaner derivation of the Green's equation was given in \cite{fra3} for models with kinetic operator $\cK_1$, but here we shall point out some subtleties in relating the Green function with the physical propagator of the theory. In order to do so, we briefly
repeat the calculation of \cite{fra3} but with the kinetic operator $\cK$ and the fractional Bessel transform \cite{frc3}, the correct generalization of the ordinary Fourier transform. The final result in momentum space will be the same as in \cite{fra3}. The steps are the same as in ordinary quantum field theory \cite{Bro92}, modulo technical differences, and begin with the partition function. At the end we will not obtain the propagator (Green function with causal prescription) but a generic Green function for the kinetic operator. This exercise is useful for sketching both the momentum structure of the actual propagator (an information sufficient to complete the power-counting-renormalizability argument) and the caveats entailed in the full derivation of the propagator itself, which will be given elsewhere.

Consider a real free scalar field with mass $m$ in a fixed-order, isotropic ($\a_0=\a$) fractional spacetime. The action is
\be
S_\a=\frac12\int\rmd\vr_\a(x)\,\phi(x)\, (\cK-m^2)\,\phi(x)\,,
\ee
where we omitted the integration domain. The free Lorentzian partition function $Z_0$ in the presence of a local source $\cJ$ is
\be\label{P07_z0}
Z_0[\cJ]:= \int [{\cal D}\phi]\,\rme^{\rmi[S_\a+\int\rmd\vr_\a(x)\, \cJ(x)\phi(x)]}=:\int [{\cal D}\phi]\,\rme^{\rmi S_\cJ}\,.
\ee
To calculate it, we move to fractional momentum space \cite{frc3}. This has the same measure $\vr_\a(k)$ as configuration space, momenta are non-negative, and there exists an invertible transform in terms of Bessel functions of the first kind. Let
\be
c_\a(k,x) := \Gamma(\a_0) (k_0 x^0)^{1-\frac{\a_0}{2}} J_{\frac{\a_0}{2}-1}(k_0 x^0)\prod_{i=1}^{D-1}\Gamma(\a) (k^i x^i)^{1-\frac{\a}{2}} J_{\frac{\a}{2}-1}(k^i x^i)
\ee
be an eigenfunction of the kinetic operator $\cK$,
\be
\cK\,c_\a(k,x) = -k^2 c_\a(k,x)\,,\qquad k^2 := k_\mu k^\mu= -(k_0)^2+ |{\bf k}|^2\,.
\ee
A reason why to choose $\cK$ instead of $\cK_1$, $\bar\cK_\a$ or $\cK_\a$ is because the momentum-space transform is expanded on a basis of $c_\a$, which are not eigenfunctions of the other kinetic operators. The transform of a function $f(x)$ and the anti-transform are \cite{frc3}
\be\label{fst1}
\tilde f(k) := \int_0^{+\infty}\rmd\vr_\a(x)\, f(x)\,c_\a(k,x)\,,\qquad
f(x)        = \int_0^{+\infty}\rmd\vr_\a(k)\,\tilde f(k)\,c_\a(k,x)\,.
\ee
Consistently, the representation of the fractional Dirac distribution in $D$ dimensions is
\be\label{dkk}
\de_\a(x,x')=\int_0^{+\infty}\rmd\vr_\a(k)\,c_\a(k,x)c_\a(k,x')\,,
\ee
stemming for the definition $\de_\a(x,x')=v_\a^{-1}(x)\de(x-x')$ and the one-dimensional closure formula $\delta(x-x')=x\int_0^{+\infty} \rmd k\,k J_\nu(kx)J_\nu(kx')$ \cite[eq.~1.17.13]{NIST}. The distribution \Eq{dkk} acts, indeed, as a delta in fractional space. Transforming both $\phi$ and $\cJ$ in eq.~\Eq{P07_z0}, one obtains
\ba
S_\cJ &=& \int\rmd\vr_\a(x)\int\rmd\vr_\a(k)\int\rmd\vr_\a(k')\,c_\a(k,x)c_\a(k',x)\nonumber\\
&&\qquad\qquad\times\left[-\frac12\tilde\phi(k)({k'}^2+m^2)\tilde\phi(k')+\tilde \cJ(k)\tilde\phi(k')\right]\nonumber\\
&\ \stackrel{\Eq{dkk}}{=}\ & \int\rmd\vr_\a(k)\left[-\frac12\tilde\phi(k)(k^2+m^2)\tilde\phi(k)+\tilde \cJ(k)\tilde\phi(k)\right]\nonumber\\
&=& \frac12\int\rmd\vr_\a(k)\left[-\tilde\vp_\a(k)(k^2+m^2)\tilde\vp(k)+\frac{\tilde \cJ(k)\tilde \cJ(k)}{k^2+m^2}\right]\,,\label{sj}
\ea
where
\be
\tilde\vp(k):= \tilde\phi(k)-\frac{\tilde \cJ(k)}{k^2+m^2}\,.
\ee
The first term in eq.~\Eq{sj} will be a normalization of the partition function. The last term can be transformed back to configuration space:
\ba
\int\rmd\vr_\a(k)\frac{\tilde \cJ(k)\tilde \cJ(k)}{k^2+m^2} &=& 
\int\rmd\vr_\a(k)\int\rmd\vr_\a(x)\int\rmd\vr_\a(x')
\frac{\cJ(x)\cJ(x')}{k^2+m^2}c_\a(k,x)c_\a(k,x')\nonumber\\
&=& -\int\rmd\vr_\a(x)\int\rmd\vr_\a(x')\, \cJ(x)\, G_\a(x,x')\, \cJ(x')\,,
\ea
where
\be\label{prop12}
G_\a(x,x';m) := -\int_0^{+\infty}\rmd\vr_\a(k)\,\frac{1}{k^2+m^2}\,c_\a(k,x)c_\a(k,x')\,.
\ee
Unlike Green functions in ordinary Minkowski spacetime, $G_\a$ does not depend on the difference of the coordinates $x$ and $x'$ of the initial and final points. This property, unnoticed in \cite{fra1}--%,fra2,
\cite{fra3} due to the use of a non-factorizable Lebesgue--Stieltjes measure and of a non-invertible transform, is a direct inheritance of the measure weight, which breaks translation invariance. On a multi-fractional geometry, translation symmetry is recovered at large scales \cite{frc3}.

The free partition function \Eq{P07_z0} becomes
\ba
Z_0[\cJ] &=& \left\{\int [{\cal D}\vp]\exp\left[
-\frac{\rmi}{2}\int_0^{+\infty}\rmd\vr_\a(k)\tilde\vp(k)(k^2+m^2)\tilde\vp(k)\right]\right\}\nonumber\\
&&\times\exp\left[-\frac{\rmi}{2}\int_0^{+\infty}\rmd\vr_\a(x)\int_0^{+\infty}\rmd\vr_\a(x')\, \cJ(x)\, G_\a(x,x';m)\, \cJ(x')\right]\nonumber\\
&=& Z_0[0]\,\exp\left[-\frac{\rmi}{2}\int_0^{+\infty}\rmd\vr_\a(x)\int_0^{+\infty}\rmd\vr_\a(x')\, \cJ(x)\, G_\a(x,x';m)\, \cJ(x')\right].\nonumber\\\label{zej}
\ea
The function $G_\a(x,x';m)$ obeys the Green equation
\be\label{green}
(\cK_x-m^2)\,G_\a(x,x';m) = \de_\a(x,x')\,,\qquad [G_\a]=D\a-2\,.
\ee
The solution given by eq.~\Eq{prop12} is not well defined because we have not specified a contour choice. In general, there are infinitely many ways to go around the poles and branch cuts of \Eq{prop12} in the complex plane $({\rm Re} k^0,{\rm Im} k^0)$, and different contour prescriptions correspond to different solutions.\footnote{An illuminating discussion on the subject can be found in \cite{BoS}.} The most general solution, in fact, is a linear combination of two solutions $G_\pm$ of the homogeneous equation and a particular solution $\bar G$ (e.g., the retarded or the advanced propagator) of the inhomogeneous equation \Eq{green}. The arbitrariness of the coefficients of the linear combination corresponds to the infinitely many possible choices of integration contour. One particular choice gives the causal propagator.

If we had used the kinetic operator $\cK_1$, not only would we have not been able to express $G_\a$ as a momentum integral, but the left-hand side of the Green equation \Eq{green} would not have corresponded to the classical equation of motion \Eq{kgeq1}, which has extra terms in the derivatives of the measure weight. Therefore, the usual definition of the Green equation as the Klein--Gordon equation in the presence of a pointwise source would have no longer been valid. The matching of the Klein--Gordon and homogeneous Green equation is related to the issue of microcausality (field observable operators commute with one another at spacelike separation, and non-local correlations do not give rise to propagation of superluminal messages). In ordinary scalar field theory, the combination of the positive- and negative-frequency Green functions $G_+(x-x')=\langle0|\phi(x)\phi(x')|0\rangle$ and $G_-(x-x')=\langle0|\phi(x')\phi(x)|0\rangle$, which are solutions to the Klein--Gordon equation, yields the Pauli--Jordan function $\rmi G_{\rm PJ}:=G_+-G_-=\langle0|[\phi(x),\phi(x')]|0\rangle$. The Pauli--Jordan function vanishes outside the light cone, thus guaranteeing that the quantum theory is causal. If the functions $G_\pm$ did not solve the classical equation of motion, the mutual relations among different Green functions and their role in causality would have been less transparent.

Both the contour prescription and causality issues will be reported elsewhere. Here, we can nevertheless extract a wealth of physical information from the Green function \Eq{prop12}.
\begin{itemize}
\item {\it Spectrum.} The physical spectrum of the theory should be extracted from the Feynman propagator, but the pole structure of the Green function,
\be
G_\a(k):=-\frac{1}{k^2+m^2}\,,\label{prop2}
\ee
already points to the final result. Since we are in the half plane ${\rm Re}\,k^0\geq 0$ instead of the full $({\rm Re}\,k^0,{\rm Im}\,k^0)$ plane, the spectrum has half the usual support at $k^2=-m^2$, corresponding to the positive pole 
\be
{\rm Re}\,k^0=\sqrt{m^2+|{\bf k}|^2}\,.
\ee
There is no $\a$-dependence, the momentum-space Green function does not change along the dimensional flow, and there is no continuum of massive modes as in \cite{fra2}.\footnote{It is expected that in fractional-symmetry scenarios $G_\a(k)$ has an algebraic branch point $k^{2\a}=-m^2_\a$, where the mass coupling has an $\a$-dependence via its scaling dimension. The associated branch cut would correspond to a continuum spectrum of massive modes.} %, somewhat resembling that of fractional powers of the d'Alembertian, $(\B-m^2)^\a$, corresponding to a propagator $\tilde G(k)\propto (k^2+m^2)^{-\a}$ \cite{BGG}--%,Gia91,BGO,doA92,BG,BOR, \cite{BBOR}.}
%momentum and mass units differ, unless one defines the mass coupling in the action to be $m^{2\a}$ instead of $m^2_\a$.
%($\g=\a\neq 1$). The complex function
%\be
%(k^0)^{2\a}=\rme^{2\a \log(k^0)} = \rme^{2\a {\rm Log}(k^0)}\rme^{4\pi\rmi\a n}\,,\qquad n\in\mathbb{Z}\,,
%\ee
%is multi-valued. Its principal value is given by $\exp(2\a {\rm Log}\, k^0)$, where Log is the principal value of the complex logarithm \cite{Gam01}. Letting $k^0=|k^0|\rme^{\rmi\theta}$, where $0<\theta<2\pi$, a continuous branch of this function is defined as $(k^0)^{2\a}=|k^0|^{2\a}\rme^{2\a\rmi\theta}$, and $G_\a(k)$ has an algebraic branch point $k^{2\a}=-m^2_\a$ and the branch cut
%\be\label{brc}
%{\rm Re}\,k^0\geq (m^2_\a+|{\bf k}|^{2\a})^{1/(2\a)}\,.
%\ee
Extending to multi-fractional spacetime, the total propagator becomes $G=\sum_\a g_\a G_\a$ or, in the external time picture, $G=\int\rmd\s g(\s) G_{\a(\s)}$. The spectrum in the integer-simmetry scenario is scale independent. %In the fractional-symmetry scenario, the branch cut changes with $\a=\a(\s)$ and the branch point in the rest frame (${\bf k}=0$) migrates from ${\rm Re}\,k^0=m^2_{1/2}$ to ${\rm Re}\,k^0=m_1$. In the theory with symmetrized kinetic term \Eq{lk2}, we expect a somewhat different spectrum (still constituted of massive poles and a continuum of modes), along the lines of \cite{fra2}.
\item {\it Scaling law and critical points.} Let $\la>0$ and consider the scaling transformation
\be
x\to \la^{\ds/\dh} x\,,\qquad k\to \la^{-\ds/\dh} k\,,\qquad m\to \la^{-\ds/\dh} m\,.
\ee
The Green function \Eq{prop12} transforms as
\be\label{scar}
G_\a\left(\la^{\ds/\dh}x,\la^{\ds/\dh}x';\la^{-\ds/\dh}m\right)= \la^{(2/\dh-1)\ds}G_\a(x,x';m)\,.
\ee
This scaling law can be also obtained by the scaling of the fractional diffusion equation \Eq{dife} \cite{frc1} and it determines the critical point in the dimensional flow at which the Green function is conformally invariant. By definition, it happens when
\be
\ds=0\qquad {\rm or}\qquad \dh=2\,.
\ee
In the first case, $\a_0=\a=0$ and there is no diffusion at all: this corresponds to Pointland, $\dh=0$. In the next section we shall extend fractional models in a non-trivial way such that even a configuration with $\a=0$ is endowed with a non-singular geometry. Therefore, eventually, one might regard this case as physical, albeit its geometry will not possess a norm. 

The second critical point is characterized by a Hausdorff dimension ${\dh}_*=2$, at the critical value
\be\label{scag}
\a_*=\frac{2}{D}\,.
\ee
This is the two-dimensional critical point advertized so far in the integer-symmetry scenario, where ${\ds}_*={\dh}_*=2$. Taking $\a_*=1/2$ (the minimum allowed value for normed fractional spaces), the topological dimension is constrained to be $D=4$. In the fractional-symmetry scenario, this critical point either does not exist for $D>2$ or it does not correspond to a normed space.
\end{itemize}

%%%%%%%%%%%%%%%%%%%%%%%%%%%%%%%%%%%%%%%%%%%%%%%%%%%%%%%%%%%%%%%%%%%%%%%%%%%%%%%%%%%%%%%%%%%%%%%%

\subsection{Superficial degree of divergence}\label{sdd}

Let us make a short summary of the characteristics of multi-fractional quantum field theory. First of all, macroscopic physics would not be described by a finite number of observables if we included the infinite class of operators $\cO_{\a,\g,n}$ with fractional derivatives, eq.~\Eq{Oab}. Whether we include these operators or not is a matter of definition of the theory, and one is entitled to opt for the formulation with good IR behaviour. In the fractional-symmetry scenario there is still a multiplicity of operators $\cO_{\a,\a,1}$, labelled by $\a$, but in the infrared the number of effective couplings is finite. In the integer-symmetry scenario there is only one kinetic term throughout the flow. Consistently, if one starts from an action with only integer derivatives, operators with fractional derivatives never appear.

Secondly, the renormalization properties of a model are dictated by the dimensionality of the operator $\cK$ in the kinetic term, so that good UV behaviour is guaranteed when the spacetime dimension is the same as the dimensionality of $\cK$. In this respect, since the harmonic structure is determined by $\cK$, the spectral and Hausdorff dimensions of the theory (related to, respectively, the harmonic and geometric structures \cite{frc1}) are equally important and yield complementary informations.

The scaling argument for the Green function, eqs.~\Eq{scar}--\Eq{scag}, and the power counting of sections \ref{friss} and \ref{frass} are in mutual agreement. In fact, the power-counting argument of section \ref{pc} applies, \emph{mutatis mutandis}, also to fractional theories. One difference is in the replacement of the topological dimension $D$ with the Hausdorff dimension $\dh$, due to the non-trivial measure obtained in momentum space: each loop integral gives $[\rmd\vr_\a(k)]=D\a$. The momentum-space propagator has the same scaling dimension of the Green function, which is $[G_\a(k)]=-2\g$ in a general fractional scenario with kinetic operator of order $2\g$. In configuration space, $G_\a$ contributes with a weight $D\a-2\g$. For the scalar field theory, interaction vertices do not carry dimensionality. Overall, the superficial degree of divergence of a Feynman graph with $L$ loops and $I$ internal lines is
\be\label{dmax}
\delta = D\a L-2\g I \leq \de_{\rm max}:=(D\a-2\g)L\,.
\ee
When $\a=1=\g$, one gets the standard result in $D$ dimensions. Otherwise:
\begin{itemize}
\item In the fractional-symmetry scenario ($\g=\a$), the maximum superficial degree of divergence is positive and the power-counting argument is inconclusive regarding the renormalizability of the model. Nevertheless, $\de_{\rm max}$ is smaller than in ordinary field theory by a factor of $\a$. Using the same line of reasoning leading to eq.~\Eq{sudedi}, one can verify that eqs.~\Eq{scap0} and \Eq{scap} are recovered from \Eq{dmax}.
\item In the integer-symmetry scenario ($\g=1$), at the critical point $\a=\a_*=2/D$ one has $\delta\leq 0$ and, at most, logarithmic divergences. If the UV fixed point had $\a<2/D$ (non-normed spaces, if $D\geq 4$) the theory is super-renormalizable. 
\end{itemize}

%%%%%%%%%%%%%%%%%%%%%%%%%%%%%%%%%%%%%%%%%%%%%%%%%%%%%%%%%%%%%%%%%%%%%%%%%%%%%%%%%%%%%%%%%%%%%%%%%%%%%%%%%%%%%%%%%%%%%%%%%%%%%%%%%%%%%%%%%%%%%%%%%%%%%%%%%%%%%%%%%%%%%%%%%%%%%%%%%%%%%%%%%%%%%%%%

\section{Complex fractional theory}\label{pola}

If we regard fractional spacetime models as effective frameworks capturing some features of quantum gravity at large, it is important to probe their capabilities in that direction, beyond the running of the effective dimension of spacetime. Fractional models represent continuum spacetimes, and one can conceive applications to regimes where the discrete nature of spacetime in quantum gravity models has been washed or zoomed away by ``hydrodynamical'' macroscopic effects. These effects are believed to take place when large ensembles of ``quanta of space'' (spin networks, complexes endowed with discrete labels, and so on) are collected together and let evolve dynamically. This evolution of a very large number of degrees of freedom is presently out of control and scantly explored in most of the theories, but it is a promising avenue leading to their yet-unclear continuum limit. Despite their intrinsically continuum structure, can fractional field theories play a role in the effective description of this transition? Surprisingly, the answer is Yes. To see this, we need a short detour.

%%%%%%%%%%%%%%%%%%%%%%%%%%%%%%%%%%%%%%%%%%%%%%%%%%%%%%%%%%%%%%%%%%%%%%%%%%%%%%%%%%%%%%%%%%%%%%%%%%%%%%%%%%%%%%%%%%%%%%%%%

\subsection{From real to complex fractional order}\label{rcfo}

A curious feature of the heat kernel trace for a Laplacian on fractals is that it displays oscillations. In a metric space of topological dimension $D$, the return probability at small diffusion time $\s$ is Weyl's expansion
\be\label{genK}
\cP(\s)=\frac{1}{(4\pi \s)^{\frac{D}{2}}}\left[1+\sum_{n=1}^{+\infty}A_n\s^n\right]\,,
\ee
where the coefficients $A_n$ depend on the background metric. From eq.~\Eq{spedi}, $\ds=D$. For non-smooth sets such as fractals, this expression is drastically modified by the presence of discrete symmetries. In particular, for deterministic fractals the counterpart of eq.~\Eq{genK} is of the form
\be\label{genKf}
\cP(\s)=\frac{1}{(4\pi \s)^{\frac{\ds}{2}}}F(\s)\,,
\ee
where $F$ is a periodic function of $\ln\s$ \cite{KiL,Kaj10}. Oscillatory behaviour has been found analytically and numerically for various fractals \cite{DIL}--%,Tep05,Akk1,
\cite{ABS}. The phenomenon of logarithmic oscillations \cite{Sor98} seems to have two origins. The high symmetry of deterministic fractal sets such as diamond fractals and the Sierpinski gasket give rise to eigenvalues of the Laplacian with unexpectedly large multiplicity; in turn, these are related to the periodicity of the counting measure \cite{KiL,Sor98,LvF}. Very recently, examples have been found (for instance, the Sierpinski carpet) where log-oscillations arise not because of large multiplicities, but because of unexpectedly large gaps in the spectrum \cite{DuT}. 

The underlying symmetry mechanism responsible for the oscillations plays a major role in the next development of fractional theory, and we wish to see how it arises in that context. We need to recall the relation between fractals and fractional calculus of real \cite{RYS}--%,YRZ,RYZLN,Yu99,QL,RQLW,
\cite{RLWQ} and complex order \cite{LMNN,NLM}. This relation was reviewed in \cite[section 4.4]{frc1}, where the proof of the following theorem was sketched: A fractional integral of real order represents either the averaging of a smooth function on a deterministic fractal, or a random fractal support. As a matter of fact, these results have been obtained only for fractals embedded in the real line ($D=1$) and, to the best of our knowledge, there is no literature on multi-dimensional embeddings. We do not see any problem in extending the theorem to fractals given by the Cartesian product of lower-dimensional fractals, while more general statements might require extra work.

Let $\cF$ be a self-similar set given by $N$ similarities \Eq{simi}. It can be expressed iteratively as an infinite intersection of \emph{pre-fractals}:
\be\label{prefr}
\cF= \bigcap_{k=1}^\infty \underbrace{\cS\circ \cdots \circ \cS}_{k~{\rm times}}  (U)\,,
\ee
where
\be
\cS(U) :=\bigcup_{i=1}^N \cS_i(U)
\ee
for any non-empty compact set $U \supset\cF$ such that $\cS_i(U)\subset U$. Consider the integral of a function $f(x)$ on a self-similar fractal set $\cF\subseteq [0,1]$,
\be\label{convo}
I_\cF(x):=\int_0^x\rmd x'\, v_\cF(x-x')f(x')\,.
\ee
We temporarily work in dimensionless units ($[x]=0$). The kernel $v_\cF$ depends on the geometry of the set and can be determined recursively at any given order of iteration. The Laplace transform of eq.~\Eq{convo} is 
\be
\hat I_\cF(p):=\int_0^{+\infty}\rmd x \rme^{-p x}I_\cF(x)=\hat v_\cF(p)\hat f(p)\,.
\ee
%$\hat I_\cF$ can be expressed iteratively as an infinite intersection of pre-fractals.
Suppose $\cF$ be composed, at the first iteration, by a number of smaller copies of length $\la$. The $k$-th iterate has Laplace-transformed kernel $\hat v_\cF^k(p)=\prod_{n=0}^{k-1} g_n(p)$, for some functions $g_n$. If all these functions are equal and with argument $g_n(p)=g(p\la^n)$, and if the asymptotics of $g$ are $g(z)\sim 1+O(z)$ for small $z$ and $g(z)\sim g_1+O(z^{-1})$ for large $z$ (all conditions fulfilled by self-similar and generalized self-similar sets; the constant $g_1$ is the first probability weight in the self-similar measure \Eq{ssim}), then \cite{LMNN,NLM}
\be\label{kK}
\lim_{k\to+\infty} \hat v^k_\cF(p) =\hat v_\cF(p) = p^{-\a} F_\a(\ln p)\,,\qquad \a=\frac{\ln g_1}{\ln\la}\,,
\ee
where $F_\a$ is a log-periodic function \cite{Sor98} of period $\ln\la$:
\be\label{pia}
F_\a(\ln p+m\ln\la)=F_\a(\ln p)= \sum_{l=-\infty}^{+\infty} c_l \exp\left(2\pi l\rmi\frac{\ln p}{\ln\la}\right)\,,
\ee
for some coefficients $c_l$. Combining \Eq{kK} with \Eq{pia},
\be\label{kap}
\hat v_\cF(p)=\sum_{l=-\infty}^{+\infty} c_l \exp\left[(\rmi\om_l-\a)\ln p\right]\,,\qquad \om_l:=\frac{2\pi l}{\ln\la}\,.
\ee
Recognizing $p^{-\a}$ as the Laplace transform of the fractional weight $v_\a(x)=x^{\a-1}/\Gamma(\a)$ and comparing eq.~\Eq{convo} with \Eq{kK}, one already sees that $I_\cF$ is quite similar to a fractional integral $I^\a$, were it not for the non-constant contribution \Eq{pia}. To complete the connection, it is sufficient to take the average of the log-periodic function $F_\a$ over the period $\ln\la$:
\be
b_\a:=\langle F_\a(\ln p)\rangle := \int_{-1/2}^{1/2}\rmd z\, F_\a(\ln p+z\ln\la)\,,
\ee
where the value $b_\a$ depends on the details of $g(p\la^n)$. The integration range is written in a conventional form and it can change according to the choice of units; once the log-period is given, the average procedure is uniquely defined. Then,
\be
\langle v_\cF(x)\rangle = b_\a \frac{x^{\a-1}}{\Gamma(\a)}\,,
\ee
and \cite{LMNN,NLM}
\be\label{aver}
\langle I_\cF f\rangle = \int_0^x\rmd x'\,\langle v_\cF(x-x')\rangle f(x') =b_\a I^\a f\,.
\ee
This relation is valid for a huge class of sets known as net fractals, and admits two other interpretations. Taking the average corresponds to a randomization of the fractal structure, where oscillations are cancelled by mutual interference. But washing oscillations away can be also seen as dropping all the modes in eqs.~\Eq{pia} and \Eq{kap} except the zero mode $l=0$. Now, the $\om_l\to 0$ limit is obtained either as a small-similarity-ratio limit, $\la\to 0$, or a large-Laplace-momentum limit in eq.~\Eq{pia}, so that eq.~\Eq{aver} can be also regarded as an approximation in Laplace momentum space, $I_\cF \sim\ b_\a I^\a$ as ${\rm Re}(p)\to +\infty$ \cite{RYS}--%,YRZ,RYZLN,Yu99,QL,RQLW,
\cite{RLWQ}.

Thus, fractional integrals of real order are associated with random fractals. In other words, fractional measures either correspond to certain random fractals or, alternatively, they approximate Borel measures of self-similar fractals in the limit of infinitely refined similarities (continuum approximation), corresponding to neglecting the oscillatory structure of fractal kernels. This was the point where the discussion in \cite{frc1} ended. However, it is not the end of the story. 

The approximation $I_\cF\sim\langle I_\cF\rangle\sim I^\a$, eq.~\Eq{aver}, can be improved by including next-to-leading oscillatory modes. Quite generally, integrals on self-similar fractals are given by an infinite series of fractional integrals of complex order. Looking at eq.~\Eq{kap}, the complex fractional measure weight $v_\cF=:\tilde v_\a$ is naturally defined as a sum (or an integral) over frequencies,
\be\label{kap2}
\tilde v_\a(x)=\sum_{\om=-\infty}^{+\infty} c_\om v_{\a,\om}(x):=\sum_{\om=-\infty}^{+\infty} c_\om \frac{x^{\a-1+\rmi\om}}{\Gamma(\a+\rmi\om)}\,,
\ee
where $c_\om$ are complex coefficients. The zero mode is the real-order measure we have considered so far in this paper and in \cite{frc1}, and the average $b_\a$ is equal to $c_0$. Complex measures are obviously not measures in the sense commonly employed by physicists: It it not positive (the measure of a set can be a non-negative number) and requires a non-trivial extension of the definition of Hausdorff dimension.\footnote{In this respect, spacetimes associated with such measures are ``pre-geometric.'' We refrain from using this adjective because the geometry of complex measures is mathematically well defined, although quite different from ordinary geometry.} To explore the properties of these objects, we pick a model with just one pair of conjugate frequencies $\pm\om$:
\be\label{kcom}
v_{\a,\om}(x) = c_0 \frac{x^{\a-1}}{\Gamma(\a)}+c_\om\frac{x^{\a-1+\rmi \om}}{\Gamma(\a+\rmi\om)}+c^*_\om\frac{x^{\a-1-\rmi \om}}{\Gamma(\a-\rmi\om)}\,,
\ee
where $c_\om=|c_\om|\rme^{\rmi\Psi}$ is a complex amplitude, $c_\om^*$ is its complex conjugate, and $\Psi\in\mathbb{R}$ is a phase. We choose to work with eq.~\Eq{kcom} rather than \Eq{kap2} only for simplicity, but there may be further justification in doing that. In fact, for Cantor sets the three-term weight \Eq{kcom} is a good approximation of the full kernel, where $\om$ is the average frequency of the leading terms \cite{NLM}. 

After some manipulations \cite{HLA}, one can see that the measure weight \Eq{kcom} is real; this happens because one is considering a conjugate combination of complex measures.\footnote{In general, a self-conjugate real measure has $c_\om=c_{-\om}$; if this condition is not satisfied, the measure and the ensuing dimension are complex \cite{LvF}.} We set $c_0=1$ without loss of generality and $\Psi=0$, commenting on this last assumption at the end. Noting that
\be\nonumber
\frac{1}{\Gamma(\a\pm\rmi\om)}={\rm Re}\left[\frac{1}{\Gamma(\a+\rmi\om)}\right]\pm\rmi {\rm Im}\left[\frac{1}{\Gamma(\a+\rmi\om)}\right]=: R_\Gamma(\a+\rmi\om)\pm\rmi I_\Gamma(\a+\rmi\om)\,,
\ee
one has
\ba
v_{\a,\om}(x) &=&  x^{\a-1}\left[\frac{1}{\Gamma(\a)}+\frac{c_\om\rme^{\rmi \om\ln x}}{\Gamma(\a+\rmi\om)}+\frac{c_\om\rme^{-\rmi \om\ln x}}{\Gamma(\a-\rmi\om)}\right]\label{temK}\\
 &=&  \frac{x^{\a-1}}{\Gamma(\a)}+x^{\a-1} c_\om R_\Gamma(\a+\rmi\om)\left(\rme^{\rmi \om\ln x}+\rme^{-\rmi \om\ln x}\right)\nonumber\\
 &&+x^{\a-1}c_\om \rmi I_\Gamma(\a+\rmi\om)\left(\rme^{\rmi \om\ln x}-\rme^{-\rmi \om\ln x}\right)\nonumber\\
 &=& x^{\a-1}\left[\frac{1}{\Gamma(\a)}+2c_\om R_\Gamma(\a+\rmi\om)\cos(\om\ln x)+2c_\om I_\Gamma(\a+\rmi\om)\sin(\om\ln x)\right]\,.\nonumber\\\label{kcom1}
\ea
The primitive of $v_{\a,\om}$ is the oscillatory extension of the measure $\vr_\a$:
\be
\vr_{\a,\om}(x) = \frac{x^\a}{\Gamma(\a+1)}\left[1+A_{\a,\om}\cos(\om\ln x)+B_{\a,\om}\sin(\om\ln x)\right]\,,\label{kcom20}
\ee
where
\ba
A_{\a,\om} &:=& \frac{2c_\om}{\a^2+\om^2}\Gamma(\a+1)[\a R_\Gamma(\a+\rmi\om)-\om I_\Gamma(\a+\rmi\om)]\,, \\
B_{\a,\om} &:=& \frac{2c_\om}{\a^2+\om^2}\Gamma(\a+1)[\om R_\Gamma(\a+\rmi\om)+\a I_\Gamma(\a+\rmi\om)]\,.
\ea
These expressions are even in $\om$, so we can restrict our attention to positive frequencies $\om>0$. In the limit $\om\to 0$, one recovers the power-law measure. Restoring dimensionful units, $\om$ remains dimensionless, but we must introduce a length scale, which we call $\ell_\infty$, in the arguments of the logarithms. Then, eq.~\Eq{kcom20} becomes
\be
\boxd{\vr_{\a,\om}(x) = \frac{x^\a}{\Gamma(\a+1)}\left[1+A_{\a,\om}\cos\left(\om\ln \frac{x}{\ell_\infty}\right)+B_{\a,\om}\sin\left(\om\ln \frac{x}{\ell_\infty}\right)\right]\,.}\label{kcom2}
\ee

%%%%%%%%%%%%%%%%%%%%%%%%%%%%%%%%%%%%%%%%%%%%%%%%%%%%%%%%%%%%%%%%%%%%%%%%%%%%%%%%%%%%%%%%%%%%%%%%%%%%%%%%%%%%%%%%%%%%%%%%%

\subsection{Discrete scale invariance}\label{lodsi}

Complex fractional models with self-conjugate measure are characterized by oscillations governed by a constant 
\be\label{scale0}
\la_\om:=\exp\left(\frac{2\pi}{\om}\right)\,.
\ee
Notice the highly non-perturbative dependence on the frequency. Asymptotically,
\bs\ba
\la_\om &\ \stackrel{\om\to 0^+}{\longrightarrow}\ &+\infty\,,\\ 
\la_\om &\ \stackrel{\om\to +\infty}{\longrightarrow}\ & 1\,.
\ea\es
$\la_\om$ defines a characteristic (as opposed to fundamental) physical scale as
\be\label{scale}
\ell_\om := \la_\om \ell_\infty>\ell_\infty\,.
\ee
The oscillatory part of the measure weight \Eq{kcom1} and of the measure \Eq{kcom2} is log-periodic under the discrete scaling transformation
\be
\ln\frac{x}{\ell_\infty}\,\to\, \ln\frac{x}{\ell_\infty}+\frac{2\pi n}{\om}=\ln\frac{x}{\ell_\infty}+n\ln\la_\om\,,\qquad n=0,\pm1,\pm2,\dots\,,%=\ln\frac{x}{\ell_\om\la_\om^{n-1}}
\ee
implying
\be\label{dsi}
\boxd{x\,\to\, \la_\om^n x\,,\qquad n=0,\pm1,\pm2,\dots\,.}
\ee
With a slight abuse of terminology, we shall call \emph{log-period} both the period $\ln\la_\om$ and the dimensionful scale $\ell_\om$.

The transformation rule \Eq{dsi} is one of the pivot results of the paper. Log-periodicity is a phenomenon intimately related to the presence of a fundamental length scale $\ell_\infty$ or, in other words, a microscopic cut-off. This happens due to a symmetry unknown to continuous systems or artificially discrete systems such as lattices. This symmetry, called \emph{discrete scale invariance} (DSI), is a dilation transformation under integer powers of a preferred, special scaling ratio $\la_\om$ \cite{Sor98,JoS,JSH}. Equation \Eq{dsi} is a discrete scale symmetry.
All deterministic fractals possess a DSI by definition \cite{Kig01,Ham08}. For example, the middle-third Cantor set is defined by the similarities $\cS_1(x)=(1/3) x$, $\cS_2(x)=(1/3) x+(2/3)$. Here, the contraction ratio is fixed once and for all: $\la_1=1/3$. The set is invariant only under contractions with ratios $\la_n=(1/3)^n$, $n$ natural. Log-periodicity and the associated DSI appear also in Laplacian growth models, rupture in heterogeneous systems, analysis of earthquakes and financial crashes, out-of-equilibrium systems, quenched disordered systems, and two-dimensional turbulence \cite{Sor98} (for early applications in the theory of phase transitions and L\'evy flights see, respectively, \cite{Nau75,JUPM} and \cite{HSM}).

Fractional calculus of complex order enjoys (slight modifications of) all the properties\footnote{In particular, left fractional derivatives of order $\a+\rmi\om$ with $0\leq\a<1$ are defined by eq.~\Eq{pan} with $\a$ replaced by $\a+\rmi\om$ \cite{KST}.} we have listed in \cite[section 2.3]{frc1} but, since the appearance of early papers on the subject \cite{Kob41,Lov71}, only recently it has received attention for its physical applications \cite{LMNN}--%,NLM,
\cite{HLA,OLMN}. Returning to fractional spacetime models, assume eq.~\Eq{kcom2} as the measure along each direction $x^\mu$. Fractional spacetimes with this measure admit two inequivalent interpretations: either as exact \emph{per se} or as next-to-leading approximations of a discrete fractal spacetime. In both cases, the extension to fractional operators of complex order produces a more complicated geometric pattern unravelling an underlying discrete scale. One renders the continuous scale invariance \Eq{Rdsimx} discrete, and oscillatory fractional measures are much closer to genuine fractals than their monotonic counterparts. 

Consistently with \Eq{dsi}, to get the continuum limit one should send the frequency to zero from above, so that the length cut-off vanishes:
\be\label{ellom}
\ell_\om\to 0\qquad {\rm as}\qquad \om\to 0^+\,.
\ee
The corresponding asymptotics of the measure weight \Eq{temK} is
\ba
v_{\a,\om} &\ \stackrel{\om\sim0}{=}\ & \frac{x^{\a-1}}{\Gamma(\a)}\left\{(1+2c_\om)+\om^2 c_\om\left[\psi'(\a)-\psi^2(\a)+2\psi(\a) \ln\frac{x}{\ell_\infty}-\left(\ln \frac{x}{\ell_\infty}\right)^2\right]\right\}\nonumber\\
&&+O(\om^4)\,,\label{asiom0}
\ea
where $\psi(x)=\p_x\Gamma(x)/\Gamma(x)$ is the digamma function. Notice that the measure weight at $\om=0$ is only proportional, and not equal, to its average $v_\a$ over a log-period.

%%%%%%%%%%%%%%%%%%%%%%%%%%%%%%%%%%%%%%%%%%%%%%%%%%%%%%%%%%%%%%%%%%%%%%%%%%%%%%%%%%%%%%%%%%%%%%%%%%%%%%%%%%%%%%%%%%%%%%%%%

\subsection{Log-oscillations}\label{logos}

In the case of real fractional order, we have excluded the case $\a=0$ because the measure becomes degenerate \cite[section 2.6]{frc1},
\be
v_0(x)=\de(x)\,,
\ee
and spacetime reduces to a structureless point. When the action is equipped with the measure \Eq{kcom2}, however, Pointland is no longer trivial. Setting $\a=0$ in eq.~\Eq{kcom1}, one obtains
\be
v_{0,\om}(x) = \de(x)+\frac{2c_\om}{x}
\left[R_\Gamma(\rmi\om)\cos\left(\om\ln \frac{x}{\ell_\infty}\right)+I_\Gamma(\rmi\om)\sin\left(\om\ln \frac{x}{\ell_\infty}\right)\right]\,.\label{ka0}
\ee
%\be
%\vr_{\a=0}(x) = c_0+\frac{2}{\om}\left[R_\Gamma(\rmi\om)\sin\left(\om\ln\frac{x}{\ell_\infty}\right)-I_\Gamma(\rmi\om)\cos\left(\om\ln\frac{x}{\ell_\infty}\right)\right]\,.\label{ka0}
%\ee
The first term would yield an integration constant (the singular measure of the real-$\a$ case) and can be ignored, but the rest has a genuine dependence on the coordinate. 

The behaviour of eqs.~\Eq{kcom1} and \Eq{kcom2} is depicted in figure \ref{fig1}. For $\om\neq 0$, and for any $\a\geq 0$, the weight $v_{\a,\om}$ is periodic with increasing period. Due to the power-law pre-factor, the amplitude decreases for $0\leq\a<1$ and is constant for $\a=1$. On the other hand, the measure $\vr_{\a,\om}$ is log-periodic with increasing period and constant amplitude for $\a=0$ and increasing amplitude for $\a>0$ (inclusive $\a=1$). The amplitudes are magnified by the choice of a large coefficient $c_\om=1$. In realistic fractals, however, $c_\om$ is very small and oscillations reduce to tiny ripples around the average (see, e.g., \cite{Akk1}).

\FIGURE{
\includegraphics[width=7.3cm]{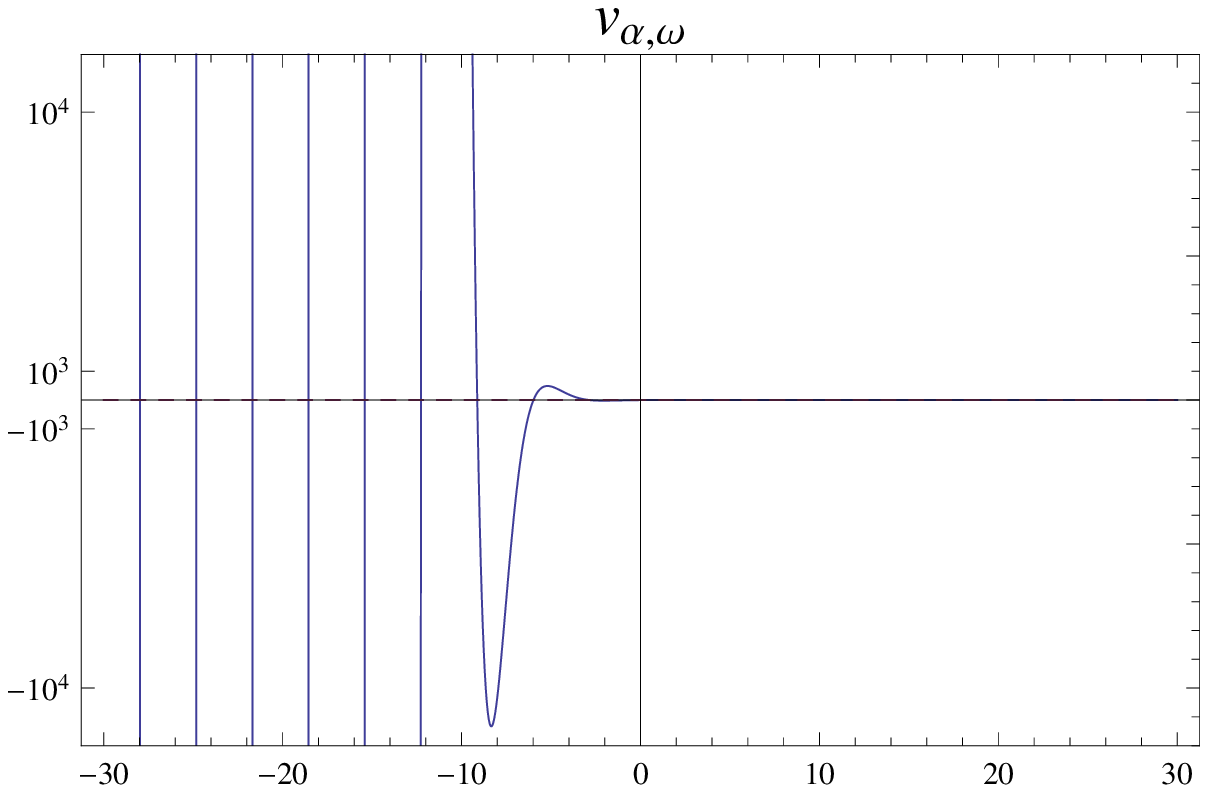}
\includegraphics[width=7.3cm]{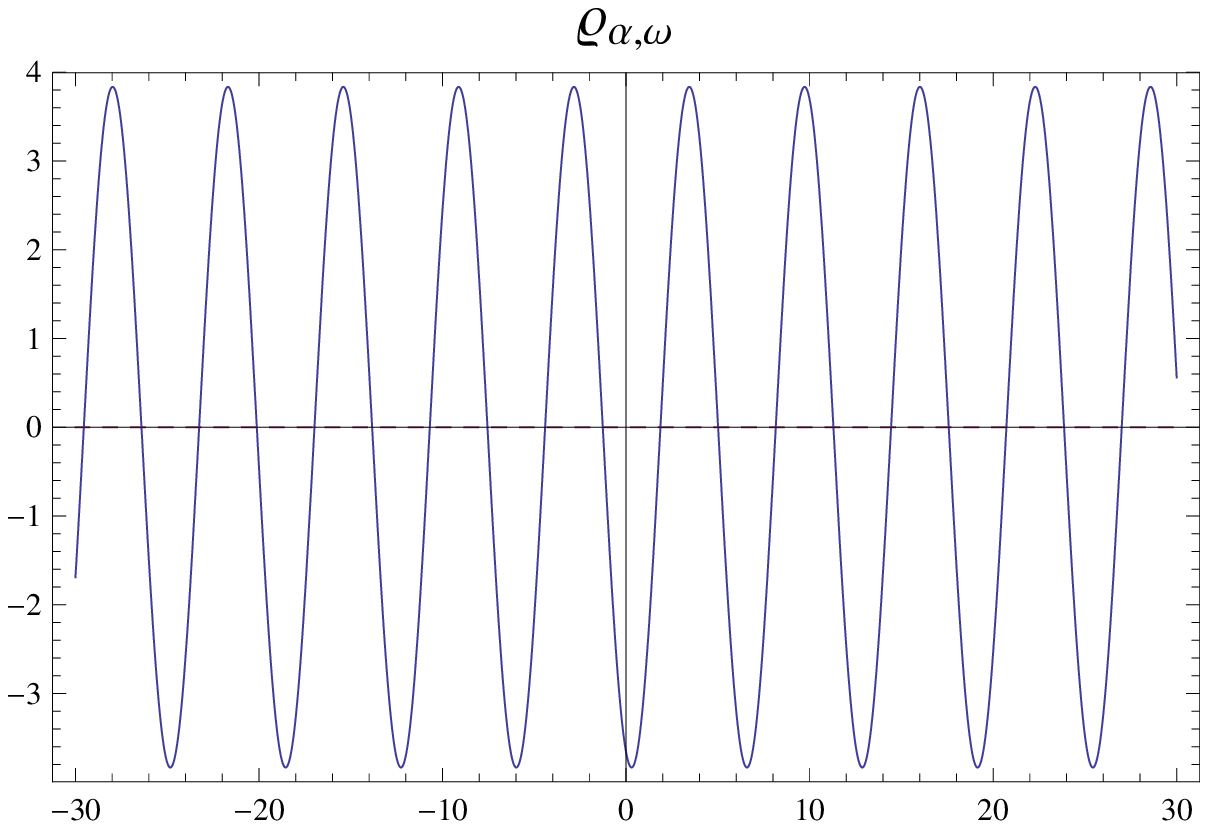}
\includegraphics[width=7.3cm]{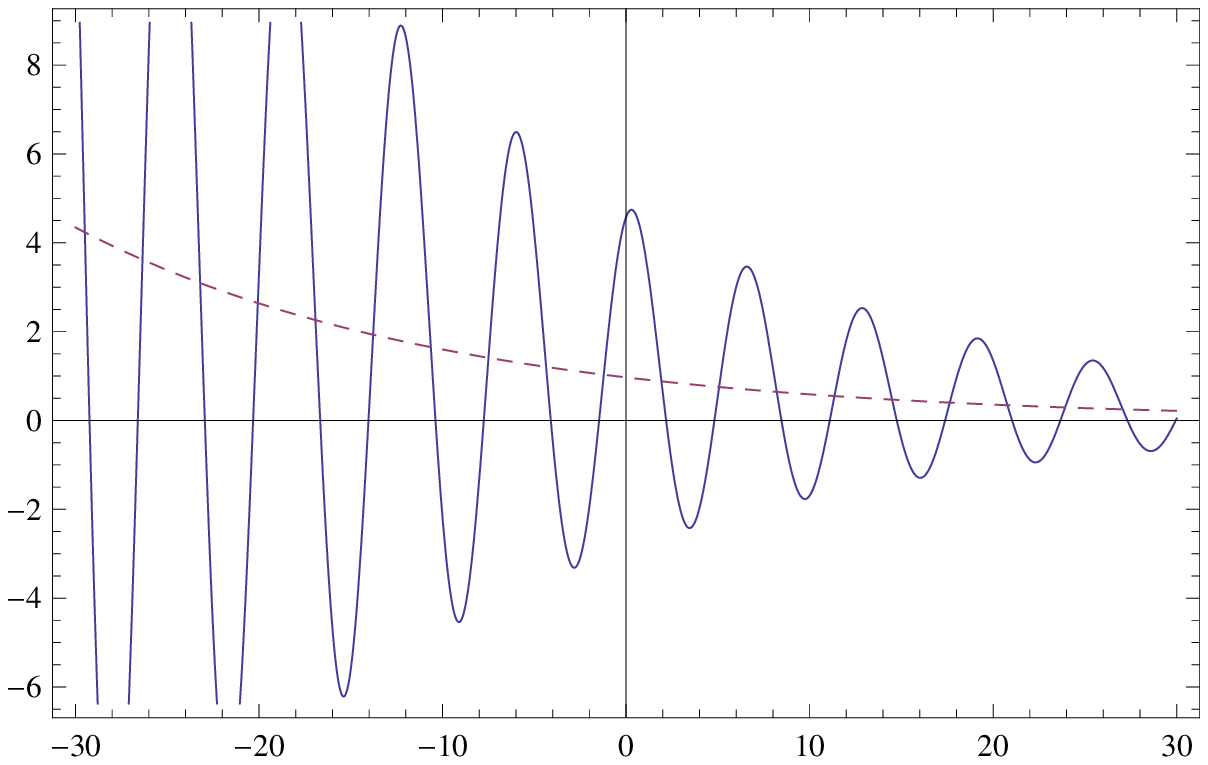}
\includegraphics[width=7.3cm]{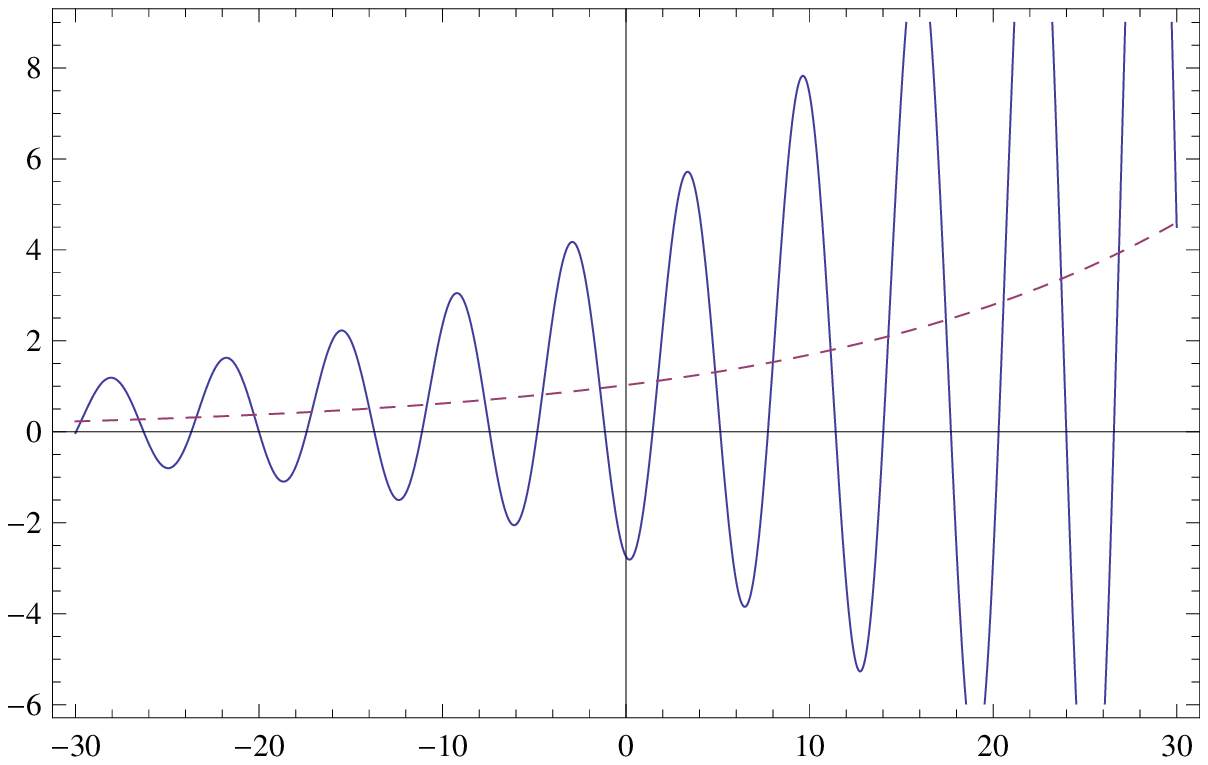}
\includegraphics[width=7.3cm]{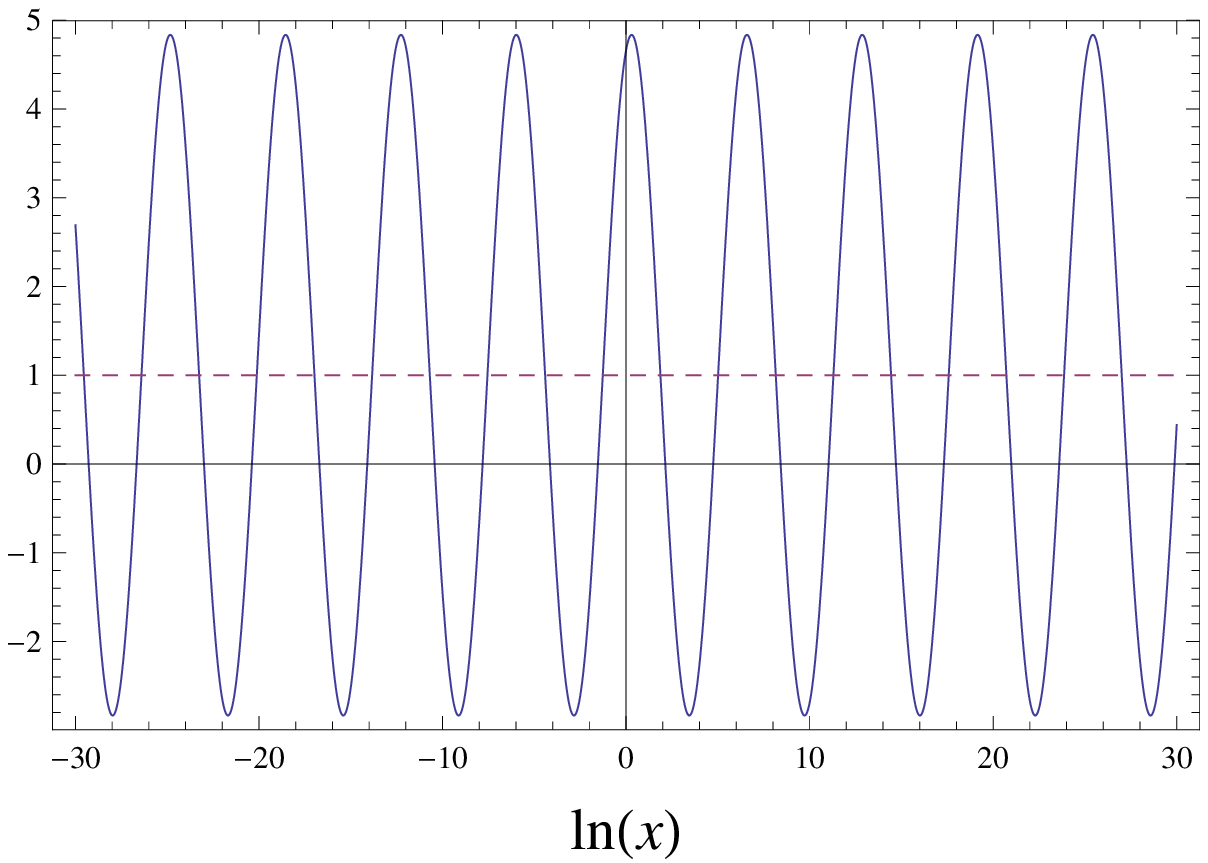}
\includegraphics[width=7.3cm]{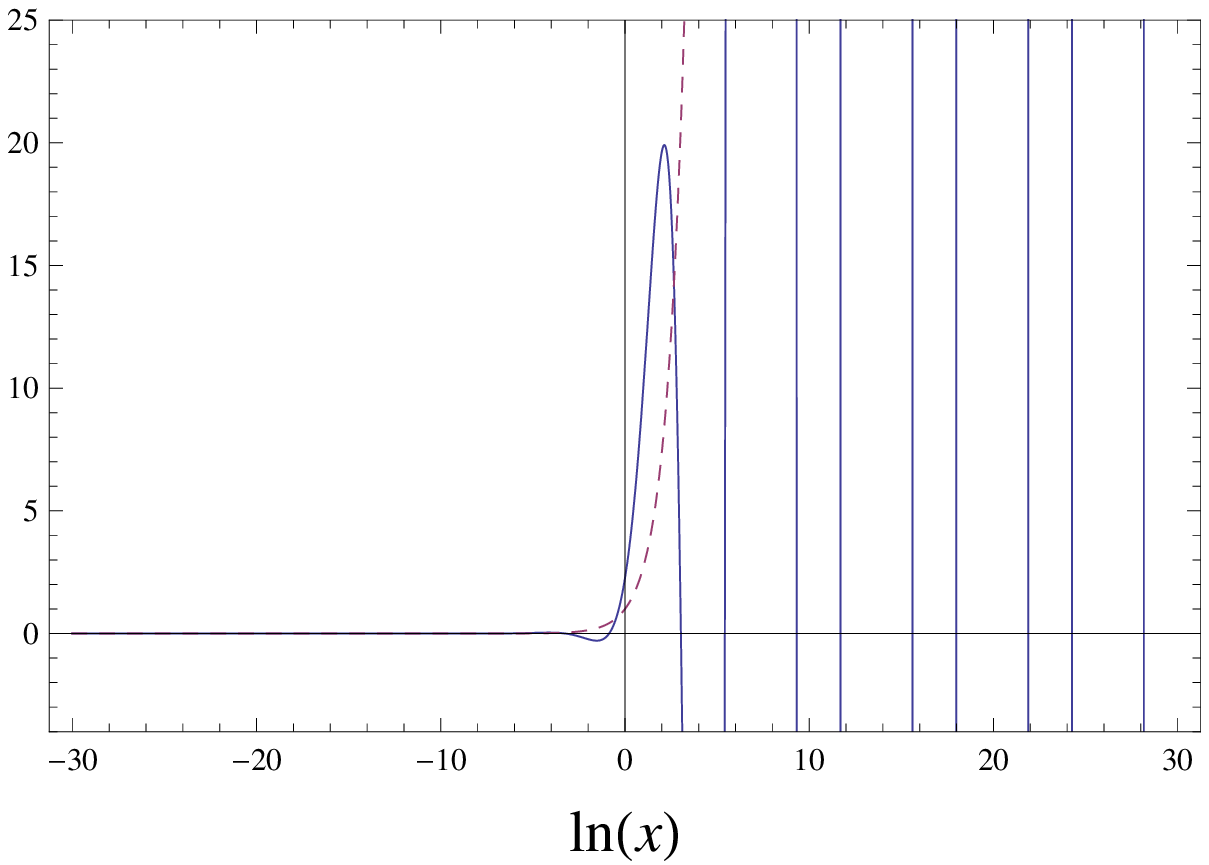}\caption{\label{fig1}
The measure weight $v_{\a,\om}$ (eq.~\Eq{kcom1}, left-side plots) and the associated measure $\vr_{\a,\om}$ (eq.~\Eq{kcom2}, right-side plots) for $c_\om=1$, a fixed value of the frequency (here, $\om=1$), and for $\a=0$ (top row), $0<\a<1$ (center row, with $\a=0.95$ for $v_{\a,\om}$ and $\a=0.05$ for $\vr_{\a,\om}$), and $\a=1$ (bottom row). The dashed curves are the averages $v_\a=\langle v_{\a,\om}\rangle$ and $\vr_\a=\langle\vr_{\a,\om}\rangle$. In the case $\a=0$, the singular term in the measure has been dropped.}}

It is important to stress a striking difference with respect to real-order multi-fractional spacetime. There, like in many other models of high-energy physics, one defines a characteristic scale $\ell_*$ (compare with section \ref{mul2}) distinguishing an exotic regime $\ell/\ell_*\gg 1$ from a classical regime $\ell/\ell_*\ll 1$, depending on the observational scale $\ell$. Here, on the other hand, these regimes are achieved in a subtler way. The characteristic scale $\ell_\om$, determined by the fundamental scale $\ell_\infty$, is the ever-present period of the oscillations, and what yields the ``classical'' result is not an analytic expansion in a small quantum parameter, but a spacetime averaging procedure. 

Lifting the assumption $\Psi=0$ does not present any difficulty. Writing $\bar\om:=\om+\Psi$, one has to replace $\om$ with $\bar\om$ everywhere (including the characteristic scale $\ell_{\bar\om}$) except in the arguments of $R_\Gamma$ and $I_\Gamma$. The measure is no longer even in $\om$, but one can still consider positive frequencies. The only non-trivial difference is in the asymptotic limit \Eq{asiom0}, where the trigonometric functions would survive with log-period $2\pi/\Psi$. Then, it is no longer true that the limit $\om\to 0$ yields a measure proportional to its average. This is because the scale $\ell_{\bar\om}$ does not vanish and the continuum limit is not recovered: eq.~\Eq{ellom} is replaced by
\be\label{ellomps}
\ell_{\bar\om}\to \ell_\Psi=\ell_\infty\exp\left(\frac{2\pi}{\Psi}\right)\qquad {\rm as}\qquad \om\to 0^+\,.
\ee
Thus, the interplay of the phase $\Psi$ and the zero mode can further enrich the physical properties of the model.

The transient nature of the discrete-to-continuum transition in simplicial quantum gravity models could be captured by complex fractional models. In fact, the lesson from fractal and chaos theory is that, while the large-Laplace-momentum approximation and operators of real fractional order are apt to describe ``static'' fractal configurations, log-periodic systems described by complex fractionality also include transient phenomena unobservable in the first case.

%%%%%%%%%%%%%%%%%%%%%%%%%%%%%%%%%%%%%%%%%%%%%%%%%%%%%%%%%%%%%%%%%%%%%%%%%%%%%%%%%%%%%%%%%%%%%%%%%%%%%%%%%%%%%%%%%%%%%%%%%

\subsection{Dimensions}\label{osdi}

The parameter $\a$ modulates the position and the height of the peaks, but it cannot remove them, even in the $\a=n$ cases corresponding to Pointland ($\a=0$) and smooth space ($\a=1$). This reflects the deep relation between genuine discrete fractals constructed via contracting maps and their continuum approximation. At higher orders in the harmonic expansion, the concept of Hausdorff dimension becomes ambiguous. If one used eq.~\Eq{opdeH} with the measure \Eq{kcom2}, the volume $\cV^{(D)}(\de)$ of a $D$-ball of radius $\de$ would strongly depend on $\de$: even tiny variations of the radius would lead to great differences in the output value $\cV^{(D)}(\de)$. This value could even be negative, and would not monotonically increase with the radius. In general, the volume would not be a power law, $\cV^{(D)}(\de)\not\sim \de^{\dh}$.

To define the Hausdorff dimension meaningfully, one has to take the average of the full measure over a period. Then, $\dh$ is simply determined, at all scales, by the scaling law of $\vr_\a=\langle\vr_{\a,\om}\rangle$. This is the leading order of the approximation of a highly non-trivial fractal measure via an effective fractional measure. In other words, the correct operational definition of $\dh$ is not \Eq{opdeH} but
\be\label{opdeH2}
\boxd{\dh := \lim_{\de\to 0}\frac{\ln \langle\cV^{(D)}(\de)\rangle}{\ln \de}\,,}
\ee
where $\cV^{(D)}(\de)$ is calculated with the measure \Eq{kcom2}. Similarly, the spectral dimension is the exponent of the leading term in the heat kernel expansion \Eq{genKf}. In the multi-fractional complex case, it is defined through the return probability \Eq{kkkin} with the measure $\vr_\a$ replaced by the full oscillatory measure in momentum space, and
\be\label{spedi2}
\boxd{\ds : = -2\frac{\rmd\ln \langle\cP(\s)\rangle}{\rmd\ln\s}\,.}
\ee
This definition replaces eq.~\Eq{spedi}. Again, using eq.~\Eq{spedi} would lead to problematic results. The greater the amplitude of the oscillations, the less meaningful the concept of spectral dimension would be. One would encounter situations where oscillations are so large that neither $\dh$ nor $\ds$ make sense any longer (a regime with geometry, yet with a very ``bad'' one), or where $\ds$ temporarily becomes greater than $\dh$ (a non-fractal regime) or $D$. 

To conclude, the Hausdorff and spectral dimensions are the same as in multi-fractional spacetimes, because their correct definition entails the average of, respectively, the volume and the heat kernel over a log-period.

%%%%%%%%%%%%%%%%%%%%%%%%%%%%%%%%%%%%%%%%%%%%%%%%%%%%%%%%%%%%%%%%%%%%%%%%%%%%%%%%%%%%%%%%%%%%%%%%%%%%%%%%%%%%%%%%%%%%%%%%%%%%%%%%%%%%%%%%%%%%%%%%%%%%%%%%%%%%%%%%%%%%%%%%%%%%%%%%%%%%%%%%%%%%%%%%

\section{Discussion and research agenda}\label{ra}

In this paper, we completed the construction of a flat spacetime endowed with certain properties typical of fractals. While in \cite{fra1}--%,fra2,
\cite{fra3} we outlined the motivations for doing so and described some results with general exotic integration measures, in \cite{frc1} we focussed on fractional measures and a space with Euclidean signature and fixed dimensionality. Here we extended that set-up to Lorentzian signature, changing dimensionality (multi-fractal geometry) and complex fractional order. A scalar field theory was given as an example of how fractal geometry deeply modifies the ultraviolet structure of a field theory, eventually softening or removing UV divergences. For simplicity, gravity was not included in the picture.

On one hand, fractional field theories can be regarded as effective theories, i.e., approximations in the continuum of a microscopic theory either with genuinely fractal properties or with a genuine (but non-fractal) dimensional flow. In this respect, the present model would propose itself as a tool to describe effective physics in some regimes stemming from fundamental models known to display dimensional flow, such as spin foams and others mentioned in the introduction. Then, one would not be interested in the renormalization properties of a given fractional action. In this case, predictions of fractional models should be associated with features of a given full theory, and there would arise the theoretical goal to obtain fractional dynamics as an emergent phenomenon. On the other hand, fractional theories may be also regarded as fundamental and unrelated to other proposals (fractal or non-fractal), in which case one should take care of their UV finiteness.

This last section is divided in three parts. In the first, we describe the hierarchy of scales implicit in the measure and the geometric regimes characterizing spacetime at different resolutions. In the second and third parts, open issues and applications to quantum gravity and non-commutative geometry are discussed.

%%%%%%%%%%%%%%%%%%%%%%%%%%%%%%%%%%%%%%%%%%%%%%%%%%%%%%%%%%%%%%%%%%%%%%%%%%%%%%%%%%%%%%%%%%%%%%%%

\subsection{From continuum to discrete geometry}\label{hier}

Gathering all the information obtained from the spacetime measure, we can summarize the emergent physical picture as follows \cite{fra4}. In complex self-conjugate fractional models, there exists a hierarchy of scales, one fundamental and the others characteristic, 
\be
\ell_\infty<\{\ell_\om\}<\ell_*\,.
\ee
In the simple model with only one frequency, three scales divide six different regimes non-perturbatively. We proceed from large to small scales and write expressions in one topological dimension, identifying the coordinate $x$ with the resolution $\ell$.
\begin{enumerate}
\item \emph{Classical regime.} At spacetime scales larger than a characteristic scale $\ell_*$, ordinary Euclidean/Minkowski geometry and ordinary field theory are recovered. The measure along a given direction is
\bs\be
\vr(x)\sim \langle\vr_{1,\om}(x)\rangle=x\,,\qquad \ell\gg\ell_*\,.
\ee
The number of dimensions can be theoretically constrained to be four.
\item \emph{Multi-fractional regime.} At mesoscopic spacetime scales around $\ell_*$, one obtains a model where one can construct field theories on a multi-fractal geometry. The scaling of volumes is anomalous and changes with the resolution. The renormalization group properties of field theory on these spacetimes are improved in the ultraviolet. The measure is given by
\be
\vr(x)\sim \sum_\a\langle\vr_{\a,\om}(x)\rangle\,,\qquad \ell\sim\ell_*\,.
\ee
A natural norm for the space exists if the fractional parameter $\a$ is comprised between $1/2$ (at the smallest scales) and $1$ (large scales).
\item\emph{Two-dimensional regime.} At microscopic scales much larger than a log-period $\ell_\om$ but smaller than $\ell_*$, spacetime is effectively two-dimensional with fractional geometry given by the measure
\be\label{2dme}
\vr(x)\sim \langle\vr_{\frac12,\om}(x)\rangle\sim x^{1/2}\,,\qquad   \ell_\om\ll\ell\lesssim\ell_*\,.
\ee
\item \emph{Oscillatory transient regime.} In the ultra-microscopic regime $\ell\sim\ell_\om$, geometrical concepts such as dimension and volumes make sense only in average (over a log-period), discrete symmetries make their appearance, and despite the continuous embedding the notion of continuous spacetime begins to blur. In this respect, scenarios with self-conjugate measures are non-perturbative, intrinsically quantum models of spacetime. The measure is
\be
\vr(x)\sim \sum_\a\vr_{\a,\om}(x)\,,\qquad \ell_\infty<\ell\lesssim\ell_\om\,.
\ee
\item \emph{Boundary-effect regime.} Here geometry is still given by a fractional continuum with discrete scale invariance, but boundary effects become important: this happens when the argument $x/\ell_\infty$ in the oscillatory part of the measure is of order unity. We associate this regime with boundary effects because, according to section \ref{boun}, a small $x$ expansion corresponds to getting close to the terminal at $x=0$. Expanding eq.~\Eq{kcom2} around the point $x/\ell_\infty=1$, we have
\be\nonumber
\vr_{\a,\om}(x) =\left[\frac{1}{\Gamma(\a)}+2R_\Gamma(\a+\rmi\om)\right]\ln \frac{x}{\ell_\infty}+O\left(\frac{x}{\ell_\infty}\right)\,,
\ee
so that, dropping immaterial constant terms, the measure becomes
\be\label{noncor}
\vr(x)\sim \ln x\,,\qquad \ell\sim\ell_\infty\,.
\ee
This is not the same as taking the limit $\om\to\infty$, which is not well defined. Notice that all information on the fractional structure of the measure has been absorbed in a finite normalization constant. We will comment later on the relation of this result with non-commutative geometry.
\item \emph{Fractal regime.} Finally, at scales $\ell<\ell_\infty$, the physics is governed by discrete symmetries and the continuum approximation breaks down. From the perspective of fractional spacetime, there is no longer a distinction between ambient space and its boundary ($x\sim 0$), meaning that the neighborhood of \emph{any} point $x$ will contain $x=0$. This description is clearly inadequate, since the support of the measure is highly disconnected (as in dust-type fractals), and the tools of fractional calculus must be abandoned definitely:
\be\label{frare}
\vr(x)=\,?\,,\qquad \ell<\ell_\infty\,.
\ee\es
\end{enumerate}
Forfeiting some of the above stages, one can even devise scenarios with a shorter hierarchy:
\begin{itemize}
\item\emph{Real-order multi-fractional models.} These were the main object of \cite{frc1} and the first four sections of this paper. The fractional integration order is real and one distinguishes between a classical and a multi-fractional regime, via the fundamental length $\ell_*$. In the deep ultraviolet, spacetime is in the continuous two-dimensional fractional regime \Eq{2dme}.
\item\emph{Pure complex models.} Setting $\a=1$, we obtain a complementary scenario with just two scales, $\ell_\om$ and $\ell_\infty$. After the oscillatory transient regime, the average over a log-period immediately yields the classical result, with no multi-fractional structure in between.
\end{itemize}
Both these possibilities have drawbacks. In fractal geometry, one expects both anomalous scaling for averaged measures and oscillatory structures. The symmetry structure of the real-order multi-fractional model is less rich and it does not distinguish different regimes in the UV. Thus, one would loose a number of interesting connections between some models of quantum spacetime and quantum gravity, as we shall comment later. So, the three-scale scenario of the complex multi-fractional theory, albeit more complicated, is more complete and, perhaps, better motivated than its reductions. 

%%%%%%%%%%%%%%%%%%%%%%%%%%%%%%%%%%%%%%%%%%%%%%%%%%%%%%%%%%%%%%%%%%%%%%%%%%%%%%%%%%%%%%%%%%%%%%%%

\subsection{Open issues}\label{opis}

\subsubsection{Fractal regime}

Fractional operators capture many features of fractals, but not all. In particular, they are not complete mathematical realizations of fractals and of diffusion on fractals. The features of a genuinely fractal background in the ultra-microscopic regime symbolized by eq.~\Eq{frare} would eventually deviate from those predicted in the continuum fractional approximation. For instance, the relation between spectral and Hausdorff dimension strongly depends on the class of fractals considered, and it quite often happens that $\dh\neq\ds$. While in fractal geometry this property is a consequence of the definition of the set, in fractional theories it is replicated only in spacetimes with anomalous diffusion \cite{frc1}. In fractional spaces, the type of diffusion depends on two choices: that of the operator $\cD_\s^\b$, entailing a certain degree of arbitrariness, and the choice of Laplacian, which is tightly related to the construction of an invertible and unitary transform in momentum space. The fractional Bessel transform \Eq{fst1} \cite{frc3} has the desired requisites and selects eq.~\Eq{ck2} as the natural Laplacian. It is to be seen whether an invertible transform also exists such that its basis functions are eigenfunctions of a fractional Laplacian of order $2\g$. Reflecting upon the interconnected issues of the form of the diffusion equation, the choice of Laplacian, and the existence of a well-defined transform will hopefully improve our understanding of the transition between fractal and fractional regimes. There do exist fractals for which $\ds=\dh$ (e.g., random walks \cite{AO} and diamond fractals \cite{Akk1,Met04}) but, even then, real-order fractional models miss the log-oscillations of the heat kernel trace. Inclusion of complex modes does improve the approximation and accounts for the oscillations. Further exploring the interrelation between complex fractional models and general results in fractal spectral theory will benefit our understanding of Fourier analysis in both frameworks \cite{frc1,Akk2}.

In alternative to what done here, one can try a bottom-up approach and attempt a brute-force construction on a genuine fractal. Models similar to \cite{Ey89a,Ey89b} can be useful to probe a deterministic-fractal regime, but the technical challenges involved therein have partially stalled progress in that direction. Nevertheless, this task may be now within our capabilities, at least in simple scenarios, thanks to the advances in fractal geometry \cite{Kig01,Str06}.\footnote{In the most general situation, the underlying texture of spacetime may be not just a fractal, but a field of fractals tiling the embedding space \cite{HaK2}. For the time being, there is no phenomenological reason to consider these configurations.}

\subsubsection{Scale hierarchy and the Planck length}

The relation between the scales $\ell_*$, $\ell_\om$, $\ell_\infty$ and the Planck scale $\lp$ in quantum gravity approaches deserves some attention. $\ell_\om$ is a scale derived from $\ell_\infty$ via a frequency relation, and it acquires infinite multiplicity when considering measures with an infinite number of Laurent modes. Therefore, it can be considered as characteristic but not fundamental in any obvious sense. This reduces the comparison of $\lp$ to $\ell_*$ and $\ell_\infty$.

Regarded as the numerical constant \Eq{plen}, the ``Planck scale'' remains a remote concept, since we have not introduced gravity in the picture. Also regarding $\lp$ as a symbol for a fundamental scale, the issue is rather undecided. On one hand, from the point of view of quantum gravity at large, dimension $\dh\sim 2$ is directly associated with the Planck scale (eq.~\Eq{cint}), thus preferring the identification of $\ell_*$ with $\lp$. On the other hand, it is natural to identify 
\be\label{lpinf}
\lp = \ell_\infty\,,
\ee
both in the perspective of discrete approaches to quantum gravity (because at $\ell_\infty$ the continuum picture breaks down, while at $\ell_*$ it does not) and in that of non-commutative spacetimes (see section \ref{ncft} and \cite{ACOS}). Equation \Eq{lpinf} is of interest also for the following reason. 

In fractal spectral theory on self-similar fractals, the period $\ln\la_\om$ of the oscillations is not an independent parameter, but it is determined by the geometry and the harmonic structure of the set \cite{Kig01} (see also \cite[section 5.1]{frc1} for a sketchy introduction). Consider a self-similar fractal with measure \Eq{ssim}. Instead of defining the fractal via eq.~\Eq{prefr}, one can adopt a formally identical equation where the similarities $\cS_i$ are replaced by injections maps $f_i(x)=r_i x+\dots$, where the coefficients $r_i$ are called resistance ratios and they are determined by the transformation property, in a subcopy $i$ of the fractal, of the Laplacian $\cK$ under the mapping $f_i$, $\cK[f_i(x)]=r_i^{-1}\cK(x)$. Summing over all the copies with the appropriate weight, one gets $\cK(x)=\sum_i g_i r_i\cK[f_i(x)]=\sum_i \g_i^2\cK[f_i(x)]$, where
\be\label{gai}
\g_i:=\sqrt{r_i g_i}\,.
\ee
The harmonic structure is characterized by the probabilities $g_i$ of the self-similar measure and the resistance ratios $r_i$, via the combination \Eq{gai}. For the simplest self-similar fractals, all $\g_i$ are equal. Now, it turns out that
the log-period of the heat kernel trace associated with these sets is, actually, $\ln\g_i$ \cite{KiL}. So, we can identify $\la_\om$ with $\g_i$, and recognize that the scale hierarchy of fractional spacetimes would depend, in a realistic fractal scenario, on the geometric and harmonic structure of the underlying set. In particular, one can contemplate the possibility to obtain the Planck length purely from symmetry and geometry. 

\subsubsection{Quantum fractional theories}\label{quantu}

Dimensional flow was only conjectured in previous papers on fractal spacetimes, due to the difficulty in understanding the dimensional properties of these models even at a fixed time (here, fixed $\a$). In the introduction, we emphasized the role of $d_{\rm H,S}=2$ in quantum gravity theories; although we have not considered curved fractional scenarios, we see the key principle in action for a scalar, via eq.~\Eq{scap}. Thus inspired, we built a theory with the aim to obtain a two-dimensional geometry at very small scales.  In \cite{fra1,fra2} it was postulated that fractal field theories with ordinary Laplacian flow from a non-trivial UV fixed point at $\dh=2$ (corresponding to what we called critical point) to an IR fixed point which has $\dh=4$ for obvious phenomenological reasons. Thanks to the requirement of triangle inequality, we have considerably restricted and better motivated, with respect to \cite{fra1,fra2}, the properties of this flow. On one hand, isotropic fractional geometries with Hausdorff dimension smaller than 2 can be realized only in embedding spaces with $D=3$, which are excluded empirically. On the other hand, if one assumes that the flow stops at $\dh=2$, then it cannot end at dimension greater than $D=4$. Embedding spacetimes with $D\geq 5$ cannot reach a two-dimensional UV fixed point. Therefore, if we exclude Kaluza--Klein scenarios, the macroscopic dimensionality of spacetime is deeply related to the microscopic geometry. Only a full renormalization-group analysis \cite{Col84}--%,Car96,BB,BTW,Pol01,
\cite{Ros10} will determine the existence and characteristics of UV fixed points and the perturbative renormalizability of multi-fractional models. The results obtained here should allow us to begin such a detailed study.

We have also considered the issue of parity and time reversal in classical fractional actions (sections \ref{emca} and \ref{frass}). The notion of fractional conjugation \cite{frc1}, 
\be
\x\to \bx\,,\qquad \p^\a\to \bp^\a\,,
\ee
is realized simply by an axis reflection from the point of view of the embedding coordinates, acting on the space points and its boundary. Just as parity is a discrete operation rigorously implemented pointwise on a local frame in ordinary curved backgrounds, in the general case the conjugation operation will have to be defined locally. Since we are not considering curved manifolds, there is a global frame for fractional conjugation analogous to the Fermi frame in ordinary Minkowski spacetime. Conjugation should also play a more involved role at the quantum level, where we expect left and right sectors to be mixed in the dynamics.

Other open issues about field theories on fractals should now be prone to direct investigation in the easier context of fractional spacetime. For instance, it is not known how to place spinors fields on fractals, a problem of interest also in low-lacunarity models \cite{Ey89a}. The existence of a vielbein formalism would allow one to ask the same question even on curved fractional backgrounds.

Finally, in quantum field theories where Lorentz invariance is broken, perturbative quantum effects can enhance classical deviations to unacceptable levels. As argued on general grounds \cite{CPSUV,CPS}, even if deviations from Lorentz invariance are classically negligible, one-loop corrections to the propagator of fields lead to violations several orders of magnitude larger than the tree-level estimate, unless the bare parameters of the model are fine tuned. For instance, this expectation \cite{Lif2} is indeed fulfilled for Lifshitz-type scalar models \cite{IRS}. One may ask whether a similar problem arises in fractional field theories, but the answer is not obvious to us. Systems with a discrete structure may also be protected from large Lorentz violations \cite{GRP}. Even if the theory is not integer-Lorentz invariant except in the infrared, the existence of other symmetries (discrete first, and then fractional Lorentz) must heavily affect loop calculations in a way quite different from a naive modification of dispersion relations.

\subsubsection{Fractional gravity}\label{gravi}

The extension of the present framework to fractional curved manifolds is obviously desirable for at least the reasons outlined in the introduction and in \cite{fra1,frc1,fra4}. A number of properties already displayed by flat fractional spacetimes should be inherited by fractional models of gravity and tighten potentially interesting relationships with theories of quantum gravity. We have seen that a discrete structure of spacetime emerges through a hierarchy of characteristic scales, and the tools of smooth integer geometry become progressively inadequate as the resolution increases. There is no obvious obstruction for this to occur also in curved scenarios, barring technical difficulties.

For instance, Tarasov \cite[section 9.2]{Tar12} noticed that failure of the Leibniz rule \Eq{leib} for differentiation in fractional calculus may make the definition of fractional differential forms on manifolds problematic. However, this obstacle should be eventually circumvented. In fact, the concepts of parallel transport and Lie derivative can be extended to fractional differential calculus \cite{Kri06}, and that of manifold exists already at the level of pure fractal geometry. Topological spaces such that the neighborhood of every point is homeomorphic to a neighborhood in a given fractal $\cF$ are called \emph{fractafolds} \cite{Str03,StT} (see also \cite{KuZ}). The existence of fractafolds is a positive indication that fractional Minkowski spacetime should admit a generalization to manifolds. As far as dynamics is concerned, recasting the Einstein equations in fractional fashion \cite{Mun10} is not sufficient by itself to define an action theory of gravitation on a fractional manifold, but it should not be difficult to derive them from the variational principle. 
 
\subsubsection{Inflation, big bang, cosmological observations}

A multi-fractional field theory of gravity can have applications in the history of the early universe and inflation, which might be explained via the alternative mechanism of dimensional flow. In ordinary geometry, the flatness and horizon problems are solved if the comoving Hubble horizon contracts while the universe expands at an accelerated rate, for a certain period just after the big bang. Inhomogeneities inside the horizon are pushed out and diluted by inflation. To realize inflation and explain the cosmological perturbation spectra, a slowly rolling scalar field is sufficient. In multi-fractional geometry, on the other hand, the effect of a contracting comoving horizon can be mimicked by a change in the spacetime geometry, without the need to invoke a matter field. The origin of perturbations is less obvious to guess, since it would also depend on the gauge symmetries of the model. This scenario will be investigated elsewhere.

Going to smaller scales and earlier times, the log-oscillatory pattern of the measure in complex self-conjugate models might shed some light into the big bang problem and, hopefully, the Belinsky--Khalatnikov--Lifshitz (BKL) conjecture \cite{BKL70,BKL82}. Near the classical big bang singularity, a number of cosmological backgrounds admit an anisotropic evolution. The simplest BKL model is characterized by three scale factors $a_i(t)$, $i=1,2,3$, which change through a sequence of epochs named after Kasner. Within a single Kasner epoch, $a_i\sim t^{p_i}$ and the powers $p_i$ obey certain conditions dictated by the dynamics. When spatial curvature effects are taken into account, one has the following behaviour. Going forward in time from the initial singularity, the universe passes through an infinite sequence of Kasner epochs where at least one direction is contracting ($p_i<0$ for one $i$), although the total spatial volume $\sqrt{-g}=a_1a_2a_3$ increases. Across transitions between one epoch and the next, the contracting direction exchanges roles with one of the other two, while the third evolves monotonically until after a certain number of epochs, constituting a ``Kasner era.'' The oscillation amplitudes and the duration of epochs increase during one era. The lengths of the Kasner eras are distributed according to stochastic laws which can be studied with the methods of chaos theory. 
In multi-fractional self-conjugate theory, we have an oscillatory regime coming neither from the metric tensor nor from the dynamics but, still, from geometry. The qualitative behaviour of the measure oscillations (increasing amplitude and period) resembles the one described above, and the interference of measure oscillations along different directions via an inhomogeneous metric might give rise to something like a BKL pattern near the putative singularity. Due to the physically different nature of the oscillatory mechanism and to the incompleteness of fractional theory in its present form, it is still premature to advance this parallelism any further.

Finally, observations of the galaxy distribution are generally regarded as compatible with the cosmological principle: at large scales (i.e., $\ell >10 h^{-1}\,{\rm Mpc}$) the universe is approximately homogeneous and isotropic. However, certain statistical analyses of the luminous matter distribution show anomalous correlations at scales $\ell \sim 10 \div 150 h^{-1}\,{\rm Mpc}$ \cite{JSGMP}--%,BYS,SVB1,SVBL,SVB2,Sy09a,
\cite{Sy09b} and, at scales up to $20 h^{-1}\,{\rm Mpc}$, matter distribution was claimed to have fractal dimension $\approx 2$ \cite{JSGMP}. (Even if anomalous scaling was present, by itself it would not mean that matter distribution is the approximation of a mathematical fractal. However, in practice, anomalous scaling and fractality are taken as synonyms.) At the time of writing, there is no consensus on these results, and there is ongoing debate about whether a fractal distribution clashes with the cosmological principle or not \cite{Sy09a}--%,Hog04,
\cite{YBK} (the upper scale at which homogeneity holds has been estimated to be as large as $260 h^{-1}\,{\rm Mpc}$ \cite{YBK}). Nevertheless, in case of confirmation it would be desirable to have a theoretical model explaining this deviation from the standard lore.\footnote{Motivated by earlier large-scale structure analyses invoking a fractal matter distribution, a mini-superspace model with varying dimension was proposed in \cite{MaN}.} The most natural explanation could be found in the details of the extreme non-linear regime of gravitational attraction at galaxy-cluster scales. However, a cosmological model of fractional gravity should predict deviations from standard geometry around scales corresponding to the Hubble horizon at and before matter-radiation equality (roughly corresponding to the scales where anomalous correlations have been allegedly detected), and it could be compared with the available data.

%%%%%%%%%%%%%%%%%%%%%%%%%%%%%%%%%%%%%%%%%%%%%%%%%%%%%%%%%%%%%%%%%%%%%%%%%%%%%%%%%%%%%%%%%%%%%%%%

\subsection{Other applications}\label{appli}

\subsubsection{Doubly special relativity}

Fractional theories are not the first models sporting a non-linear modification of Lorentz transformations. If the Planck
length is a fundamental building block of a theory of quantum gravitation, one may wonder what its significance is in the context of special relativity: If it is a minimal length smearing spacetime, should not different inertial observers measure the same value $\lp$? To do so, they should agree on an invariant energy/length scale, but ordinary Lorentz transformations act on \emph{any} length-type quantity. With quantum gravity in mind, one can then try to modify Lorentz transformations so that the Planck length be observer independent. It turns out that the new transformation rules are non-linear in the spacetime coordinates and they are parametrized by two invariants: the speed of light and $\lp$. Frameworks implementing the principle of Planck-length invariance are collectively called doubly special relativity (DSR) \cite{DSR1}--%,DSR2,Mag01,BAK,DSR3,LuN1,DSR4,
\cite{Mag02}. Global Lorentz invariance in the usual sense is only an accidental symmetry of Nature (as first conceived in \cite{Pav67,ChN}) in the classical limit $\lp\to 0$. 

If we compare DSR with fractional theories, we notice some common features. Lorentz symmetries are deformed and non-linear, and a fundamental length is present therein. We have not specified the parameter dependence of the fractional Lorentz matrices $\tilde\Lambda_\nu^\mu$ in eq.~\Eq{fpotra}, but they should naturally depend on $\ell_\infty$. In fact, as a fundamental building block of the measure, $\ell_\infty$ appears in the fractional coordinates
\be
\x^\mu_\om:=\vr_{\a,\om}(x^\mu)\,.
\ee
At any fixed $\a$ and $\om$, the measure can be written in terms of these $\x^\mu_\om$ (rather than the coordinates $\x^\mu=\langle \x^\mu_\om\rangle$), and is invariant under eq.~\Eq{fpotra}. However, differences between doubly special relativity and fractional models soon become apparent. In the first case only boosts are deformed, while here all transformations are affected. In both cases, Lorentz symmetry is an emergent, non-fundamental property of spacetime, but in fractional theory symmetries are more structured throughout the scale range, from the fundamental DSI of the oscillatory era to fractional Lorentz symmetries of the zero mode of the averaged measure, up to ordinary Lorentz invariance in the infrared. Also, while DSR is meant to describe a quantum world, fractional theories reach that conclusion from a different route. Fractional geometry automatically encodes quantum features, such as a discrete structure at ultra-microscopic scales (where DSI emerges) and a natural multi-scale structure equivalently prescribed by RG arguments and the lessons of multi-fractal geometry. 

Despite these differences, it may be possible that the DSR paradigm is implicit or natural in fractal scenarios. In fact, DSR can arise as a statistical phenomenon from a spacetime with a stochastic/fractal structure \cite{JiS}. Further study of fractional special relativity, which we have not completely formulated here, will be necessary to clarify this quite promising point.

\subsubsection{$\kappa$-Minkowski}\label{ncft}

In $D$ embedding dimensions with integer time direction ($\a_0=1$), the integration measure of the action in the boundary-effect regime \Eq{noncor} is
\be\label{kminm}
\rmd\vr_{\a,\om}\ \stackrel{\ell\sim\ell_\infty}{\sim}\ v_{\rm BE}(x)\,\rmd^D x:=\frac{\rmd^D x}{x_1\cdots x_{D-1}}\,.
\ee
In order to recover this measure in a real-order fractional action with measure $\vr_\a$, one should send $\a$ to zero in the measure weight $v_\a$ and formally keep the leading term in the expansion $v_\a(x^\mu)\sim \a/x^\mu$, getting
\be\label{fokmin}
\rmd\vr_\a\ \stackrel{\a\to0}{\sim}\ \a^{D-1} v_{\rm BE}(x)\,\rmd^D x\,.
\ee
However, in the real-order multi-fractional scenario we know that the case $\a=0$ has a pathological geometric structure (a zero-dimensional object): the correct limit in the sense of distributions is $v_\a\sim\delta(x)$, and the formal inverse-power limit \Eq{fokmin} is at least doubtful. One could simply absorb the vanishing constant $\a^{D-1}$ into a new normalization $c_0$ for the action, but the geometric considerations leading to eq.~\Eq{newra} would still lead to tension.

The inclusion of the log-oscillations is crucial not only to obtain a \emph{finite} result, but also to give it a completely different physical interpretation. The integration measure weight $v_{\rm BE}$ can be regarded either as a log-oscillating weight in the boundary-effect regime, or as an averaged measure $v_\a=\langle v_{\a,\om}\rangle$ in the limit $\a\to 0$. Only the first case is well defined. Equation \Eq{kminm} opens up the possibility to link together fractional and non-commutative manifolds, in particular $\kappa$-Minkowski. Imposition of a cyclicity-inducing measure in $\kappa$-Minkowski yields the condition \cite{AAAD}
\be
\p_i[x^i v_\kappa(x)]=0\,,\qquad \p_t v_\kappa(x)=0\,.
\ee
If one further imposes rotational symmetry, in $D-1$ dimensions one obtains $v_\kappa(x)=|\mathbf{x}|^{1-D}$ \cite{AAAD}. However, this is not motivated by strict physical arguments, so another solution is $v_\kappa(x)=v_{\rm BE}(x)$.

At this point, it is natural to conjecture a relation between $\kappa$-Minkowski and fractional models in the boundary-effect regime, with integer time direction. The fundamental scale of $\kappa$-Minkowski (what non-commutativists would call ``the Planck length'') is then identified with $\ell_\infty$. Many details should be considered. First, eq.~\Eq{dstar} implies that the critical point with lowest integer Hausdorff dimension in a $D=4$ ambient space with integer time has $\dh=3$. This may seem to be in agreement with the fact that the spectral dimension of $\kappa$-Minkowski is 3 \cite{Ben08}, but the result of \cite{Ben08} relies on the non-cyclic-invariant measure $v_\kappa=1$. One can also try to extend the field of investigation and ask whether a mapping exists between general fractional models with power-law measure $\vr_\a$ and non-commutative spacetimes with algebras more general than $\kappa$-Poincar\'e. It turns out that what we called fractional or geometric coordinates $q^\mu$ are nothing but coordinates obeying a canonical (Heisenberg) algebra onto which non-linear algebras can be mapped. A consequence of this identification is that fractional spacetimes are in one-to-one correspondence with a certain class of non-commutative spaces. All this is discussed in a separate publication \cite{ACOS}.

\subsubsection{Phases of quantum gravity}\label{fuzz}

Oscillatory measures can have concrete applications in quantum gravity approaches. Typically, a quantum spacetime endowed with a fundamental length $\ell_*$ undergoes a series of regimes characterized by different spectral and Hausdorff dimensions \cite{NN,MoN}. Taking the coarse-graining of spacetime texture as physical (i.e., considering a quantum non-commutative manifold), the initial condition for the diffusion equation of the heat kernel is no longer a Dirac distribution but a Gaussian of width $\sim\ell_*$ \cite{SSN}. In the absence of gravity, the scale-dependent spectral dimension of this object is \cite{MoN}
\be
\ds(\ell)=\frac{\ell^2}{\ell^2+\ell_*^2}D\,.
\ee
At large scales/diffusion times, $\ell\gg\ell_*$, the spectral dimension coincides with the topological dimension of spacetime. At scales near the fundamental scale, $\ell\sim\ell_*$, $\ds\sim D/2$, and if $D=4$ one has a two-dimensional regime. At ultra-microscopic scales, $\ds\sim 0$ and spacetime dissolves into a zero-dimensional configuration. This ``trans-Planckian'' fuzzy regime was recognized as difficult to interpret in \cite{MoN}. In the light of our results, we can provide a description of this regime and it becomes clear why we denoted the fundamental scale of \cite{MoN} with $\ell_*$. Assuming normal diffusion in fractional spacetime, the two-dimensional regime corresponds to the $\ds\sim 2$ special point in the multi-fractional/RG flow. In a real-order fractional theory, the flow can go beyond this point into a no-norm regime, down to the limit $\a\to 0$. In contrast, complex fractional theories allow us to pass the artificial barrier at $\ds\sim 2$ and probe spacetime at much smaller scales, until the continuum approximation breaks down. Estimates of the spectral dimension are not enough to describe the ultra-microscopic geometry of spacetime, because they are based on the average return probability, eq.~\Eq{spedi2}. Crucially, by looking only at $\ds$ one misses the oscillations of the spectral function we do expect in a truly fractal quantum gravity approach.

As already stated in the introduction and in the previous papers on the subject, starting from \cite{fra1}, the logic motivating Lebesgue--Stieltjes field theories is the following. (i) At first, one notices that most theories of quantum gravity display universal properties of dimensional flow. (ii) Then, one can ask whether and how these properties are related to the problem of UV finiteness. (iii) To this purpose, we constructed a geometry and field theory where dimensional flow is an intrinsic (not indirect) property, and where one can check the UV finitess explicitly. In this respect, fractional theories are independent from other models of quantum gravity and forcing two independent models to fit one another (say, fractional theories versus quantum Einstein gravity, or versus Ho\v{r}ava--Lifshitz, or versus CDT) may cause misleading interpretations of otherwise constructive independent insights. Nevertheless, it may be possible and very instructive to attempt to establish such a connection. With this attitude in mind, in the rest of the section we shall briefly mention other approaches.

The fractal properties of quantum gravity theories in $D$ dimensions have been explored in several contexts. At first, renormalizability of perturbative gravity at and near two topological dimensions drew much interest into $D=2+\epsilon$ models, with the hope to understand the $D=4$ case better \cite{GKT}--%,ChD,Wei79,KN,JJ,KKN,
\cite{AKNT}. Even in exactly two dimensions, there exists a rich fractal structure: short-scale spacetime fluctuations become self-similar, while matter field mass has anomalous scaling governed by an order parameter (the ``fractal dimension'' of spacetime) \cite{KN,KPZ}. The fractal geometry of two-dimensional quantum gravity is 
difficult to probe in the continuum formulation (Liouville theory \cite{KPZ}--%,Dav92,ABNRW,Dav88,DiK,DHK,DuS1,
\cite{DuS2}), but further progress was made in the discretized setting of $D=2$ dynamical triangulations \cite{ADJ}--%,KKSW,KKMW,AW,AJW1,AJW2,Amb96,Amb98
\cite{ABNRW}. These studies showed the emergence of a branched polymer phase in $D=2$ \cite{ADJ,KKMW}--%,AW,AJW1,
\cite{AJW2,ABNRW}, as well as in $D=4$ Euclidean simplicial gravity \cite{AJJK}--%,CKR,dBS,AmJ,
\cite{ETY} (dynamical triangulations \cite{Lol98}) and CDT \cite{AJL5,AGJJL} (see also \cite{AMM}). Branched polymers \cite{Dav92} are a special case of a class of random fractals including tree graphs and random combs. These objects have $\dh=2$, while their spectral dimension is bounded both from above and from below, according to the relation $2\dh/(\dh+1)\leq \ds\leq \dh$ \cite{JW}--%,CW,DD,DJW1,
\cite{DJW2} (see also \cite{JoS2}). The model of \cite{AGW} is one of the very few instances with scale-dependent spectral dimension; the scale hierarchy constructed there bears some resemblance with the multi-fractional flow of section \ref{mul3}.

Studying how fractional calculus encodes these random fractals would help to clarify whether a polymeric phase can correspond to scales $\ell_\infty<\ell<\ell_*$ or lower, where the continuum approximation breaks down. From the theoretical observations we have collected so far, it seems to us that fractional spacetime models might be able, by themselves, to provide interesting geometrical insights. Consider, in particular, the crumpled phase (phase B) and the two-dimensional branched-polymeric phase (phase A) of the CDT phase diagram in $3+1$ dimensions \cite{AGJJL}. In phase B, the concepts of dimension, metric and volume seem not to play a major role, and topology should become important \cite{AmJ}. In CDT, a phase-B universe has no extension in any direction,\footnote{Numerical simulations in $2+1$ dimensions are visually compatible with the presence of oscillations in the spectral dimension at small diffusion scales \cite{BeH}, but this is a discrete lattice effect regarded as unphysical \cite{Lol11}. For the sake of completeness, we mention that, in CDT, fractional derivatives have been used as a technical device to calculate the area-to-area propagator \cite{BLZ}.} while phase A is characterized by a connected structure with $\dh\sim 2$. A third phase (dubbed C) corresponds to a semi-classical four-dimensional universe. These regimes coexist in the phase plane and are not sequential in a history of quantum evolution. Yet, the regimes of fractional spacetimes can be related to phase-space regions as effective ``snapshot'' descriptions. A comparison with complex fractional models indicates that phases A and B of (causal) dynamical triangulations correspond, respectively, to the near-boundary regime (formally similar to the structureless limit $\a\to 0$ in the averaged measure, and where topology should become important) and to the oscillatory regime, stuck at the $\dh=2$ fixed point. %Then, the scale hierarchy of multi-fractional models suggests a natural flow of the system from phase B to phase A, to the semiclassical phase C.
 It is also clear why random combs \cite{DJW1,DJW2,AGW} cannot be associated with phase B: oscillations are washed away in random structures. % As remarked in \cite{SVW1}, the flow of the spectral dimension in Ho\v{r}ava--Lifshitz gravity and CDT is the same at scales where discreteness effects are negligible (phase A). However, we see now that there is little hope for Ho\v{r}ava--Lifshitz gravity to fit also the phase-B regime: In order for oscillations in $\ds$ to arise, there must be an explicit discrete symmetry at work in the diffusion equation.
  The hybrid character of fractional theories could make them promising candidates for data fits in causal dynamical triangulations.

Last, the fractional models considered in this paper and in \cite{frc1} are characterized by a Hausdorff dimension $\dh$ smaller than or equal to the topological dimension $D$ of embedding spacetime, $\dh\leq D$. However, there are other scenarios of quantum gravity where $\dh>D$. An instance is two-dimensional quantum gravity (Liouville theory and dynamical triangulations), where typically $\dh>2$ \cite{KKSW}--%,KKMW,AW,AJW1,AJW2,Amb96,
\cite{Amb98} (for instance, for pure gravity, zero central charge, $\dh=4$ \cite{KKMW,AW}). Is it possible to accommodate these geometries in multi-fractional theories? Technically, the answer is rather simple. While fractional models such that $\a_0,\a\leq 1$ describe spacetimes where $D\geq\dh$ (eq.\ \Eq{dgd}), those with fractional charges greater than 1 are characterized by a Hausdorff dimension greater than the topological dimension (including the $D=2$ case). The details of these models are also governed by the type of diffusion prescribed for the fractional manifold which, in turn, dictates the relation between $\ds$ and $\dh$. Just to give an example, consider a fractional spacetime in $D=2$, eq.\ \Eq{newspe}: $\ds=\b\dh$, $\dh=\a_0+\a$. To get a model with $\ds=2$ and $\dh=4$, mimicking the fractal properties of two-dimensional quantum gravity in vacuum, it is sufficient to set $\a_0+\a=4$ and $\b=1/2$ (fractional anomalous diffusion). Such configuration, if desired, could be regarded as asymptotic in a multi-fractional setting. This is not sufficient to establish a physical equivalence between these fractional models and a particular (and, in this case, toy-model) alternative theory of quantum gravity. An in-depth study of the relation between fractional theories as effective models and other theories of quantum gravity goes beyond the purpose of this work. However, we point out that more precise comparative analyses should be well within our capabilities.

%%%%%%%%%%%%%%%%%%%%%%%%%%%%%%%%%%%%%%%%%%%%%%%%%%%%%%%%%%%%%%%%%%%%%%%%%%%%%%%%%%%%%%%%%%%%%%%%
%%%%%%%%%%%%%%%%%%%%%%%%%%%%%%%%%%%%%%%%%%%%%%%%%%%%%%%%%%%%%%%%%%%%%%%%%%%%%%%%%%%%%%%%%%%%%%%%

\begin{acknowledgments}
The author thanks D.~Benedetti, G.~Dunne, S.~Gielen, J.~Magueijo, L.~Modesto, G.~Nardelli, D.~Oriti, J.~Th\"urigen, and S.~Vacaru for useful discussions.

%\bigskip

%\noindent {\bf Open Access.} This article is distributed under the terms of the Creative Commons
%Attribution Noncommercial License which permits any noncommercial use, distribution,
%and reproduction in any medium, provided the original author(s) and source are credited.
\end{acknowledgments}

%%%%%%%%%%%%%%%%%%%%%%%%%%%%%%%%%%%%%%%%%%%%%%%%%%%%%%%%%%%%%%%%%%%%%%%%%%%%%%%%%%%%%%%%%%%%%%%%%%%%%%%%%%%%%%%%%%%%%%%%%%%%%%%%%%%%%%%%%%%%%%%%%%%%%%%%%%%%%%%%%%%%%%%%%%%%%%%%%%%%%%%%%%%%%%%%%%%%%%%%%%%%%%%%%%%%%%%%%%%%%%%%%%%%%%%%%%%%%%%%%%

\end{document}